\documentclass[fleqn,usenatbib,onecolumn]{mnras} 
\usepackage{newtxtext,newtxmath}
\usepackage[T1]{fontenc}
\usepackage{ae,aecompl}

\usepackage{mathtools} 
\usepackage{amsfonts} 
\usepackage{mathrsfs} 
\usepackage{graphicx} 
\usepackage{subfig}
\usepackage{float} 
\usepackage{caption} 
\usepackage{bigstrut} 
\usepackage{tabularx} 
\usepackage{cancel} 
\usepackage{soul}
\usepackage{upgreek}
\usepackage{slashed} 
\usepackage{epstopdf} 
\newcommand\Lagr{\mathcal{L}} 

\newcommand\Lie{\pounds} 
\usepackage{breqn} 
\usepackage[subnum]{cases} 
\usepackage{xcolor} 
  {\color{red}}%
  {}
\allowdisplaybreaks

\title[MHD stability of magnetars I]{Magnetohydrodynamic stability of magnetars in the ultrastrong field regime I: The core}

\author[P. B. Rau and I. Wasserman]{
Peter B. Rau\thanks{E-mail: pbr44@cornell.edu}
and Ira Wasserman
\\
Cornell Center for Astrophysics and Planetary Science and Department of Astronomy, Cornell University, Ithaca, New York 14853, USA
}

\date{}

\pubyear{2020}

\begin{document}
\label{firstpage}
\pagerange{\pageref{firstpage}--\pageref{lastpage}}
\maketitle


\begin{abstract}
We study magnetohydrodynamic stability of neutron star core matter composed of neutrons, protons and leptons threaded by a magnetar-strength magnetic field $10^{14}$--$10^{17}$ G, where quantum electrodynamical effects and Landau quantization of fermions are important. Stability is determined using the Friedman--Schutz formalism for the canonical energy of fluid perturbations, which we calculate for a magnetizable fluid with $H\neq B$. Using this and the Euler--Heisenberg--Fermi--Dirac Lagrangian for a strongly magnetized fluid of Landau-quantized charged fermions, we calculate the local stability criteria for a neutron star core with a spherical axisymmetric geometry threaded by a toroidal field, accounting for magnetic and composition gradient buoyancy. We find that, for sufficiently strong fields $B\gtrsim10^{15}$ G, the magnetized fluid is unstable to a magnetosonic-type instability with growth times of order $10^{-3}$ s. The instability is triggered by sharp changes in the second-order field derivative of the Euler--Heisenberg--Fermi--Dirac Lagrangian which occur where additional Landau levels start being populated. These sharp changes are divergent at zero temperature, but are finite for nonzero temperature, so realistic neutron star core temperatures $5\times10^7$ K$<T<5\times10^8$ K are used. We conjecture that this mechanism could promote the formation of magnetic domains as predicted by Blandford and Hernquist (1982) and Suh and Mathews (2010).
\end{abstract}


\section{Introduction}

The problem of magnetohydrodynamic (MHD) stability is of great importance to the study of magnetars and compact stars in general, since their magnetic field configurations must be stable over timescales much longer than their dynamical timescales. Determining which types of field configuration are allowed by stability considerations is of particular interest following the discovery of a multipolar field configuration in a neutron star by the NICER experiment~\citep{Bilous2019,Riley2019}; the dipolar field model often assumed is clearly too simple. Understanding the field configuration of magnetars and how they could be destabilized is also of fundamental interest in helping to understand their emission mechanisms, including the theoretical explanation for soft gamma repeaters (SGRs) as caused by fractures in the magnetar crust~\citep{Thompson1995,Heyl2005} or field reconnection in the magnetosphere~\citep{Lyutikov2006}. Magnetars are also a leading candidate for the source of fast radio bursts (FRBs)~\citep{Popov2010,Lyubarsky2014,Beloborodov2017,Lu2018,Metzger2019,Lyubarsky2020}, and a recent detection by the CHIME radio telescope of an FRB originating from a magnetar in our galaxy~\citep{CHIMEFRB2020} has provided evidence that this could be the case. MHD instabilities within the star could power magnetic outbursts which deposit energy into the magnetar magnetosphere, which in turn powers the different proposed FRB mechanisms.

Though the complicated nature of MHD has made aspects of MHD stability intractable to analytic study, many useful results have been proven analytically and later confirmed numerically for the stability of stellar magnetic fields. A purely toroidal stellar field is known to be unstable along the axis of symmetry to sausage (e.g. interchange) and kink instabilities~\citep{Tayler1973}, while a purely poloidal field with closed field lines within the star is also unstable to sausage and kink instabilities where the field vanishes~\citep{Markey1973,Wright1973}.~\citet{Flowers1977} showed that a star with a purely poloidal field is unstable even if no field lines are closed within it.~\citet{Markey1973} and~\citet{Wright1973} both suggested that a mixed poloidal-toroidal field configuration could be stable, which was demonstrated numerically~\citep{Braithwaite2004,Braithwaite2006,Yoshida2006,Duez2010} for near equal-strength poloidal and toroidal fields. Simulations by~\citet{Braithwaite2009} showed that mixed toroidal-poloidal configurations can be stable with a much weaker poloidal component, which was confirmed analytically by~\citet{Akgun2013}. Stable stratification of a star, represented mathematically by a positive square Brunt--V\"{a}is\"{a}l\"{a} frequency, has been shown analytically~\citep{Tayler1973} and numerically~\citep{Braithwaite2006} to stabilize the magnetic field, and in general it allows for a greater variety of possible field configurations since the field no longer needs to be a solution of the Grad--Shafranov equation~\citep{Reisenegger2009}. Using numerical simulations,~\citet{Mitchell2015} found that different field configurations in barotropic stars always decayed away, giving further evidence to the idea that stable stratification must be included to obtain a stable field configuration. Stable stratification alone may not be sufficient to stabilize a fluid with a magnetic field that diminishes quickly enough with increasing height, which can be unstable to magnetic buoyancy~\citep{Parker1955,Acheson1979}. Additionally, as discussed by~\citet{Reisenegger2009}, the erosion of stable stratification by dissipative processes (weak decays and ambipolar diffusion in neutron stars) could lead to field rearrangement and perhaps instability.

Previous analyses of the stability of stellar magnetic fields have made the often reasonable assumption that the stellar medium through which the field is threaded is not magnetizable i.e. that $B=H$. This is clearly not the case for superfluid-superconducting neutron stars, nor is it true for extremely strong magnetic fields. The field strengths attained in magnetars are up to $10^{15}$ G at the surface and perhaps one to two orders of magnitude greater in the core~\citep{Mereghetti2015,Turolla2015,Kaspi2017,Uryu2019}. Since these fields exceed the quantum critical field $B_{\text{crit}}=m_{\text{e}}^2/e=4.4\times10^{13}$ G, quantum electrodynamic effects are relevant and could be important to magnetar stability. In the vacuum case, nonlinear electromagnetic effects are encoded in the Euler--Heisenberg Lagrangian~\citep{Heisenberg1936}. In matter, this must be supplemented with terms accounting for the interaction of fermions with the magnetic field. These additional Lagrangian terms have been computed for a charged Fermi gas at zero temperature $T$~\citep{Chodos1990} and at finite $T$~\citep{Elmfors1993,Persson1995}. In the presence of such strong fields, the charged fermions in a neutron star will undergo Landau quantization, which modifies the equation of state (EOS)~\citep{Lai1991,Broderick2000,Mao2003,Chamel2012,Sinha2013,Chamel2020a}, magnetization, and transport properties~\citep{Potekhin1999a,Potekhin2001,Potekhin2015} of the star. The effects of $B\lesssim 10^{18}$ G on the EOS are generally quite small at typical core densities, and $B\approx H$ to within a few percent. However, the analysis of MHD stability requires not only examining the first order partial derivatives of the magnetic free energy, but its second order derivatives with respect to $B$ and the density $\rho$. These derivatives have been studied in the context of magnetic domain formation in strong fields by~\citet{Blandford1982} and~\citet{Suh2010}, but as far as the authors are aware, the implications of strong-field quantum mechanical effects on \textit{MHD stability} have not been examined in the literature.

In this paper, we study magnetohydrodynamic stability including nonlinear, non-vacuum electromagnetism appropriate for magnetar-strength magnetic fields. We employ the canonical energy approach to stability analysis~\citep{Bernstein1958} and in particular follow closely the coordinate basis version of the nonrelativistic fluid perturbation theory expounded by~\citet{Friedman1978}. We extend the nonrelativistic magnetohydrodynamic perturbation theory of~\citet{Glampedakis2007} to $B\neq H$ to allow us to consider the effect of the medium and vacuum magnetization on MHD stability. We also consider non-barotropic EOS, and employ a Brunt--V\"{a}is\"{a}l\"{a} frequency accounting for both neutron-proton fraction buoyancy~\citep{Reisenegger1992} and leptonic buoyancy~\citep{Kantor2014,Passamonti2016,Yu2017,Rau2018}, in contrast to e.g.~\citet{Akgun2013}, who only included the neutron-proton fraction buoyancy in their analysis and were considering the global instability of specific axisymmetric fields. We consider the local stability of  strongly magnetized neutron star core fluid in a planar geometry, in which the effects on stability of nonlinear electromagnetism and an accurate buoyant force, including magnetic buoyancy, are more easily understood than in the spheroidal star case. After reviewing the electromagnetic Lagrangian and the relevant partial derivatives of it which determine the stability, we conclude by discussing its numerical application to the stability criterion for the planar fluid case. The background magnetic field in the neutron star crust obeys a fundamentally different constraint equation compared to ideal MHD~\citep{Cumming2004,Gourgouliatos2013}, and the stability analysis is fundamentally different~\citep{Lyutikov2013}; we thus leave this subject to a subsequent paper.

In Section ~\ref{sec:Perturbations} we introduce the canonical energy for $B\neq H$ MHD, with most details of the derivation left to Appendix~\ref{app:FluidPerturbationTheory}. Section~\ref{sec:StabilityAnalysis} derives the stability criteria using the canonical energy. In Section~\ref{sec:Thermodynamics} the pressure and energy density for strong fields and Landau quantized fermions are discussed in detail and the required thermodynamic derivatives are derived, and the background stellar model is described. Section~\ref{sec:Results} describes the numerical results for the stability criteria, and Section~\ref{sec:Conclusion} discusses their observational implications. Appendix~\ref{app:Thermodynamics} gives explicit expressions for thermodynamic partial derivatives used in evaluating the stability criteria. We work in Gaussian units and set $c=\hbar=1$. We also employ the Einstein summation convention using Latin letters as spatial indices $i=1,2,3$. The letter $a$ is reserved as a subscript to denote particle species $a=\textrm{n},\textrm{p},\textrm{e},\text{m}$, with sums over particle species always denoted explicitly.

\section{MHD equations and canonical energy}
\label{sec:Perturbations}

We consider a nonrotating neutron star core composed of neutrons $\text{n}$, protons $\text{p}$, electrons $\text{e}$ and, at sufficiently high densities, muons $\text{m}$. We assume the magnetic field is strong enough to destroy proton superconductivity, and ignore neutron superfluidity. We also assume that the collisional coupling time between the fluids is short and that they all comove-- this would not be the case if the neutrons were a superfluid, but the charged fluids are expected to comove generally. We work in the ideal magnetohydrodynamic approximation of zero net electric charge density $\rho_e=0$ and infinite conductivity. 

In nonrelativistic magnetohydrodynamics for comoving fluids and zero temperature, the appropriate independent thermodynamic variables to work with are the mass density $\rho$, magnetic field $B^i$, and species fractions $Y_a$. The total internal energy density $u$ for this fluid is thus
\begin{equation}
u(B,\rho,Y_a)=\frac{B^2}{8\pi}+u_M(B,\rho,Y_a)=\frac{B^2}{8\pi}-\Lagr_{\text{EH}}(B)+u_{\text{mat}}(B,\rho,Y_a),
\label{eq:InternalEnergyDensity}
\end{equation}
where $B=\sqrt{g_{ij}B^iB^j}$ is the magnitude of the magnetic field and $g_{ij}$ is the (Euclidean, flat space) metric tensor. $u_M$ includes the (negative) standard vacuum Euler--Heisenberg Lagrangian $-\Lagr_{\text{EH}}$, plus the matter contribution to the energy density $u_{\text{mat}}$: combined these are responsible for the magnetization. The exact form of $u_{\text{mat}}$ will be discussed in Section~\ref{sec:Thermodynamics}: microscopically, it will also depend on (mean) meson fields responsible for nuclear interactions. Eq.~(\ref{eq:InternalEnergyDensity}) and standard thermodynamic relations imply that the magnetic $H$-field is~\citep{Landau1960}
\begin{equation}
H_i=4\pi\left.\frac{\partial u}{\partial B^i}\right|_{\rho,Y_a}=B_i+4\pi\left.\frac{\partial u_M}{\partial B^i}\right|_{\rho,Y_a}=B_i-4\pi\frac{\partial\Lagr_{\text{EH}}}{\partial B^i}+4\pi\left.\frac{\partial u_{\text{mat}}}{\partial B^i}\right|_{\rho,Y_a}.
\label{eq:HDef}
\end{equation}

For a magnetizable medium in ideal MHD, the Euler equation takes the form
\begin{align}
\rho\left(\partial_t v_i+(v^j\nabla_j)v_i\right)+\nabla_iP+\rho\nabla_i\Phi=\nabla^jT^B_{ij},
\label{eq:EulerEquation}
\end{align}
where $v^i$ is the common fluid velocity, $P$ is the total matter pressure, including meson fields responsible for nuclear interactions (see Section~\ref{sec:Thermodynamics} for further details of this) and magnetic-field dependence, and $\Phi$ is the gravitational potential. $T^B_{ij}$ is the magnetic stress tensor for a magnetizable medium~\citep{Easson1977}:
\begin{equation}
T^B_{ij}=\left[\frac{1}{8\pi}B^2-\Lagr_{\text{EH}}-\frac{1}{4\pi}B^kH_k\right]g_{ij}+\frac{1}{4\pi}H_iB_j.
\label{eq:MagneticStressTensor}
\end{equation}
$T^B_{ij}$ is symmetric since $u_M$ only depends on $B$ and hence $B^i$ and $H^i$ are aligned. We thus have
\begin{equation}
\frac{\partial u_M}{\partial B^k}=\hat{B}_k\frac{\partial u_M}{\partial B}, \rightarrow H=4\pi\left.\frac{\partial u}{\partial B}\right|_{\rho,Y_a},
\label{eq:duMdBi}
\end{equation}
where $\hat{B}^i=B^i/B$. The total mass density of the fluid is
\begin{equation}
\rho=\sum_{a=\text{n,p,e,m}}m_an_a\approx m_{\text{N}}n_b
\label{eq:MassDensity}
\end{equation}
where $m_a$ and $n_a$ are the mass per particle and number density of species $a$, $m_{\text{N}}=938.92$ MeV is the average nucleon mass and $n_{\text{b}}=n_{\text{n}}+n_{\text{p}}$ is the total baryon number density. We could hence replace $\rho$ with $n_{\text{b}}$ as an independent variable. The species fractions $Y_a$ can be represented using two quantities: the proton fraction (of total baryons) $Y$ and the electron fraction (of total leptons) $f$, defined by
\begin{equation}
Y=\frac{n_{\text{p}}}{n_{\text{b}}}, \qquad f=\frac{n_{\text{e}}}{n_{\text{p}}}=\frac{n_{\text{e}}}{n_{\text{e}}+n_{\text{m}}}.
\end{equation}

For simplicity, we have assumed zero temperature, and hence zero entropy, in the equations of motion. The $T=0$ limit is a very good approximation for the neutron star core where the fermion chemical potentials $\mu_a$ satisfy $\mu_a\gg k_BT$. An exception is made for certain second-order partial derivatives of $u_{\text{mat}}$, which we discuss in Section~\ref{sec:Thermodynamics}. In these terms, we treat the temperature as a fixed parameter and do not concern ourselves with the dynamics of the entropy.

We now derive the canonical energy for a magnetizable fluid, which is used to study the fluid's MHD stability. We follow the definitions of the Lagrangian perturbations of fluid quantities of~\citet{Friedman1978}, which has antecedents in~\citet{Taub1969},~\citet{Carter1973},~\citet{Friedman1975} and~\citet{Bardeen1977}, and which has been applied to MHD by~\citet{Glampedakis2007}. In these definitions, we work in a coordinate basis and hence the components of (contravariant) vectors and covariant vectors are in general distinct. The perturbation theory is reviewed in Appendix~\ref{app:FluidPerturbationTheory} and then applied to Eq.~(\ref{eq:EulerEquation}) and used to compute the canonical energy of the perturbations.

Using results from Appendix~\ref{app:FluidPerturbationTheory}, the full expression for the canonical energy $E_c[\xi]$ for perturbation with Lagrangian displacement field $\xi^i$ is shown to be 
\begin{align}
E_c[\xi]=\frac{1}{2}\int \text{d}V\Bigg[{}&\rho|\partial_t\xi^i|^2-\rho|v^j\nabla_j\xi^i|^2+\text{Re}\left[(\xi^i)^*\xi^j\right]\nabla_i\nabla_j\left(P+\mathcal{P}_B\right)
+\left(\gamma P+\frac{B^2}{4\pi}+2\rho B\frac{\partial^2 u_M}{\partial \rho\partial B}+B^2\frac{\partial^2 u_M}{\partial B^2}\right)|\nabla_j\xi^j|^2
\nonumber
\\
{}&+\frac{BH}{4\pi}\left|\hat{B}^j\nabla_j\xi^i\right|^2
+\left(\frac{\partial^2 u_M}{\partial B^2}-\frac{1}{B}\frac{\partial u_M}{\partial B}\right)\left| B^i\hat{B}^j\nabla_j\xi_i \right|^2 
+ 2\left(\frac{1}{B}\frac{\partial u_M}{\partial B}-\frac{\partial^2 u_M}{\partial B^2}-\frac{\rho}{B}\frac{\partial^2 u_M}{\partial\rho\partial B}\right)B^kB^j\text{Re}[\nabla_k(\xi_j)^*\nabla_i\xi^i]
\nonumber
\\
{}&+\frac{1}{2\pi}B^j\text{Re}\left[\nabla_i(\xi^i)^*(\xi^k\nabla_jH_k-H_k\nabla_j\xi^k)\right]-2\rho\left(v^k\nabla_kv_i+\nabla_i\Phi\right)\text{Re}\left[(\xi^i)^*\nabla_j\xi^j\right]+\rho\text{Re}\left[(\xi^i)^*\xi^j\right]\nabla_i\nabla_j\Phi
\nonumber
\\
{}&-\frac{1}{4\pi G}|\nabla^i\delta_{\xi}\Phi|^2\Bigg]+\frac{1}{2}\int_x\text{d}V\left(\frac{g_{ij}}{4\pi}-\frac{\partial^2 u_M^x}{\partial B^i_x\partial B^j_x}\right)\delta_{\xi^*}B^i_x\delta_{\xi}B^j_x
\nonumber
\\
{}&\hspace{-12mm}+\frac{1}{2}\oint \text{d}S\Bigg[|\hat{n}_i\xi^i|^2\hat{n}^j\nabla_j\left(\left\langle P+\mathcal{P}_B\right\rangle-\frac{1}{4\pi}\hat{n}_{k}B^k\hat{n}^{\ell}\langle H_{\ell}\rangle\right)-\frac{1}{4\pi}\hat{n}_iB^i\text{Re}\left[(\xi^j)^*\xi^k\right]\nabla_kH^x_j
+\frac{1}{4\pi}\hat{n}_i\nabla_j\left(\text{Re}\left[\xi^i(\xi^j)^*\right]\hat{n}_{\ell}B^{\ell}\hat{n}^k\langle H_k\rangle\right)
\nonumber
\\
{}&\hspace{2mm}-\frac{1}{2\pi}\hat{n}_iB^i\langle H_j\rangle\text{Re}\left[(\xi^k)^*\nabla_k\xi^j\right]+\frac{1}{4\pi G}\nabla_i\left(\delta_{\xi}\Phi\nabla^i\delta_{\eta}\Phi\right)\Bigg].
\label{eq:CanonicalEnergy}
\end{align}
In this expression $\hat{n}^i$ is the unit normal to the surface enclosing the fluid, $\gamma$ is the adiabatic index defined as
\begin{equation}
\gamma\equiv\frac{\rho}{P}\left.\frac{\partial P}{\partial\rho}\right|_{s,Y_a,B}=\frac{\rho}{P}c_s^2,
\label{eq:AdiabaticIndex}
\end{equation}
for adiabatic, constant magnetic field sound speed $c_s$, and $\delta_{\xi}$ is the Eulerian perturbation associated with the Lagrangian displacement field $\xi^i$. $^*$ indicates complex conjugation, and we have defined the magnetic pressure $\mathcal{P}_B$
\begin{equation}
\mathcal{P}_B\equiv-\frac{B^2}{8\pi}+\Lagr_{\text{EH}}+\frac{1}{4\pi}HB=\frac{B^2}{8\pi}+\Lagr_{\text{EH}}+B\frac{\partial u_M}{\partial B},
\label{eq:MagneticPressure}
\end{equation}
which reduces to $B^2/(8\pi)$ in the vacuum, low field limit as expected. The first volume integral in Eq.~(\ref{eq:CanonicalEnergy}) is over the fluid i.e. the star, the second is over the exterior of the fluid, denoted with a superscript or subscript $x$, and the final term is the surface term. In the surface term, angled brackets denote the difference between the quantity outside and inside the star. We neglect to explicitly denote that $B$, $\rho$, $Y$ and/or $f$ are held constant in the partial derivatives of $u_M$. The surface term in $E_c$, which is unimportant for our purposes, was calculated by making the simplifying assumption that the exterior of the star is vacuum threaded by a magnetic field, as opposed to a realistic plasma-filled magnetosphere-- this was also the choice made by~\citet{Glampedakis2007}. In the exterior region we still allow $H\neq B$, but only the vacuum Euler–-Heisenberg Lagrangian is included in $u_M$ there.

In the rest of the paper, we employ the following abbreviations for partial derivatives of $u_M$:
\begin{align}
u_B\equiv\frac{\partial u_M}{\partial B}, \quad u_{\rho B}\equiv\frac{\partial^2 u_M}{\partial\rho\partial B}, \quad u_{BB}\equiv\frac{\partial^2 u_M}{\partial B^2}, \quad u_{BY}\equiv\frac{\partial^2 u_M}{\partial B\partial Y}, \quad u_{Bf}\equiv\frac{\partial^2 u_M}{\partial B\partial f}.
\label{eq:LagrangianPartialAbbreviations}
\end{align}

\section{Stability analysis}
\label{sec:StabilityAnalysis}

Evaluating the integral Eq.~(\ref{eq:CanonicalEnergy}) requires either solving for the global structure of the star and its quasinormal modes, or guessing trial canonical data $\xi^i$, though the result in the latter case could be far from correct if the initial guess is not reasonable. Instead, in the remainder of this paper we investigate the local stability. We consider the example system of an infinite slab of fluid extending in the $x$--$y$ plane and stratified in the $z$-direction, with magnetic field varying in $z$ and directed in the plane of the fluid. This allows us to derive the stability criterion for magnetic buoyancy. This system is used to approximate the local stability in a star where the $z$-direction replaces the radial direction, the magnetic field is toroidal, and where the curvature orthogonal to this direction is ignored.

The problem of magnetic buoyancy has been extensively investigated in the $B=H$ case~\citep{Parker1955,Newcomb1961,Gough1966,Schubert1968,Acheson1979}. We follow the canonical energy approach of~\citet{Newcomb1961} and~\citet{Gough1966} to derive the magnetic buoyancy stability criteria for the $B\neq H$ case. The background magnetic field and gravitational field are
\begin{equation}
B^i=B(z)\delta^i_x, \qquad H^i=H(z)\delta^i_x, \qquad-\nabla_i\Phi=g_i=-g\delta_i^z.
\end{equation}
The pressure $P=P(z)$, density $\rho=\rho(z)$ and species fractions $Y=Y(z)$ and $f(z)$ are only functions of $z$. The only nonzero component of the background Euler equation is
\begin{equation}
\frac{\text{d}}{\text{d}z}\left(P+\mathcal{P}_B\right)+\rho g=0.
\label{eq:MBEuler}
\end{equation}
We drop all surface and exterior vacuum terms, work in the Cowling approximation and assume zero background velocity $v^i=0$. Define twice the canonical energy per unit mass $\mathcal{E}_c$ via
\begin{equation}
E_c\equiv\frac{1}{2}\int \text{d}V\rho\mathcal{E}_c.
\label{eq:CanonicalEnergyPerUnitMass}
\end{equation}
Since there are no dissipation mechanisms included in this analysis, $E_c$ is a conserved quantity, and hence the unstable modes must have $E_c=0$~\citep{Friedman1978}. Since the kinetic energy term $\propto|\partial_t\xi^i|^2$ is clearly positive definite, unstable modes are possible if the remainder of $E_c$ is negative. We henceforth drop the uninteresting kinetic energy term from $\mathcal{E}_c$. Since we study local stability here, we look for possible instability by examining the conditions for which $\mathcal{E}_c<0$. Taking $\xi^i\propto\exp(i(k_xx+k_yy))$, Eq.~(\ref{eq:CanonicalEnergy}) and~(\ref{eq:CanonicalEnergyPerUnitMass}) give
\begin{align}
\mathcal{E}_c[\xi]={}&\left(v_{\text{A}}^2k_x^2-g\frac{\text{d}\ln\rho}{\text{d}z}\right)\left|\xi^{z}\right|^2
+\left(c_s^2+\frac{B^2}{4\pi\rho}+\frac{1}{\rho}\left[B^2u_{BB}+2B\rho u_{\rho B}\right]\right)|\partial_j\xi^j|^2
+v_{\text{A}}^2k_x^2\left|\xi^y\right|^2
+v^2_Bk_x^2\left|\xi^x\right|^2
\nonumber
\\
{}&-2\left(Bu_{\rho B}+v^2_B\right)k_x\text{Re}[\partial_i(\xi^i)^*i\xi^x]
-2g\text{Re}\left[\partial_j(\xi^j)^*\xi^z\right],
\label{eq:CanonicalEnergyMB}
\end{align}
where we used Eq.~(\ref{eq:MBEuler}) and have defined
\begin{align}
v_{\text{A}}^2\equiv{}&\frac{BH}{4\pi\rho},
\label{eq:AlfvenV}
\\
v_{B}^2\equiv{}&\frac{B^2}{4\pi\rho}\left(1+4\pi u_{BB}\right).
\label{eq:VelocityB}
\end{align}
$v_{\text{A}}$ is the Alfv\'{e}n velocity, and in the limit $H=B$, $v_A=v_B$.  We now consider the separate cases of no undulations in the direction of the magnetic field $k_x=0$, and with undulations permitted in this direction $k_x\neq0$. 

\subsection{Case 1: $k_x=0$}
\label{sec:KxZeroStabilityCriteria}

Setting $k_x=0$ in Eq.~(\ref{eq:CanonicalEnergyMB}) and expanding out the divergence $\partial_j\xi^j=\partial_z\xi^z+ik_y\xi^y$ gives
\begin{align}
\mathcal{E}_c[\xi]={}&V^2\left|\partial_z\xi^z+ik_y\xi^y\right|^2-g\frac{\text{d}\ln\rho}{\text{d}z}\left|\xi^{z}\right|^2
-2g\text{Re}\left[\partial_z\xi^z(\xi^z)^*\right]-2gk_y\text{Re}\left[i\xi^y(\xi^z)^*\right],
\label{eq:CanonicalEnergyMBKxZero}
\end{align}
where we define a commonly-used velocity squared
\begin{equation}
V^2\equiv c_s^2+v_B^2+2Bu_{\rho B}.
\label{eq:V2}
\end{equation}
Completing the square to move all dependence on $\partial_z\xi^z$ into a single positive definite term, we obtain
\begin{align}
\mathcal{E}_c[\xi]={}&V^2\left|\partial_z\xi^z+ik_y\xi^y-\frac{g\xi^z}{V^2}\right|^2
-g\left[\frac{\text{d}\ln\rho}{\text{d}z}+\frac{g}{V^2}\right]\left|\xi^{z}\right|^2.
\label{eq:CanonicalEnergyMBKxZero2}
\end{align}
For the canonical energy to be positive, it is sufficient that $V^2>0$ and the prefactor of $|\xi^z|^2$ is positive definite i.e.
\begin{align}
-\frac{\text{d}\ln\rho}{\text{d}z}-\frac{g}{V^2}{}&>0.
\label{eq:kxZeroStabilityCriterion}
\end{align}
Using the definition of the Brunt--V\"{a}is\"{a}l\"{a} frequency motivated by Eq.~(\ref{eq:PressureEulerianPerturbation2})
\begin{equation}
N^2\equiv g\left(\frac{1}{\gamma}\frac{\text{d}\ln P}{\text{d}z}-\frac{\text{d}\ln\rho}{\text{d}z}-\frac{\zeta B}{\gamma P}\frac{\text{d}\ln B}{\text{d}z}\right)=\frac{g}{\rho c_s^2}\left(\frac{\partial P}{\partial Y}\frac{\text{d}Y}{\text{d}z}+\frac{\partial P}{\partial f}\frac{\text{d}f}{\text{d}z}\right),
\label{eq:BVFrequency}
\end{equation}
and Eq.~(\ref{eq:MBEuler},\ref{eq:VelocityB}), Eq.~(\ref{eq:kxZeroStabilityCriterion}) can be rewritten as
\begin{equation}
\text{SC1}\equiv\frac{c_s^2N^2}{g}+\frac{B^2}{4\pi\rho}\left(1+4\pi u_{BB}\right)\frac{\text{d}}{\text{d}z}\ln\left(\frac{B}{\rho}\right)-Bu_{\rho B}\frac{\text{d}\ln\rho}{\text{d}z}+\frac{B}{\rho}u_{BY}\frac{\text{d}Y}{\text{d}z}+\frac{B}{\rho}u_{Bf}\frac{\text{d}f}{\text{d}z}>0.
\label{eq:SC1}
\end{equation}
This is analogous to the criterion for stability against magnetic buoyancy derived in~\citet{Schubert1968} but generalized to include $B\neq H$. The second term on the right is the $B\neq H$ analog of the $\text{d}/\text{d}z\ln(B/\rho)$ term in the usual $k_x=0$ magnetic buoyancy stability criterion (e.g. Eq.~(1.2) in~\citet{Acheson1979}). The species fraction dependence of $H$ also contributes new terms proportional to $\textrm{d}Y/\textrm{d}z$ and $\textrm{d}f/\textrm{d}z$-- this is analogous to the usual composition gradient Brunt--V\"{a}is\"{a}l\"{a} frequencies, whose effects are included in $N^2$. This motivates a redefinition of the 
Brunt--V\"{a}is\"{a}l\"{a} frequency to absorb these magnetic contributions:
\begin{equation}
\tilde{N}^2\equiv\frac{g}{\rho c_s^2}\left[\left(\frac{\partial P}{\partial Y}+Bu_{BY}\right)\frac{\text{d}Y}{\text{d}z}+\left(\frac{\partial P}{\partial f}+Bu_{Bf}\right)\frac{\text{d}f}{\text{d}z}\right].
\label{eq:MagneticBVFrequency}
\end{equation}
The two stability criterion, $V^2>0$ and SC1$>0$, are both sufficiency conditions; a fluid configuration may still be stable globally if either one of them is violated locally. 

\subsection{Case 2: $k_x$ \texorpdfstring{$\neq$}{not equals} $0$}
\label{sec:KxNotZeroStabilityCriteria}

Allowing $k_x\neq0$ in Eq.~(\ref{eq:CanonicalEnergyMB}) and expanding out $\partial_j\xi^j$ gives
\begin{align}
\mathcal{E}_c[\xi]={}&V^2|\partial_z\xi^z|^2+\left(v_{\text{A}}^2k_x^2-g\frac{d\ln\rho}{dz}\right)\left|\xi^{z}\right|^2
+\left(V^2k_y^2+v_{\text{A}}^2k_x^2\right)\left|\xi^y\right|^2
+c^2_sk_x^2\left|\xi^x\right|^2+2k_y\left(c_s^2+v_B^2\right)\text{Re}[\partial_z(\xi^z)^*i\xi^y]
\nonumber
\\
{}&+2k_xk_y\left(c_s^2+Bu_{\rho B}\right)\text{Re}[\xi^x(\xi^y)^*]+2k_x\left(c_s^2+Bu_{\rho B}\right)\text{Re}[\partial_z(\xi^z)^*i\xi^x]
-2gk_x\text{Re}\left[i\xi^x(\xi^z)^*\right]
-2gk_y\text{Re}\left[i\xi^y(\xi^z)^*\right]
\nonumber
\\
{}&-2g\text{Re}\left[\partial_z\xi^z(\xi^z)^*\right],
\label{eq:CanonicalEnergyMBKxNonzero1}
\end{align}
To simplify this somewhat, we can complete the square in an analogous manner to the $k_x=0$ case. Combining all terms depending on $\xi^x$ into a single positive definite contribution to $\mathcal{E}_c[\xi]$, we obtain
\begin{align}
\mathcal{E}_c[\xi]={}&c^2_s\left|ik_x\xi^x+\frac{(c_s^2+Bu_{\rho B})(\partial_z\xi^z+ik_y\xi^y)-g\xi^z}{c^2_s}\right|^2
+K_1|\partial_z\xi^z|^2
+(K_1k_y^2+v^2_{\text{A}}k_x^2)\left|\xi^y\right|^2
+(K_2+v_{\text{A}}^2k_x^2)\left|\xi^{z}\right|^2
\nonumber
\\
{}&+2L_1k_y\text{Re}[\partial_z(\xi^z)^*i\xi^y]
+2L_2\text{Re}\left[\partial_z\xi^z(\xi^z)^*\right]
+2L_2k_y\text{Re}\left[i\xi^y(\xi^z)^*\right].
\label{eq:CanonicalEnergyMBKxNonzero2}
\end{align}
where we define
\begin{align}
K_1\equiv{}&V^2-\frac{\left(c_s^2+Bu_{\rho B}\right)^2}{c^2_s},
\label{eq:K1}
\\
K_2\equiv{}&-g\frac{d\ln\rho}{dz}-\frac{g^2}{c_s^2},
\\
L_1\equiv{}&c_s^2+v_B^2-\frac{\left(c_s^2+Bu_{\rho B}\right)^2}{c_s^2}=K_1-2Bu_{\rho B},
\\
L_2\equiv{}&g\frac{Bu_{\rho B}}{c_s^2}.
\end{align}
Completing the square once again to combine the $\partial_z\xi^z$ terms outside the first term of Eq.~(\ref{eq:CanonicalEnergyMBKxNonzero2}) into a single term, we obtain
\begin{align}
\mathcal{E}_c[\xi]={}&c^2_s\left|ik_x\xi^x+\frac{(c_s^2+Bu_{\rho B})(\partial_z\xi^z+ik_y\xi^y)-g\xi^z}{c^2_s}\right|^2
+K_1\left|\partial_z\xi^z+\frac{L_1ik_y\xi^y+L_2\xi^z}{K_1}\right|^2
+\left[\left(K_1-\frac{L_1^2}{K_1}\right)k_y^2+v^2_{\text{A}}k_x^2\right]\left|\xi^y\right|^2
\nonumber
\\
{}&+\left[K_2-\frac{L_2^2}{K_1}+v_{\text{A}}^2k_x^2\right]\left|\xi^{z}\right|^2
+2k_yL_2\left(1-\frac{L_1}{K_1}\right)\text{Re}\left[i\xi^y(\xi^z)^*\right].
\label{eq:CanonicalEnergyMBKxNonzero3}
\end{align}
We see that one stability criterion is $c_s^2>0$. This could only possibly be violated by magnetic terms, and we show that this is not the case, at least in a neutron star core: $c_s^2$ is always positive there. We can now take the $k_x\rightarrow 0$ limit since the $v^2_{\text{A}}k_x^2$ contributions to the prefactors of $\left|\xi^y\right|^2$ and $\left|\xi^z\right|^2$ will only help to stabilize the system, and the $k_x$ contribution to the first term will always be positive as long as $c_s^2>0$. After taking this limit, if both $c_s^2>0$ and $K_1>0$, Eq.~(\ref{eq:CanonicalEnergyMBKxNonzero3}) consists of positive definite terms plus a quadratic form in $ik_y\xi^y$ and $\xi^z$. This quadratic form is positive definite as long as the following three criteria are satisfied:
\begin{subequations}
\begin{align}
\text{SC2}{}&\equiv K_1-\frac{L_1^2}{K_1}=4Bu_{\rho B}\left(1-\frac{Bu_{\rho B}}{K_1}\right)>0,
\label{eq:SC2}
\\
\text{SC3}{}&\equiv K_2-\frac{L_2^2}{K_1}>0,
\label{eq:SC3}
\\
\text{SC4}{}&\equiv K_1\left(1-\frac{L_1^2}{K_1^2}\right)\left(K_2-\frac{L_2^2}{K_1}\right)-L_2^2\left(1-\frac{L_1}{K_1}\right)^2>0.
\label{eq:SC4}
\end{align}
\end{subequations}
SC2 is a purely magnetohydrodynamic stability criterion since it is independent of the gravitational acceleration, while SC3 and SC4, like SC1, are associated with gravity and hence buoyancy. 

The $K_1>0$ stability condition can be rewritten as
\begin{equation}
K_1=\frac{B^2}{4\pi\rho}\left(1+4\pi u_{BB}\right)-\frac{B^2u^2_{\rho B}}{c_s^2}>0.
\label{eq:MPR1}
\end{equation}
This is identical to Eq.~(122) of~\citet{Akgun2008}, though note that we define the magnetic free energy in a different way here. The instability associated with this criterion not being met is the Muzikar--Pethick--Roberts (MPR) instability~\citep{Muzikar1981,Roberts1981}, a magnetosonic-type instability first derived in the context of type-II superconducting fluids. In such systems the instability acts to attract flux tubes together-- this suggests that, in the normal fluid case, it will concentrate magnetic field lines and create regions of higher and lower magnetic flux. Strictly speaking, the original MPR instability criterion is associated with the second term in Eq.~(\ref{eq:MPR1}) being larger in magnitude than the first term, while in this paper it will turn out that the first term is of greater interest as a potential source of instability. The connection of SC3 and SC4 to buoyancy can be made clear by rewriting $K_2$ similarly to Eq.~(\ref{eq:SC1}) using
\begin{equation}
-c_s^2\frac{d\ln\rho}{dz}-g=\frac{c_s^2\tilde{N}^2}{g}+\left(\frac{B^2}{4\pi\rho}+\frac{B^2}{\rho}u_{BB}+Bu_{\rho B}\right)\frac{d\ln B}{dz}-Bu_{\rho B}\frac{d}{dz}\ln\left(\frac{B}{\rho}\right),
\label{eq:kxNonzeroStabilityCriterionBuoyancy}
\end{equation}
which can be used where $K_2$ appears in Eq.~(\ref{eq:SC3}--\ref{eq:SC4}). This is the analog to the $k_x\neq 0$ magnetic buoyancy in Eq.~(1.4) of~\citet{Acheson1979}.
 
\section{Thermodynamics}
\label{sec:Thermodynamics}

The quantum mechanical effects of strong magnetic fields on the MHD are incorporated within an electromagnetic Lagrangian density computed in quantum electrodynamics (QED). In the vacuum case, this is the Euler--Heisenberg Lagrangian~\citep{Heisenberg1936} computed by integrating out the charged fermion species from the QED action. In a neutron star core at densities of order the nuclear saturation density $n_0=0.16$ fm$^{-3}$ populated by degenerate charged fermion species, a vacuum background cannot be assumed. The nonvacuum thermal background of fermions is incorporated by computing the effective action in an analogous manner to the usual Euler--Heisenberg Lagrangian but including a fermion chemical potential (and finite temperature if desired): we refer to this background as a Fermi--Dirac vacuum. The resulting Euler--Heisenberg--Fermi--Dirac Lagrangian was computed by~\citet{Elmfors1993}. The finite density corrections to the vacuum Euler--Heisenberg Lagrangian can be written as separate terms in the Lagrangian, and combined~\citep{Persson1995} with the zero-field fermion Lagrangian to give the usual Lagrangian (or pressure) of Landau-quantized fermions. It will be the Landau quantization that has the most important effect on the MHD stability as we later show. We ignore anomalous magnetic moments; the effect of the nucleon anomalous magnetic moments on stellar structure is not significant until $B\gtrsim10^{18}$ G~\citep{Broderick2000}, and we will only examine fields an order of magnitude below this.

To be consistent with a realistic stellar core and the specified particle content of our hydrodynamics, we employ an equation of state including neutrons, protons, electrons and muons in beta equilibrium and with interactions between the neutrons and protons. The proton-neutron interactions are important for obtaining realistic values for the $n_a$ and hence $Y$ and $f$ at a given chemical potential (e.g. muons appearing when the total number density is $\approx0.8$ times nuclear saturation density). We mostly work in the temperature $T=0$ limit, which is a very good approximation for a neutron star core for most of the star's life. An exception is made only for certain second-order partial derivatives of the pressure, which we discuss briefly in this section.

There is a straightforward way to generalize the pressure of Landau-quantized fermions to include proton-neutron interactions. This is to use the $\sigma\omega\rho$ nuclear mean field theory equation of state~(e.g. \citet{Walecka1995,Glendenning1997}), which has been generalized to the case of strong magnetic fields~\citep{Broderick2000,Mao2003,Sinha2013}. In effect, this is accomplished by replacing the mass and chemical potentials of the protons and neutrons in the non-interacting theory with their effective values $m_a^*$ and $\mu_a^*$ computed in the $\sigma\omega\rho$ model and simultaneously solving the self-consistency equation for the mean field value of the scalar meson field $\sigma$. The mass terms for the meson fields must also be added to the pressure, and we include cubic and quartic interactions for the $\sigma$ meson.

Following e.g.~\cite{Broderick2000}, the total matter pressure in the strong-field, $\sigma\omega\rho$ mean field model at zero temperature is
\begin{align}
P(B,\mu_a,\sigma,\omega_0,\rho_0^3)=\sum_a P_{\text{f},a}(B,\mu_a,\sigma,\omega_0,\rho_0)-\frac{1}{2}m_{\sigma}^2\sigma^2-\frac{1}{3}b_{\sigma}m_{\text{N}}(g_{\sigma}\sigma)^3-\frac{1}{4}c_{\sigma}(g_{\sigma}\sigma)^4+\frac{1}{2}m_{\omega}^2\omega_0^2+\frac{1}{2}m_{\rho}^2(\rho^3_0)^2,
\label{eq:Pressure}
\end{align}
where $P_{\text{f},a}(B,\mu_a,\sigma,\omega_0,\rho_0)$ is the fermion pressure for each species, $\mu_a$ are the (bare) chemical potentials, $\sigma$, $\omega_0$ and $\rho^3_0$ are the mean field values of the mesons~\footnote{The zeroth spacetime component of the $\omega$ meson and the zeroth spacetime component of the $I_3$ isospin component of the $\rho$ meson-- the ``3'' superscript on $\rho^3_0$ denotes the isospin component and not exponentiation.}, $g_{\sigma}$, $g_{\omega}$ and $g_{\rho}$ are the coupling constants between the baryons and the mesons, $m_{\omega}$, $m_{\rho}$ and $m_{\sigma}$ are the meson masses, and $b_{\sigma}$ and $c_{\sigma}$ are the self-coupling constants for the $\sigma$ meson. While in the full $\sigma\omega\rho$ model the $\rho$ meson is an isovector of meson fields with charges $0,\pm 1$ and thus the interaction of this meson with the electromagnetic field would be included, in the mean field model only the neutral meson $\rho^3$ has a nonzero expectation value. 

The baryon fermion pressure depends on the meson fields through the baryon effective mass $m^*$ and effective chemical potentials $\mu_a^*$:
\begin{align}
m_* = m_{\text{N}}-g_{\sigma}\sigma, \qquad
\mu_{\text{p}}^* = \mu_{\text{p}} - g_{\omega}\omega_0 - \frac{1}{2}g_{\rho}\rho_0^3, \qquad
\mu_{\text{n}}^* = \mu_{\text{n}} - g_{\omega}\omega_0 + \frac{1}{2}g_{\rho}\rho_0^3.
\label{eq:EffectiveMandMu}
\end{align}
The mean field values for $\omega_0$ and $\rho^3_0$ in terms of the neutron and proton number densities are
\begin{equation}
\omega_0=\frac{g_{\omega}n_{\text{b}}}{m^2_{\omega}}, \qquad \rho^3_0=\frac{g_{\rho}(n_{\text{p}}-n_{\text{n}})}{2m^2_{\rho}}.
\label{eq:MeanFieldMesonFields}
\end{equation}
Note that these are equations of motion that hold in equilibrium and must only be imposed after taking the desired partial derivatives of the thermodynamic potential of interest.

Since we ignore anomalous magnetic moments, the neutron pressure has no $B$-dependence and is
\begin{equation}
P_{\text{f},\text{n}}(\mu_n,\sigma,\omega_0,\rho_0^3)=\frac{\mu_n^*(2\mu_n^{*2}-m_*^2)\sqrt{\mu_n^{*2}-m_*^2}}{24\pi^2}+\frac{m_*^4}{8\pi^2}\ln\left[\frac{\mu_n^*+\sqrt{\mu_n^{*2}-m_*^2}}{m_*}\right].
\label{eq:PressureNeutrons}
\end{equation}
For the Landau-quantized charged fermions $a=\text{p},\text{e},\text{m}$, the pressure is given as a sum over occupied Landau levels:
\begin{align}
{}&P_{\text{f},p}(\mu_p,B,\sigma,\omega_0,\rho_0^3)=\frac{eB}{4\pi^2}\sum_{n=0}^{n_{\text{max}}}\gamma_n\left[\mu_p^*\sqrt{\mu_p^{*2}-m_*^2-2eBn}-(m_*^2+2eBn)\ln\left(\frac{\mu_p^*+\sqrt{\mu_p^{*2}-m_*^2-2eBn}}{\sqrt{m_*^2+2eBn}}\right)\right],
\label{eq:PressureProtons}
\\
{}&P_{\text{f},a}(\mu_a,B)=\frac{eB}{4\pi^2}\sum_{n=0}^{n_{\text{max}}}\gamma_n\left[\mu_a\sqrt{\mu_a^2-m_a^2-2eBn}-(m_a^2+2eBn)\ln\left(\frac{\mu_a+\sqrt{\mu_a^2-m_a^2-2eBn}}{\sqrt{m_a^2+2eBn}}\right)\right], \quad a\in\{\text{e,m}\},
\label{eq:PressureLeptons}
\end{align}
where $\gamma_n=2-\delta_{n,0}$ is the degeneracy factor of Landau level $n$ and $n_{\text{max}}=\lfloor(\mu_a^{2}-m_a^2)/(2eB)\rfloor$. 

The appearance of $n_{\text{p}}$ and $n_{\text{n}}$ in the expressions for $\mu_{\text{p}}^*$ and $\mu_{\text{n}}^*$ means that the equations for $n_{\text{p}}$ and $n_{\text{n}}$ must be solved simultaneously with the self-consistency equation for $\sigma$, which is
\begin{align}
0{}&=-m_{\sigma}^2\sigma-b_{\sigma}m_{\text{N}}g_{\sigma}^3\sigma^2-c_{\sigma}g_{\sigma}^4\sigma^3+\frac{g_{\sigma}m_*}{2\pi^2}\left[\mu^*_n\sqrt{\mu_n^{*2}-m_*^2}-m_*^2\ln\left(\frac{\mu_n^*+\sqrt{\mu_n^{*2}-m_*^2}}{m_*}\right)+eB\sum_{n=0}^{n_{\text{max}}}\gamma_n\ln\left(\frac{\mu_p^*+\sqrt{\mu_p^{*2}-m_*^2-2neB}}{\sqrt{m_*^2+2neB}}\right)\right]
\nonumber
\\
{}&=-m_{\sigma}^2\sigma-b_{\sigma}m_{\text{N}}g_{\sigma}^3\sigma^2-c_{\sigma}g_{\sigma}^4\sigma^3-g_{\sigma}\left.\frac{\partial P}{\partial m_*}\right|_{\mu_a,B,\phi}.
\label{eq:SigmaFieldSelfConsistency}
\end{align}

For the charged fermions, the second-order partial derivatives that we require are divergent at zero temperature-- this can be seen from Eq.~(\ref{eq:d2Pdmu2}--\ref{eq:d2PdB2}). Instead, we must use the finite temperature versions of $P_{\text{f},a}$ to compute these partial derivatives: it is given by (e.g.~\citet{Persson1995})
\begin{equation}
P_{\text{f},a}(\mu_a,B,T,\sigma,\omega_0,\rho_0^3)=\frac{eB}{2\pi^2}\sum_{n=0}^{\infty}\int_{m_a^2+2eBn}^{\infty}dE\sqrt{E^2-m_a^2-2eBn}f_a(E),
\label{eq:FiniteTPressureFermion}
\end{equation}
where $m_a\rightarrow m_*$ for protons. $f_a(E)$ is the Fermi--Dirac distribution
\begin{equation}
f_{a}(E)=\frac{\theta(E)}{\exp(\beta(E-\mu_a))+1},
\label{eq:FermiDiracDistribution}
\end{equation} 
where $\beta=(k_BT)^{-1}$, $\theta(x)$ is the Heaviside step function, and $\mu_a\rightarrow\mu_p^*$ for $a=\text{p}$. We have dropped the anti-fermion contribution from $f_a(E)$ since these species will not be present in neutron stars. Even including the finite temperature, the second-order partial derivatives of $P_{\text{f},a}$ for the charged fermions will still exhibit strongly peaked behaviour where new Landau levels start being populated i.e. where $(\mu_a^2-m_a^2)/(2eB)$ takes integer values, and thus will play a dominant role in the local stability criteria discussed later. Since we are not interested in the dynamics of heat flow inside the star, $T$ is treated as a fixed parameter in Eq.~(\ref{eq:FiniteTPressureFermion}), and we assume it is held constant in all thermodynamic partial derivatives.

\begin{figure}
\center
\includegraphics[width=0.7\linewidth]{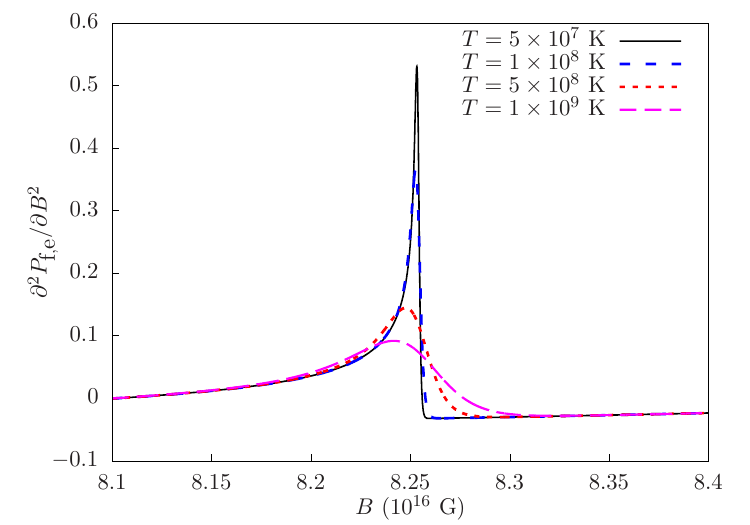}
\caption[Effect of finite temperature on $\partial^2 P_{\text{f},\text{e}}/\partial B^2$]{\label{fig:SinglePeakPlot} $\partial^2 P_{\text{f},\text{e}}/\partial B^2$ in units of $c^2$ near where a new Landau level begins being filled for $\mu_{\text{e}}=125$ MeV, computed using the finite temperature-including form of $P_{\text{f},a}$, Eq.~(\ref{eq:FiniteTPressureFermion}). The explicit expression for $\partial^2 P_{\text{f},\text{e}}/\partial B^2$ at finite $T$ is given by Eq.~(\ref{eq:d2PdB2FiniteT}). The damping effect of increasing temperature is clearly demonstrated.}
\end{figure}

The regulating effect of finite temperature on eliminating the divergences when a new Landau level begins to be populated $(\mu_a^2-m_a^2)/(2eB)=$ an integer is illustrated in Figure~\ref{fig:SinglePeakPlot}. The temperature dependence of these functions will have an important role in the MHD instabilities we show later in the paper, and sufficiently high temperatures can stabilize the magnetized fluid in regions of parameter space that would otherwise be unstable.

For $g_{\sigma}/m_{\sigma}$, $g_{\omega}/m_{\omega}$, $g_{\rho}/m_{\rho}$, $b_{\sigma}$, $c_{\sigma}$, we use the tabulated parameters for the nuclear compressibility $K=300$ MeV, $m_*/m_{\text{N}}=0.78$ at saturation density $\sigma\omega\rho$ model from Table 5.5 of~\citet{Glendenning1997}; these are $g_{\sigma}/m_{\sigma}=3.024$ fm, $g_{\omega}/m_{\omega}=2.195$ fm, $g_{\rho}/m_{\rho}=2.189$ fm, $b_{\sigma}=3.478\times10^{-3}$, $c_{\sigma}=1.328\times10^{-2}$. Unlike the reference, we do not include hyperons as our stellar model does not reach the densities at which they appear. The maximum TOV mass possible with this EOS is only $\approx1.7M_{\odot}$, so it does not cover the entire range of observed neutron star masses, but it does allow us to discuss MHD stability inside a reasonable model of a neutron star core.

The matter pressure must be supplemented by the vacuum magnetic field contributions, giving the grand potential density $\Omega_G$
\begin{equation}
\Omega_G(B,\mu_a,\sigma,\omega_0,\rho_0^3)=-P(B,\mu_a,\sigma,\omega_0,\rho_0^3)+\frac{B^2}{8\pi}-\Lagr_{\text{EH}}(B),
\label{eq:GrandPotentialDensity}
\end{equation}
where $\Lagr_{\text{EH}}$ is the vacuum Euler--Heisenberg Lagrangian for only a magnetic field
\begin{equation}
\Lagr_{\text{EH}}=-\sum_{a=\text{e,p,m}}\frac{m_a^4}{8\pi^2}\int^{\infty}_0\frac{dx}{x^3}e^{-x}\left[x\frac{eB}{m_a^2}\text{coth}\left(x\frac{eB}{m_a^2}\right)-1-\frac{1}{3}\left(x\frac{eB}{m_a^2}\right)^2\right].
\label{eq:EHLagrangianVacuum}
\end{equation}
$\Lagr_{\text{EH}}$ is significant (e.g. has magnitude greater than 5\% of the linear term $-B^2/8\pi$) for species $a$ when $B\gtrsim 1.26\times10^{17}m^2_{a,\text{MeV}}$ G. This corresponds to $B\gtrsim 3.3\times10^{16}$ G for electrons and magnetic fields far stronger than any field expected to exist even within magnetars for the protons and muons, so we can safely drop $\Lagr_{\text{EH}}$ for the latter two species. 

The first law of thermodynamics for $\Omega_G$ is
\begin{align}
\text{d}\Omega_G={}&\sum_a\left.\frac{\partial\Omega_G}{\partial\mu_a}\right|_{\mu_b\neq\mu_a,B,\phi}\text{d}\mu_a+\left.\frac{\partial\Omega_G}{\partial B}\right|_{\mu_a,\phi}\text{d}B+\left.\frac{\partial\Omega_G}{\partial\sigma}\right|_{\mu_a,B,\omega_0,\rho_0^3}\text{d}\sigma+\left.\frac{\partial\Omega_G}{\partial\omega_0}\right|_{\mu_a,B,\sigma,\rho_0^3}\text{d}\omega_0+\left.\frac{\partial\Omega_G}{\partial\rho_0^3}\right|_{\mu_a,B,\sigma,\omega_0}\text{d}\rho_0^3
\nonumber
\\
={}&-\sum_a\left.\frac{\partial P_{\text{f},a}}{\partial\mu_a}\right|_{B,\phi}\text{d}\mu_a
+\left.\frac{\partial\Omega_G}{\partial B}\right|_{\mu_a,\phi}\text{d}B+\left.\frac{\partial P}{\partial\sigma}\right|_{\mu_a,B,\omega_0,\rho_0^3}\text{d}\sigma+\left.\frac{\partial P}{\partial\omega_0}\right|_{\mu_a,B,\sigma,\rho_0^3}\text{d}\omega_0+\left.\frac{\partial P}{\partial\rho_0^3}\right|_{\mu_a,B,\sigma,\omega_0}\text{d}\rho_0^3,
\label{eq:dOmegaG}
\end{align}
where $\phi$ is used to denote all the meson fields being held constant during differentation. Using Eq.~(\ref{eq:MeanFieldMesonFields}--\ref{eq:SigmaFieldSelfConsistency}) and standard thermodynamic definitions~\citep{Landau1960}, we have
\begin{align}
n_a={}&\left.\frac{\partial P_{\text{f},a}}{\partial\mu_a}\right|_{B,\phi},
\label{eq:NumberDensityDef}
\\
H={}&4\pi\left.\frac{\partial \Omega_G}{\partial B}\right|_{\mu_a,\phi}=B-4\pi\frac{\partial \Lagr_{\text{EH}}}{\partial B}-4\pi\left.\frac{\partial P}{\partial B}\right|_{\mu_a,\phi},
\\
0={}&\left.\frac{\partial P}{\partial\sigma}\right|_{\mu_a,B,\omega_0,\rho_0^3}=\left.\frac{\partial P}{\partial\omega_0}\right|_{\mu_a,B,\sigma,\rho_0^3}=\left.\frac{\partial P}{\partial\rho_0^3}\right|_{\mu_a,B,\sigma,\omega_0},
\end{align}
where $H$ is the magnetic $H$-field and is identical to Eq.~(\ref{eq:HDef}) as we show in the next subsection. Hence when computing the first order partial derivatives of $\Omega_G$ or $P$ with respect to their natural (independent) variables $\mu_a$ and $B$, we can hold the meson fields constant.

\subsection{Changes of variable and connection to magnetohydrodynamics}
\label{sec:COV}

Because it is derived in the framework of quantum statistical mechanics using the grand canonical ensemble, the expression for $P$ is in terms of chemical potentials of the fermions as opposed to their number densities. This is inconvenient for hydrodynamics where we work with conserved currents and fixed (number) densities and not at fixed chemical potentials. To compute the thermodynamic partial derivatives $u_B$, $u_{BB}$, $u_{\rho B}$, $u_{BY}$ and $u_{Bf}$ which appear in the stability criteria and which are computed at fixed $\rho$, $Y$, $f$ and/or $B$, we must first relate the pressure $P$, or more generally $\Omega_G$ of Eq.~(\ref{eq:GrandPotentialDensity}), to the internal energy density defined in Eq.~(\ref{eq:InternalEnergyDensity}). 

We start by computing the internal energy density corresponding to Eq.~(\ref{eq:GrandPotentialDensity}) using the Legendre transformation
\begin{align}
u(n_a,B,\sigma,\omega_0,\rho_0^3){}&=\sum_an_a\mu_a+\Omega_G(\mu_a,B,\sigma,\omega_0,\rho_0^3)=-\sum_a\mu_a\left.\frac{\partial P_{\text{f},a}}{\partial\mu_a}\right|_{B,\phi}+\Omega_G(\mu_a,B,\sigma,\omega_0,\rho_0^3)
\nonumber
\\
{}&=\frac{B^2}{8\pi}-\Lagr_{\text{EH}}(B)+u_{\text{mat}}(n_a,B,\sigma,\omega_0,\rho_0^3).
\label{eq:InternalEnergyDensityMicro}
\end{align}
The internal energy density of the matter $u_{\text{mat}}(n_a,B,\sigma,\omega_0,\rho_0^3)$ including meson fields is
\begin{align}
u_{\text{mat}}(n_a,B,\sigma,\omega_0,\rho_0^3)={}&\sum_au_{\text{f},a}(n_a,B,\sigma,\omega_0,\rho_0^3)+\frac{1}{2}m_{\sigma}^2\sigma^2+\frac{1}{3}b_{\sigma}m_{\text{N}}(g_{\sigma}\sigma)^3+\frac{1}{4}c_{\sigma}(g_{\sigma}\sigma)^4-\frac{1}{2}m_{\omega}^2\omega_0^2-\frac{1}{2}m_{\rho}^2(\rho^3_0)^2
\nonumber
\\
{}&+g_{\omega}\omega_0n_{\text{b}}+\frac{1}{2}g_{\rho}\rho^3_0(n_{\text{p}}-n_{\text{n}}),
\label{eq:InternalEnergyDensityFull}
\end{align}
where $u_{\text{f},a}$ is the internal energy for the fermions of species $a$ and is expressed in terms of $P_{\text{f},a}$ as
\begin{equation}
u_{\text{f},a}=\mu_a\left.\frac{\partial P_{\text{f},a}}{\partial\mu_a}\right|_{B,\phi}-P_{\text{f},a}.
\label{eq:FermionInternalEnergyDensity}
\end{equation}
Eq.~(\ref{eq:InternalEnergyDensityFull}) can be simplified at equilibrium using Eq.~(\ref{eq:MeanFieldMesonFields}). Since we can easily switch between $\rho,Y,f,\rightarrow n_{\text{n}},n_{\text{p}},n_{\text{e}},n_{\text{m}}$, Eq.~(\ref{eq:InternalEnergyDensity}) and Eq.~(\ref{eq:InternalEnergyDensityMicro}) are identical except for the dependence on the meson fields in $u_{\text{mat}}$. Since the meson fields are in a sense ``microscopic'' variables that should not affect the macroscopic fluid flow except through their influence on the baryons with which they interact, we do not hold them constant when calculating $u_B$, $u_{BB}$, etc. Hence $u_B$, $u_{BB}$, etc. may depend on partial derivatives of the meson fields, as we now show.

Taking the differential of $n_a=n_a(\mu_a,B,\sigma,\omega_0,\rho_0)$ (the meson field dependence can be neglected for the leptons), we can write the following matrix equation
\begin{equation}
\left(\begin{array}{c} \text{d}n_a \\ \text{d}B \\ \text{d}\sigma \\ \text{d}\omega_0 \\ \text{d}\rho_0^3 \end{array}\right)=\left(\begin{array}{ccccc} \partial n_a/\partial \mu_a & \partial n_a/\partial B & \partial n_a/\partial\sigma & \partial n_a/\partial\omega_0 & \partial n_a/\partial\rho_0^3\\ 0 & 1 & 0 & 0 & 0 \\ 0 & 0 & 1 & 0 & 0 \\ 0 & 0 & 0 & 1 & 0 \\ 0 & 0 & 0 & 0 & 1\end{array}\right)\left(\begin{array}{c} \text{d}\mu_a \\ \text{d}B \\ \text{d}\sigma \\ \text{d}\omega_0 \\ \text{d}\rho_0^3\end{array}\right),
\end{equation}
where all variables except that being differentiated with respect to are implicitly held constant. This can be inverted to give
\begin{equation}
\text{d}\mu_a=\left(\left.\frac{\partial n_a}{\partial\mu_a}\right|_{B,\phi}\right)^{-1}\left[\text{d}n_a-\left.\frac{\partial n_a}{\partial B}\right|_{\mu_a,\phi}\text{d}B-\left.\frac{\partial n_a}{\partial\sigma}\right|_{\mu_a,B,\omega_0,\rho_0^3}\text{d}\sigma-\left.\frac{\partial n_a}{\partial\omega_0}\right|_{\mu_a,B,\sigma,\rho_0^3}\text{d}\omega_0-\left.\frac{\partial n_a}{\partial\rho_0^3}\right|_{\mu_a,B,\sigma,\omega_0}\text{d}\rho_0^3\right]
\label{eq:dmuPreliminary}
\end{equation}
where the meson field-dependent terms only contribute to $\text{d}\mu_n$ and $\text{d}\mu_p$. Using Eq.~(\ref{eq:Pressure}--\ref{eq:EffectiveMandMu}),
\begin{subequations}
\begin{align}
\left.\frac{\partial n_a}{\partial\sigma}\right|_{\mu_a,B,\omega_0,\rho_0^3}{}&=\frac{\partial m_*}{\partial\sigma}\left.\frac{\partial n_a}{\partial m_*}\right|_{\mu_a,B,\phi}=g_{\sigma}\frac{m_*}{\mu_a^*}\left.\frac{\partial n_a}{\partial\mu_a}\right|_{B,\phi}, 
\\
\left.\frac{\partial n_a}{\partial\omega_0}\right|_{\mu_a,B,\sigma,\rho_0^3}{}&=\left.\frac{\partial n_a}{\partial\mu_a^*}\right|_{\mu_a,B,\phi}\left.\frac{\partial \mu_a^*}{\partial\omega_0}\right|_{\mu_a,B,\phi}=-g_{\omega}\left.\frac{\partial n_a}{\partial\mu_a}\right|_{\mu_a,B,\phi}, 
\\
\left.\frac{\partial n_a}{\partial\rho_0^3}\right|_{\mu_a,B,\sigma,\omega_0}{}&=\left.\frac{\partial n_a}{\partial\mu_a^*}\right|_{\mu_a,B,\phi}\left.\frac{\partial \mu_a^*}{\partial\rho_0^3}\right|_{\mu_a,B,\phi}=-g_{\rho}I_3\left.\frac{\partial n_a}{\partial\mu_a}\right|_{\mu_a,B,\phi},
\end{align}
\end{subequations}
where $I_3=\pm1/2$ is the third component of isospin for protons/neutrons. Inserting these into Eq.~(\ref{eq:dmuPreliminary}) gives
\begin{equation}
\text{d}\mu_a=\left(\left.\frac{\partial n_a}{\partial\mu_a}\right|_{B,\phi}\right)^{-1}\text{d}n_a-\left(\left.\frac{\partial n_a}{\partial\mu_a}\right|_{B,\phi}\right)^{-1}\left.\frac{\partial n_a}{\partial B}\right|_{\mu_a,\phi}\text{d}B-g_{\sigma}\frac{m_*}{\mu_a^*}\text{d}\sigma+g_{\omega}\text{d}\omega_0+g_{\rho}I_3\text{d}\rho_0^3.
\label{eq:dmu}
\end{equation}
Inserting this into Eq.~(\ref{eq:dOmegaG}) and using Eq.~(\ref{eq:GrandPotentialDensity}) to isolate $\text{d}P$, we obtain
\begin{align}
\text{d}P={}&\sum_an_a\left(\left.\frac{\partial n_a}{\partial\mu_a}\right|_{B,\phi}\right)^{-1}\text{d}n_a+\sum_a\left[\left.\frac{\partial P_{\text{f},a}}{\partial B}\right|_{\mu_a,\phi}-n_a\left(\left.\frac{\partial n_a}{\partial\mu_a}\right|_{B,\phi}\right)^{-1}\left.\frac{\partial n_a}{\partial B}\right|_{\mu_a,\phi}\right]\text{d}B+g_{\omega}n_{\text{b}}\text{d}\omega_0+\frac{1}{2}g_{\rho}(n_{\text{p}}-n_{\text{n}})\text{d}\rho_0^3
\nonumber
\\
{}&-g_{\sigma}m_*\left(\frac{n_{\text{n}}}{\mu_{\text{n}}^*}+\frac{n_{\text{p}}}{\mu_{\text{p}}^*}\right)\text{d}\sigma.
\label{eq:dP}
\end{align}
This expression is necessary to compute the partial derivatives of the pressure with respect to $\rho=m_{\text{N}}n_{\text{b}}$, $B$, $Y$ and/or $f$ with the other variables held constant as needed in the hydrodynamics.

The first law of thermodynamics for $u(n_a,B,\sigma,\omega_0,\rho_0^3)$ is most conveniently found by taking the differential of Eq~.(\ref{eq:InternalEnergyDensityMicro}). Doing so and using Eq.~(\ref{eq:dmu}--\ref{eq:dP}) gives
\begin{align}
\text{d}u=\sum_a\left.\frac{\partial u}{\partial n_a}\right|_{n_c\neq n_a,B,\phi}\text{d}n_a+\left.\frac{\partial u}{\partial B}\right|_{n_a,\phi}\text{d}B+\left.\frac{\partial u}{\partial\sigma}\right|_{n_a,B,\phi}\text{d}\sigma+\left.\frac{\partial u}{\partial\omega_0}\right|_{n_a,B,\phi}\text{d}\omega_0+\left.\frac{\partial u}{\partial\rho_0^3}\right|_{n_a,B,\phi}\text{d}\rho_0^3=\sum_a\mu_a \text{d} n_a+\left.\frac{\partial u}{\partial B}\right|_{n_a,\phi}\text{d}B,
\label{eq:du}
\end{align}
We see that, like for $\Omega_G$ and $P$, when computing the first order partial derivatives of $u$ with respect to its natural variables $n_a$ and $B$ the meson fields can be held constant. Taking the differential of the first equality in Eq~.(\ref{eq:InternalEnergyDensityMicro}) and using Eq.~(\ref{eq:dOmegaG}) gives $u_B$:
\begin{equation}
\left.\frac{\partial\Omega_G}{\partial B}\right|_{\mu_a,\phi}=\left.\frac{\partial u}{\partial B}\right|_{n_a,\phi}=\frac{H}{4\pi}\rightarrow u_B=\left.\frac{\partial u_M}{\partial B}\right|_{\rho,Y_a}=-\left.\frac{\partial(P+\Lagr_{\text{EH}})}{\partial B}\right|_{\mu_a,\phi}=\left.\frac{\partial(u_{\text{mat}}-\Lagr_{\text{EH}})}{\partial B}\right|_{n_a,\phi}.
\label{eq:uB}
\end{equation}
The required second-order partial derivatives can all be computed starting from $u_B$. Since it is most convenient to compute thermodynamic derivatives starting from $P$ or $\Omega_G$, we take the differential
\begin{align}
\text{d}u_B=\text{d}\left(\left.\frac{\partial P}{\partial B}\right|_{\mu_a,\phi}\right)={}&-\left.\frac{\partial^2(P+\Lagr_{\text{EH}})}{\partial B^2}\right|_{\mu_a,\phi}\text{d}B-\sum_a\left.\frac{\partial^2P}{\partial B\partial\mu_a}\right|_{\mu_b\neq\mu_a,\phi}\text{d}\mu_a-\left.\frac{\partial^2P}{\partial B\partial\sigma}\right|_{\mu_a,\omega_0,\rho_0^3}\text{d}\sigma-\left.\frac{\partial^2P}{\partial B\partial\omega_0}\right|_{\mu_a,\sigma,\omega_0}\text{d}\omega_0
\nonumber
\\
{}&-\left.\frac{\partial^2P}{\partial B\partial\rho_0^3}\right|_{\mu_a,\sigma,\omega_0}\text{d}\rho_0^3.
\end{align}
Using Eq.~(\ref{eq:dmu}) and
\begin{subequations}
\begin{align}
\left.\frac{\partial^2P}{\partial B\partial\sigma}\right|_{\mu_a,\omega_0,\rho_0^3}{}&=-g_{\sigma}\left.\frac{\partial^2P}{\partial B\partial m_*}\right|_{\mu_a,\phi},
\\
\left.\frac{\partial^2P}{\partial B\partial\omega_0}\right|_{\mu_a,\sigma,\omega_0}{}&=\left.\frac{\partial^2P}{\partial B\partial\mu_a^*}\right|_{\mu_a,\phi}\left.\frac{\partial\mu_a^*}{\partial\omega_0}\right|_{\mu_a,B,\sigma,\rho_0^3}=-g_{\omega}\left.\frac{\partial^2P}{\partial B\partial\mu_a}\right|_{\mu_b\neq\mu_c,\phi}=-g_{\omega}\left.\frac{\partial n_a}{\partial B}\right|_{\mu_a,\phi},
\\
\left.\frac{\partial^2P}{\partial B\partial\rho_0^3}\right|_{\mu_a,\sigma,\omega_0}{}&=\left.\frac{\partial^2P}{\partial B\partial\mu_a^*}\right|_{\mu_a,\phi}\left.\frac{\partial\mu_a^*}{\partial\rho_0^3}\right|_{\mu_a,B,\sigma,\omega_0}=-g_{\rho}I_3\left.\frac{\partial^2P}{\partial B\partial\mu_a}\right|_{\mu_b\neq\mu_c,\phi}=-g_{\rho}I_3\left.\frac{\partial n_a}{\partial B}\right|_{\mu_a,\phi},
\end{align}
\end{subequations}
we find (using Eq.~(\ref{eq:NumberDensityDef}) to replace $n_a$)
\begin{align}
\text{d}u_B={}&\left[-\left.\frac{\partial^2(P+\Lagr_{\text{EH}})}{\partial B^2}\right|_{\mu_a,\phi}+\sum_a\left(\left.\frac{\partial^2 P_{\text{f},a}}{\partial\mu_a^2}\right|_{B,\phi}\right)^{-1}\left.\frac{\partial^2 P_{\text{f},a}}{\partial B\partial\mu_a}\right|_{\phi}\right]\text{d}B
-\sum_a\left(\left.\frac{\partial^2 P_{\text{f},a}}{\partial\mu_a^2}\right|_{B,\phi}\right)^{-1}\left.\frac{\partial^2P_{\text{f},a}}{\partial B\partial\mu_a}\right|_{\phi}\text{d}n_a
\nonumber
\\
{}&+g_{\sigma}\left[\left.\frac{\partial^2P_{\text{f},\text{p}}}{\partial B\partial m_*}\right|_{\mu_{\text{p}},\phi}+\frac{m_*}{\mu_{\text{p}}^*}\left.\frac{\partial^2P_{\text{f},\text{p}}}{\partial B\partial\mu_{\text{p}}}\right|_{\phi}\right]\text{d}\sigma.
\label{eq:duB}
\end{align}
This expression is required to compute $u_{BB}$, $u_{\rho B}$, $u_{BY}$ and $u_{Bf}$ as found in the hydrodynamics. Since we know $P$ and $\Lagr_{\text{EH}}$ we can compute all of the coefficients appearing in this expression.

In the nonrelativistic, non-magnetic limit, Eq.~(\ref{eq:dmu}) gives the differential mass density $d\rho$
\begin{align}
\text{d}u{}&=\sum_a\mu_a\text{d}n_a=[\mu_{\text{n}}(1-Y)+\mu_{\text{p}}Y+\mu_{\text{e}}Yf+\mu_{\text{m}}Y(1-f)]\text{d}n_{\text{b}}+[\mu_{\text{p}}-\mu_{\text{n}}+\mu_{\text{e}}f+\mu_{\text{m}}(1-f)]n_{\text{b}}\text{d}Y+(\mu_{\text{e}}-\mu_{\text{m}})n_{\text{b}}Ydf=\mu_{\text{n}}\text{d}n_{\text{b}}
\nonumber
\\
{}&\rightarrow \text{d}\rho = m_{\text{N}}\text{d}n_{\text{b}},
\label{eq:drho}
\end{align}
where we used beta equilibrium $\mu_{\text{n}}=\mu_{\text{p}}+\mu_{\text{e}}$ and $\mu_{\text{e}}=\mu_{\text{m}}$. So the differential $\text{d}n_{\text{b}}$ can be replaced with $\text{d}\rho/m_{\text{N}}$. The nonrelativistic, non-magnetic limit of Eq.~(\ref{eq:InternalEnergyDensityFull}) is
\begin{align}
u_{\text{mat}}(B=0)={}&m_*n_{\text{b}}+m_{\text{e}}n_{\text{e}}+m_{\text{m}}n_{\text{m}}+\frac{1}{2}m_{\sigma}^2\sigma^2+\frac{1}{3}b_{\sigma}m_{\text{N}}(g_{\sigma}\sigma)^3+\frac{1}{4}c_{\sigma}(g_{\sigma}\sigma)^4+\frac{1}{2}m_{\omega}^2\omega_0^2+\frac{1}{2}m_{\rho}^2(\rho^3_0)^2,
\label{eq:InternalEnergyDensityNR}
\end{align}
which is within a few percent of $\rho=m_{\text{N}}n_{\text{b}}$. We hence use the latter expression throughout the rest of the paper as we discussed in Section~\ref{sec:Perturbations}. The similarity of $\rho=m_{\text{N}}n_{\text{b}}$ and $u_{\text{mat}}(B=0)$ is shown in Figure~\ref{fig:CoreStructure}.

Returning to Eq.~(\ref{eq:duB}) and using Eq.~(\ref{eq:drho}), we compute the required second-order partial derivatives for computing the stability criteria
\begin{subequations}
\begin{align}
u_{BB}={}&\left.\frac{\partial^2 u_M}{\partial B^2}\right|_{\rho,Y,f}=-\left.\frac{\partial^2(P+\Lagr_{\text{EH}})}{\partial B^2}\right|_{\mu_a,\phi}+\sum_a\left(\left.\frac{\partial^2 P_{\text{f},a}}{\partial\mu_a^2}\right|_{B,\phi}\right)^{-1}\left.\frac{\partial^2 P_{\text{f},a}}{\partial B\partial\mu_a}\right|_{\phi}+g_{\sigma}\left[\left.\frac{\partial^2P_{\text{f},\text{p}}}{\partial B\partial m_*}\right|_{\mu_p,\phi}+\frac{m_*}{\mu_{\text{p}}^*}\left.\frac{\partial^2P_{\text{f},\text{p}}}{\partial B\partial\mu_{\text{p}}}\right|_{\phi}\right]\left.\frac{\partial\sigma}{\partial B}\right|_{\rho,Y,f},
\label{eq:uBB}
\\
u_{\rho B}={}&\left.\frac{\partial^2 u_M}{\partial B\partial\rho}\right|_{Y,f}=-\frac{1}{m_{\text{N}}}\sum_a\left(\left.\frac{\partial^2 P_{\text{f},a}}{\partial\mu_a^2}\right|_{B,\phi}\right)^{-1}\left.\frac{\partial^2P_{\text{f},a}}{\partial B\partial\mu_a}\right|_{\mu_b\neq\mu_a,\phi}+\frac{g_{\sigma}}{m_{\text{N}}}\left[\left.\frac{\partial^2P_{\text{f},\text{p}}}{\partial B\partial m_*}\right|_{\mu_p,\phi}+\frac{m_*}{\mu_{\text{p}}^*}\left.\frac{\partial^2P_{\text{f},\text{p}}}{\partial B\partial\mu_{\text{p}}}\right|_{\phi}\right]\left.\frac{\partial\sigma}{\partial n_{\text{b}}}\right|_{Y,f,B},
\label{eq:urhoB}
\\
u_{BY}={}&\left.\frac{\partial^2 u_M}{\partial B\partial Y}\right|_{\rho,f}=-n_{\text{b}}\left[\left(\left.\frac{\partial^2 P_{\text{f},\text{p}}}{\partial\mu_{\text{p}}^2}\right|_{B,\phi}\right)^{-1}\left.\frac{\partial^2P_{\text{f},\text{p}}}{\partial B\partial\mu_{\text{p}}}\right|_{\phi}+f\left(\left.\frac{\partial^2 P_{\text{f},\text{e}}}{\partial\mu_{\text{e}}^2}\right|_{B,\phi}\right)^{-1}\left.\frac{\partial^2P_{\text{f},\text{e}}}{\partial B\partial\mu_{\text{p}}}\right|_{\phi}+(1-f)\left(\left.\frac{\partial^2 P_{\text{f},\text{m}}}{\partial\mu_{\text{m}}^2}\right|_{B,\phi}\right)^{-1}\left.\frac{\partial^2P_{\text{f},\text{m}}}{\partial B\partial\mu_{\text{m}}}\right|_{\phi}\right]
\nonumber
\\
{}&\qquad\qquad\qquad-g_{\sigma}m_*n_{\text{b}}\left(\frac{1}{\mu_{\text{n}}^*}-\frac{1}{\mu_{\text{p}}^*}\right)\left.\frac{\partial\sigma}{\partial Y}\right|_{\rho,f,B}
\label{eq:uBY}
\\
u_{Bf}={}&\left.\frac{\partial^2 u_M}{\partial B\partial f}\right|_{\rho,Y}=-n_{\text{b}}Y\left[\left(\left.\frac{\partial^2 P_{\text{f},\text{e}}}{\partial\mu_{\text{e}}^2}\right|_{B,\phi}\right)^{-1}\left.\frac{\partial^2P_{\text{f},\text{e}}}{\partial B\partial\mu_{\text{p}}}\right|_{\phi}-\left(\left.\frac{\partial^2 P_{\text{f},\text{m}}}{\partial\mu_{\text{m}}^2}\right|_{B,\phi}\right)^{-1}\left.\frac{\partial^2P_{\text{f},\text{m}}}{\partial B\partial\mu_{\text{m}}}\right|_{\phi}\right]
\label{eq:uBf}
\end{align}
\end{subequations}
where we have used that $\partial\sigma/\partial f|_{\rho,Y,B}=0$. The partial derivatives of $\sigma$ must be computed implicitly using the $\sigma$ meson self-consistency condition Eq.~(\ref{eq:SigmaFieldSelfConsistency}). Starting with $\partial\sigma/\partial n_{\text{b}}|_{Y,f,B}$
\begin{align}
0=\left.\frac{\partial\sigma}{\partial n_{\text{b}}}\right|_{Y,f,B}\left[-m_{\sigma}^2-2b_{\sigma}m_{\text{N}}g_{\sigma}^3\sigma-3c_{\sigma}g_{\sigma}^4\sigma^2+g_{\sigma}^2\left.\frac{\partial^2 P}{\partial m_*^2}\right|_{\mu_a,B,\phi}\right]-g_{\sigma}\left.\frac{\partial^2 P_{\text{f},\text{n}}}{\partial m_*\partial\mu_{\text{n}}^*}\right|_{B,\phi}\left.\frac{\partial\mu_{\text{n}}^*}{\partial n_{\text{b}}}\right|_{Y,f,B}-g_{\sigma}\left.\frac{\partial^2 P_{\text{f},\text{p}}}{\partial m_*\partial\mu_{\text{p}}^*}\right|_{B,\phi}\left.\frac{\partial\mu_{\text{p}}^*}{\partial n_{\text{b}}}\right|_{Y,f,B}.
\end{align}
Using Eq.~(\ref{eq:EffectiveMandMu}) and~(\ref{eq:dmu}), we have
\begin{equation}
\text{d}\mu_a^*=\left(\left.\frac{\partial n_a}{\partial\mu_a}\right|_{B,\phi}\right)^{-1}\text{d}n_a+\left(\left.\frac{\partial n_a}{\partial\mu_a}\right|_{B,\phi}\right)^{-1}\left.\frac{\partial n_a}{\partial B}\right|_{\mu_a,\phi}\text{d}B-g_{\sigma}\frac{m_*}{\mu_a^*}\text{d}\sigma,
\end{equation}
and hence it can be shown that
\begin{equation}
\left.\frac{\partial\sigma}{\partial n_{\text{b}}}\right|_{Y,f,B}=\frac{g_{\sigma}m_*}{W}\left(\frac{n_{\text{n}}}{\mu_{\text{n}}^*}+\frac{n_{\text{p}}}{\mu_{\text{p}}^*}\right),
\label{eq:dsigmadnb}
\end{equation}
where $W$ is
\begin{align}
W\equiv{}& m_{\sigma}^2+2b_{\sigma}m_{\text{N}}g_{\sigma}^3\sigma+3c_{\sigma}g_{\sigma}^4\sigma^2+\frac{g_{\sigma}^2}{2\pi^2}\left[\frac{(2m_*^2+\mu_{\text{n}}^*)\sqrt{\mu_{\text{n}}^{*2}-m_*^2}}{\mu_{\text{n}}^*}-3m_*^2\ln\left(\frac{\mu_{\text{n}}^*+\sqrt{\mu_{\text{n}}^{*2}-m_*^2}}{m_*}\right)\right]
\nonumber
\\
{}&+\frac{g_{\sigma}^2eB}{2\pi^2}\sum_{n=0}^{n_{\text{max}}}\gamma_n\left[\ln\left(\frac{\mu_{\text{p}}^*+\sqrt{\mu_{\text{p}}^{*2}-m_*^2-2eBn}}{\sqrt{m_*^2+2eBn}}\right)-\frac{m_*^2\sqrt{\mu_{\text{p}}^{*2}-m_*^2-2eBn}}{\mu_{\text{p}}^*(m_*^2+2eBn)}\right].
\end{align}
Analogous calculations for $\partial\sigma/\partial B|_{\rho,Y,f}$ and $\partial\sigma/\partial Y|_{\rho,f,B}$ give
\begin{align}
\left.\frac{\partial\sigma}{\partial B}\right|_{\rho,Y,f}={}&\frac{g_{\sigma}}{W}\left(\left.\frac{\partial^2P_{\text{f},\text{p}}}{\partial B\partial m_*}\right|_{\mu_p,\phi}+\frac{m_*}{\mu_{\text{p}}^*}\left.\frac{\partial^2P_{\text{f},\text{p}}}{\partial B\partial\mu_{\text{p}}}\right|_{\phi}\right),
\label{eq:dsigmadB}
\\
\left.\frac{\partial\sigma}{\partial Y}\right|_{\rho,Y,B}={}&\frac{g_{\sigma}m_*n_{\text{b}}}{W}\left(\frac{1}{\mu_{\text{p}}^*}-\frac{1}{\mu_{\text{n}}^*}\right).
\label{eq:dsigmadY}
\end{align}
Expressions for the partial derivatives of $P$ with respect to $\mu_a$, $B$ and $m_*$ are relegated to Appendix~\ref{app:Thermodynamics}. 

\subsection{Background stellar model}
\label{sec:BackgroundModel}
 
The evaluation of the stability criteria in Section~\ref{sec:StabilityAnalysis} requires a background equilibrium fluid configuration which is perturbed according to the Lagrangian displacement field $\xi^i$. For this background model we use a canonical neutron star with mass $M_*=1.4M_{\odot}$ and structure determined by the $\sigma\omega\rho$ EOS discussed earlier. The presence of the magnetic field changes the stellar equilibrium, since the equation of hydrostatic balance for the background star is Eq.~(\ref{eq:EulerEquation}) with $v^i=0$. If the magnetic forces are small compared to the mechanical pressure, which is true for $P\gg BH/(8\pi)$, the magnetic forces will have a negligible effect on the stellar structure. For even the strongest fields $B=10^{17}$ G that we employ in this paper, this will be approximately true, with $(10^{17}\text{ G})^2/(8\pi)$ at most a few percent of the central pressure of the stellar model. Hence, when computing the background stellar model we use to find the gravitational acceleration $g$ and the species gradients that appear in the Brunt--V\"{a}is\"{a}l\"{a} frequency, we ignore the effect of the magnetic field. 


Under the assumptions described above, the background star will be spherically symmetric and its structure determined by solving the nonrelativistic hydrostatic balance equations
\begin{equation}
\frac{dP}{dr}=-\frac{G\rho M(r)}{r^2}, \qquad \frac{dM(r)}{dr}=4\pi\rho r^2,
\label{eq:NRHydrostaticBalance}
\end{equation}
where the mass enclosed in radius $r$ is $M(r)$. $P$ in the stellar core is given by Eq.~(\ref{eq:Pressure}) but using the non-magnetic fermion pressure (Eq.~(\ref{eq:PressureNeutrons})) for all fermion species and using the mean-field values of $\omega_0$ and $\rho_0^3$ given by Eq.~(\ref{eq:MeanFieldMesonFields}) and the self-consistently calculated value of $\sigma$ found using Eq.~(\ref{eq:SigmaFieldSelfConsistency}). The mass density is $\rho=m_{\text{N}}n_{\text{b}}$. The muon threshold density for this EOS is $\rho=0.79\rho_0$. To extend the star out to the exterior vacuum we use the use the BPS EOS as tabulated in~\cite{Glendenning1997}. This provides only a small contribution to the overall mass and radius of the star, and does not affect the properties of the outer core like the local gravitational acceleration or gradients of the species fractions.

\begin{figure}
\center
\includegraphics[width=0.7\linewidth]{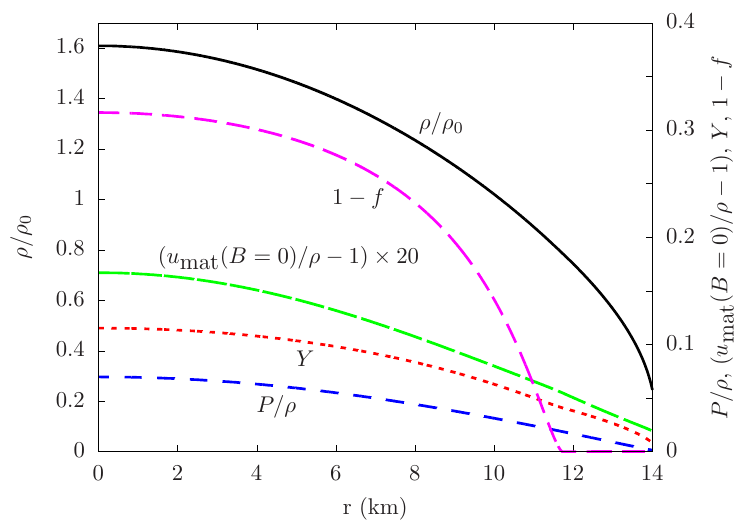}
\caption[Equation of state for stellar model used to analyze MHD stability in ultrastrong field regime]{\label{fig:CoreStructure} Properties of the neutron star model used to analyse the MHD stability: the mass density $\rho$ divided by nuclear saturation density $\rho_0=2.7\times10^{14}$ g/cm$^3$, pressure $P$ divided by $\rho$, the fractional difference between the internal energy density of the matter $u_{\text{mat}}$ at $B=0$ and $\rho$ (scaled by a factor of 20), the proton fraction $Y$ and one minus the electron fraction $f$. Only the properties within $14$ km ($\rho>0.25\rho_0$) are shown, as the crust-core transition is expected to occur before these densities and a different EOS would be needed in this region.}
\end{figure}

For a canonical neutron star with $M_{\star}=M(R_{\star})=1.4M_{\odot}$, the required central pressure and density for our choice of EOS are $P_c=2.71\times 10^{34}$ dyn/cm$^2$ and $\rho_c=4.35\times10^{14}$ g/cm$^3$, and the stellar radius is $R_{\star}=15.56$ km. This radius is unrealistically large because we used the nonrelativistic equation for hydrostatic balance-- if we had used the TOV equation, the $M_*=1.4M_{\odot}$ star would have $R_{\star}= 13.35$ km. Solving Eq.~(\ref{eq:NRHydrostaticBalance}) determines the gravitational acceleration $g$ and radial derivatives of $Y_a$ and $\rho$ as functions of distance from the centre of the star, or equivalently, as functions of $\rho$. In beta equilibrium we have 
\begin{equation}
\mu_{\text{n}}=\mu_{\text{p}}+\mu_{\text{e}},\qquad \mu_{\text{e}}=\mu_{\text{m}},
\end{equation}
which, along with the requirement of local charge neutrality
\begin{equation}
n_{\text{p}}=n_{\text{e}}+n_{\text{m}},
\end{equation}
constrains all four chemical potentials. Using this and Eq.~(\ref{eq:NumberDensityDef}), for a particular value of $\mu_{\text{e}}$ and $B$, we can solve for all of the number densities $n_a$ and hence the total mass density $\rho$, so we can find the corresponding value of $g$, $\text{d}Y/\text{d}r$, etc. by interpolation of the stellar model. $\rho$, $P$, $Y$ and $1-f$ for the stellar model used to study the MHD stability are given as a function of stellar radius in Figure~\ref{fig:CoreStructure}. 

\begin{figure}
\center
\includegraphics[width=0.6\linewidth]{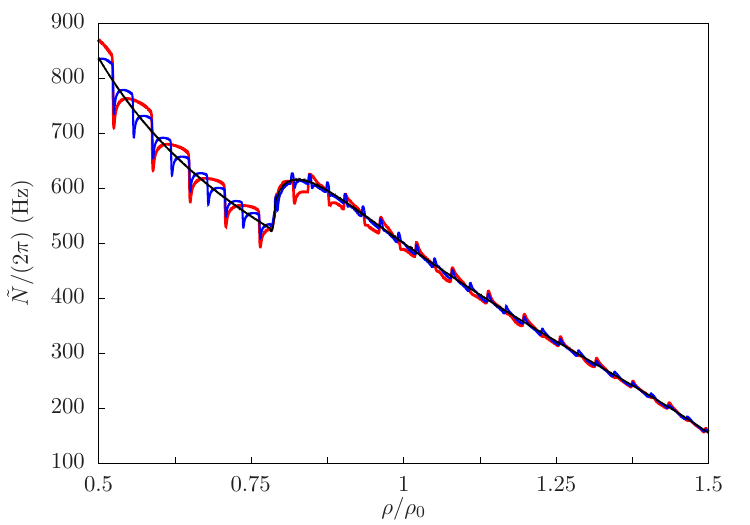}
\caption[Brunt--V\"{a}is\"{a}l\"{a} frequency in ultrastrong magnetic field]{\label{fig:N2} Brunt--V\"{a}is\"{a}l\"{a} frequency with magnetic contributions $\tilde{N}$ as a function of density for three different magnetic field strengths: $B=10^{15}$ G (black), $B=5\times10^{16}$ G (blue), $B=10^{17}$ G (red). $\rho_0=2.7\times10^{14}$ g/cm$^3$. The bump starting at $\rho=0.79\rho_0$ is the leptonic buoyancy contribution which occurs where muons are present in the star.}
\end{figure}

The other required stellar properties are the sound speed $c_s$ and the
Brunt--V\"{a}is\"{a}l\"{a} frequency $\tilde{N}^2$. The former is computed using Eq.~(\ref{eq:dP}) and by differentiating~(\ref{eq:MeanFieldMesonFields}):
\begin{equation}
c_s^2=\left.\frac{\partial P}{\partial\rho}\right|_{B,Y,f}=\frac{1}{m_{\text{N}}}\left[\sum_an_a\left(\left.\frac{\partial^2P_{\text{f},a}}{\partial\mu_a^2}\right|_{B,\phi}\right)^{-1}\left.\frac{\partial n_a}{\partial n_{\text{b}}}\right|_{Y,f,B}+\left(\frac{g_{\omega}}{m_{\omega}}\right)^2n_b+\frac{1}{4}\left(\frac{g_{\rho}}{m_{\rho}}\right)^2(2Y-1)^2n_b-g_{\sigma}m_*\left(\frac{n_p}{\mu^*_{\text{p}}}+\frac{n_n}{\mu^*_{\text{n}}}\right)\frac{\partial\sigma}{\partial n_b}\right].
\label{eq:SoundSpeed}
\end{equation}
This expression is given exclusively in terms of partial derivatives computed in Section~\ref{sec:COV} or listed in Appendix~\ref{app:Thermodynamics}. Similarly, $\tilde{N}^2$ as defined in Eq.~(\ref{eq:MagneticBVFrequency}) requires 
\begin{align}
\left.\frac{\partial P}{\partial Y}\right|_{\rho,f,B}={}&n_{\text{b}}^2\left[Y\left(\left.\frac{\partial^2P_{\text{f},\text{p}}}{\partial\mu_{\text{p}}^2}\right|_{B,\phi}\right)^{-1}-(1-Y)\left(\left.\frac{\partial^2P_{\text{f},\text{n}}}{\partial\mu_{\text{n}}^2}\right|_{B,\phi}\right)^{-1}+Yf\left(\left.\frac{\partial^2P_{\text{f},\text{e}}}{\partial\mu_{\text{e}}^2}\right|_{B,\phi}\right)^{-1}+Y(1-f)\left(\left.\frac{\partial^2P_{\text{f},\text{m}}}{\partial\mu_{\text{m}}^2}\right|_{B,\phi}\right)^{-1}
\right]
\nonumber
\\
{}&+\frac{1}{2}\left(\frac{g_{\rho}}{m_{\rho}}\right)^2n_b^2(2Y-1)-g_{\sigma}m_*\left(\frac{n_{\text{n}}}{\mu^*_{\text{n}}}+\frac{n_{\text{p}}}{\mu^*_{\text{p}}}\right)\frac{\partial\sigma}{\partial Y},
\\
\left.\frac{\partial P}{\partial f}\right|_{\rho,Y,B}={}&Y^2n_{\text{b}}^2\left[f\left(\left.\frac{\partial^2P_{\text{f},\text{e}}}{\partial\mu_{\text{e}}^2}\right|_{B,\phi}\right)^{-1}-(1-f)\left(\left.\frac{\partial^2P_{\text{f},\text{m}}}{\partial\mu_{\text{m}}^2}\right|_{B,\phi}\right)^{-1}\right],
\end{align}
which are also computed using Eq.~(\ref{eq:dP}) and~(\ref{eq:MeanFieldMesonFields}). $u_{BY}$ and $u_{Bf}$, also required in $\tilde{N}^2$, are given by Eq.~(\ref{eq:uBY}--\ref{eq:uBf}). $\tilde{N}$ is plotted in Figure~\ref{fig:N2} for three different magnetic field strengths, showing how the magnetic field-dependent terms modify the usual neutron-proton and leptonic buoyancy.

The functional form of the background magnetic field, as opposed to solely the field strength, is required to compute the magnetic buoyancy stability criteria. We assume a completely toroidal field: though this is not stable globally, we are only interested in the local stability here. The toroidal field in a stable, combined poloidal-toroidal field configuration inside a neutron star can also be two orders of magnitude stronger than the poloidal field~\citep{Akgun2013}, providing justification for us to ignore the latter as a first approximation. Being perpendicular to the gravitational field, a toroidal field is the analog to the field in the $x$--direction used in deriving the magnetic buoyancy stability criterion in Section~\ref{sec:StabilityAnalysis}. Following e.g.~\citet{Lander2009}, we use
\begin{equation}
B^i=B_0\frac{\rho}{\rho_cR_{\star}}\delta^i_{\phi}\rightarrow B=\sqrt{B^iB_i}=B_0\frac{\rho}{\rho_c}\frac{\varpi}{R_{\star}},
\label{eq:BForm}
\end{equation}
where $B_0$ is a constant and $\varpi=r\sin\theta$ the cylindrical radius ($\theta=$ polar angle). This field vanishes at the surface of the star due to its $\rho$ dependence. This is unstable to the sausage and kink instability along the $z$-axis, but could be stabilized by a weaker poloidal field which we ignore. When looking at the stability properties of the star, the $z$-axis is excised since the field vanishes here and the strong-field MHD in which we are interested is not relevant.

We convert this field to Cartesian coordinates in line with our local analysis: the $\phi$-direction is changed to the $x$-direction and the (spherical) radial direction is changed to the $z$-direction. Eq.~(\ref{eq:BForm}) hence implies $\text{d}/\text{d}z\ln(B/\rho)=1/z$ and $\text{d}/\text{d}z\ln B=\text{d}\ln\rho/\text{d}z+1/z$. Since we are free to choose $B_0$, we will pick values of $B$ independent of the radial position and density with the understanding that $B_0$ can be chosen appropriately to satisfy Eq.~(\ref{eq:BForm}). Based on the density profile in Figure~\ref{fig:CoreStructure}, the maximum value of $B$ within the core is $\approx 0.42B_0$.

\section{Numerical calculation of stability criteria}
\label{sec:Results}

\begin{figure}
\center
\includegraphics[width=0.92\linewidth]{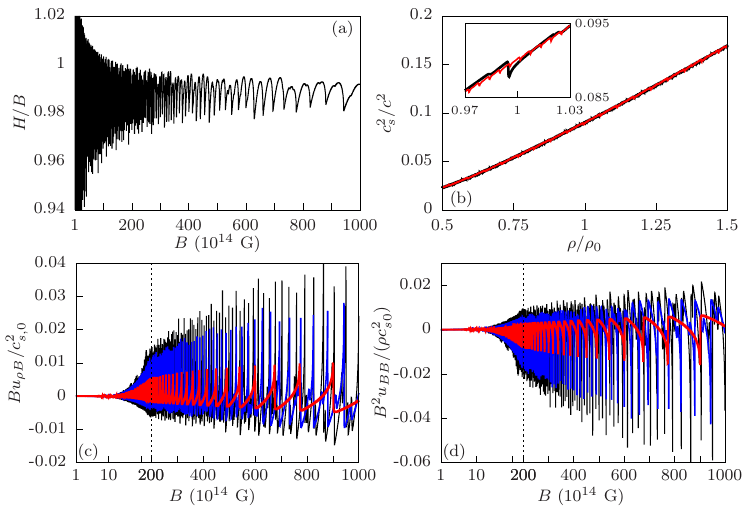}
\caption[$H/B$, $c_s^2$, $u_{\rho B}$ and $u_{BB}$ in ultrastrong magnetic fields and typical core densities]{\label{fig:LagrangianPartials} (a) $H/B$ as a function of $B$- this is comparable to~\citet{Broderick2000}, Figure 2. The fine details visible at $B\gtrsim 7\times 10^{16}$ G are due to including multiple fermion species; the overall curve is dominated by the Landau quantization of the electrons. (b) $c_s^2/c^2$ as a function of mass density for $B=10^{16}$ G (red) and $B=10^{17}$ G (black). The inset shows the small fluctuations in $c_s^2$ due to Landau quantization. (c) $Bu_{\rho B}/c_{s,0}^2$ as a function of $B$, where $c_{s,0}^2\equiv 0.09c^2$ is the approximate sound speed squared at nuclear saturation density $\rho_0$. The three curves are computed for fixed $\mu_{\text{e}}=80$ MeV (red), $125$ MeV (blue) and $150$ MeV (black), corresponding to $\rho\approx0.56\rho_0,\rho_0$ and $1.34\rho_0$ respectively. (d) $B^2u_{BB}/(\rho c_{s,0}^2)$ as a function of $B$. The three curves are computed for the same values of $\mu_e$ as panel (d). Note the change from logarithmic to linear scaling of the horizontal axis at $B=2\times10^{16}$ G, separated by a dotted line, in (c) and (d). All curves shown were computed using $T=10^8$ K.}
\end{figure}

The properties of the stellar model discussed in Section~\ref{sec:BackgroundModel} and the thermodynamic derivatives of $u_M$ discussed in Section~\ref{sec:Thermodynamics} are the only inputs required to compute the local stability criteria from Section~\ref{sec:StabilityAnalysis}. Before doing so, we compute the first and second order partial derivatives of $u_M$ with respect to $B$ and $\rho$. In Figure~\ref{fig:LagrangianPartials}, we plot $H$, $c_s^2$, $B\rho u_{\rho B}$ and $B^2u_{BB}/\rho$ as a function of $B$ or $\rho$, for a representative core temperatures for a magnetar over a few hundred years old~\citep{Potekhin2018} of $10^8$ K. Panel (a) shows that $B\approx H$ as mentioned in the introduction, and prominently shows the de Haas--van Alphen oscillations of $H$, resulting from the population of successive Landau levels as $B$ is decreased, where a new Landau level starts being populated significantly at $B=(\mu_a^2-m_a^2)/(2en)$ for integer $n$. At $T=0$, the slope of $H(B)$ is discontinuous at $B=(\mu_a^2-m_a^2)/(2en)$ for integer $n$; nonzero $T$ restores continuity. However, the derivatives of $H(B)$ fluctuate rapidly with large positive and negative values near $B=(\mu_a^2-m_a^2)/(2en)$. As the inset in (b) shows, the Landau quantization of the fermions also has a small effect on the sound speed, causing fluctuations of a few percent around its zero field value. Landau quantization also results in the spikes that are observed in $Bu_{\rho B}$ and $B^2u_{BB}/\rho$ for $B\gtrsim 10^{15}$. $Bu_{\rho B}$ and $B^2u_{BB}/\rho$ can clearly be negative, which will lead to potential instabilities. Since their sign change is associated with Landau quantization, we conclude that this will be the dominant mechanism in any potential instability that will occur as opposed to the vacuum Euler--Heisenberg Lagrangian. However, $Bu_{\rho B}$ and $B^2u_{BB}/\rho$ are much smaller than $c_s^2$, so it is clear that $V^2$ defined in Eq.~(\ref{eq:V2}) will be positive, and so for the $k_x=0$ perturbations to be unstable will require the $k_x=0$ magnetic buoyancy criterion SC1 (Eq.~(\ref{eq:SC1})) to be negative.

\begin{figure}
\center
\includegraphics[width=0.95\linewidth]{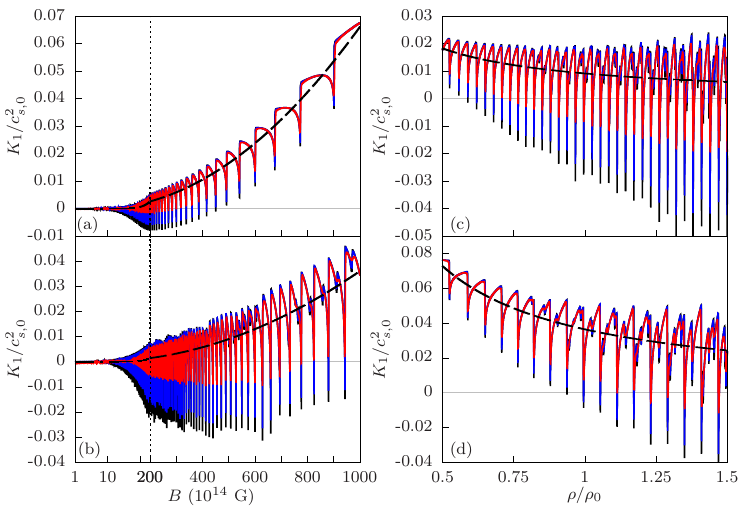}
\caption[MPR instability criterion in ultrastrong magnetic fields and typical core densities]{\label{fig:MPRFig} The MPR stability criterion $K_1$ defined in Eq.~(\ref{eq:V2}) divided by $c_{s,0}^2$ as a function of $B$ (a,b) and $\rho$ (c,d) for $T=5\times 10^7$ (black), $10^8$ K (blue) and $5\times 10^8$ K (red). Values below the thin grey line (i.e. negative values) are unstable. $\mu_{\text{e}}=80$ MeV ($\rho\approx0.56\rho_0$) and $\mu_{\text{e}}=125$ MeV ($\rho\approx\rho_0$) are used in (a) and (b) respectively; note the change from logarithmic to linear scaling of the horizontal axis at $B=2\times10^{16}$ G, separated by a dotted line. $B=5\times 10^{16}$ and $10^{17}$ G are used in (c) and (d). The Alfv\'{e}n velocity squared $BH/(4\pi\rho)\approx B^2/(4\pi\rho)$, also divided by $c_{s,0}^2$, is shown as a dot-dashed line in each panel. The irregular-looking behaviour which occurs in (c) and (d) for $\rho>0.79\rho_0$ is due to the presence of the muons, which misaligns the Fermi momenta of the protons and electrons. Additional evidence of this effect can be seen by comparing (a) and (b), since (a) is at a density below the muon threshold and (b) is above it.}
\end{figure}

The quantity $K_1$ defined in Eq.~(\ref{eq:K1}) and rewritten in Eq.~(\ref{eq:MPR1}) is plotted in Figure~\ref{fig:MPRFig} for a range of reasonable neutron star core temperatures $T=5\times10^7,10^8,5\times10^8$ K. The damping effect of increasing temperature is clearly demonstrated, but is important only where new Landau levels become populated i.e. where the second-order partial derivatives of $P$ would be divergent at zero temperature. Increasing the temperature stabilizes the fluid; as the temperature is increased, fewer ``spikes'' in $K_1$ become negative. The unstable regions are very narrow in $B$-space for constant $\rho$ and in $\rho$-space for constant $B$. 

Because $c_s^2$ can be so large, the behaviour of $K_1$ is dominated by the first term in Eq.~(\ref{eq:MPR1}). This term is approximately the Alfv\'{e}n velocity squared, multiplied by a factor which can change sign because $u_{BB}$ can be negative (Figure~\ref{fig:LagrangianPartials} (d)). The fluctuations of $u_{BB}$ are responsible for $K_1<0$ in parts of $B$--$\rho$ parameters space, indicating a possible MPR instability. Increasing $B$ at fixed density makes $K_1$ more stable, while increasing density at fixed $B$ leads to further unstable regions in parameter space. Recall that Alfv\'{e}n waves have magnetic tension as their restoring force, and that likewise there is a magnetic pressure contribution to the restoring force for standard magnetosonic waves. Intuitively, the instability occurs because the $(1+4\pi u_{BB})$ factor multiplying $B^2/\rho\propto v_{\text{A}}^2$ becoming negative corresponds to an effective negative magnetic tension/pressure. Thus changes in the field strength or direction in these unstable regions are energetically favourable, if the (always positive) sound speed (i.e. the matter pressure) is unable to stabilize the fluid. We discuss ways around stabilization by the sound speed at the conclusion of this section.

\begin{figure}
\center
\includegraphics[width=0.77\linewidth]{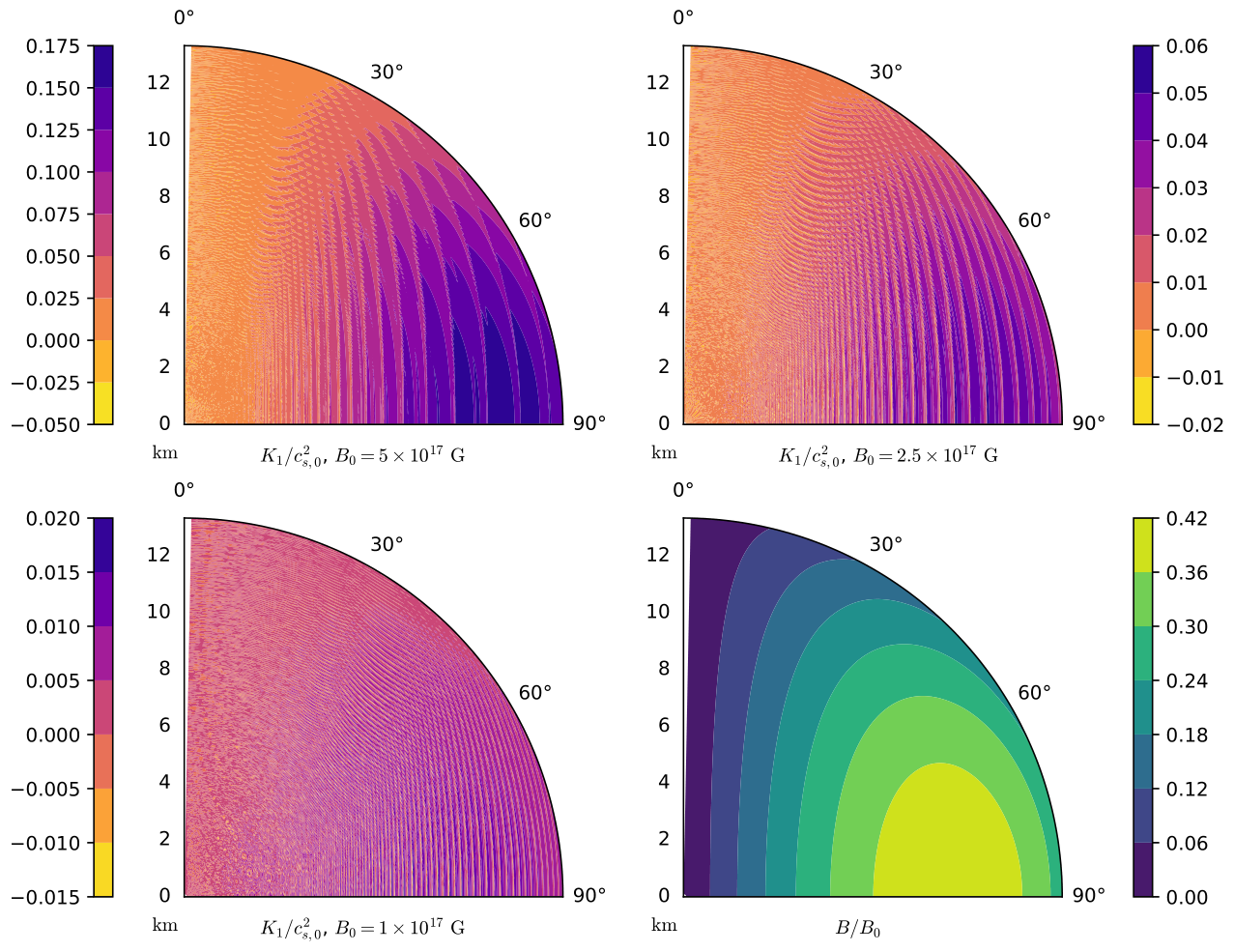}
\caption[MPR instability criterion inside a magnetar core model with toroidal magnetic field]{\label{fig:K1HeatMap} $K_1/c_{s,0}^2$ for different values of $B_0$, and $B/B_0$ (lower right), for a cross section of the model stellar core at $T=5\times10^{7}$ K. Negative values correspond to unstable regions. Instability is associated with negative values which occur near lines of $(\mu_{\text{e}}^2-m_{\text{e}}^2)/(2eB)=$ an integer. Note that some of the small-scale structure at high densities is spurious and due to interpolation based on a limited sampling in $B$--$\rho$ parameter space, and that the region along the symmetry axis $\theta<1^{\circ}$ was excluded as $B$ vanishes here.}
\end{figure}

In Figure~\ref{fig:K1HeatMap}, $K_1$ is plotted in the neutron star core for the field in Eq.~(\ref{eq:BForm}) with specific values of $B_0$. The maximum field inside the star is $\approx0.42B_0$. For fixed $B$, the higher density regions are more unstable (have more negative $K_1$), and as $B$ is increased, higher densities are required for the fluid to become unstable at all (c.f. Figure~\ref{fig:MPRFig}(c)--(d)). For fixed $\rho$, instability can occur at intermediate fields-- as $B\rightarrow0$, the fluid is stable, and for sufficiently high fields it can also become stable (c.f. Figure~\ref{fig:MPRFig}(a)--(b)), with the field required to stabilize the fluid increasing with increasing density.

\begin{figure}
\center
\centering
  \subfloat[]{\includegraphics[width=0.5\linewidth]{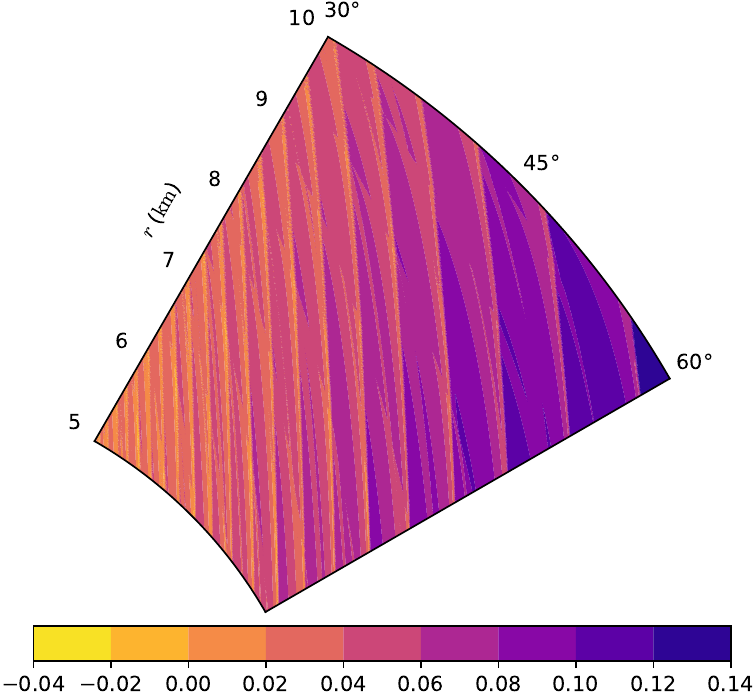}}
  \hfill
  \subfloat[]{\includegraphics[width=0.5\linewidth]{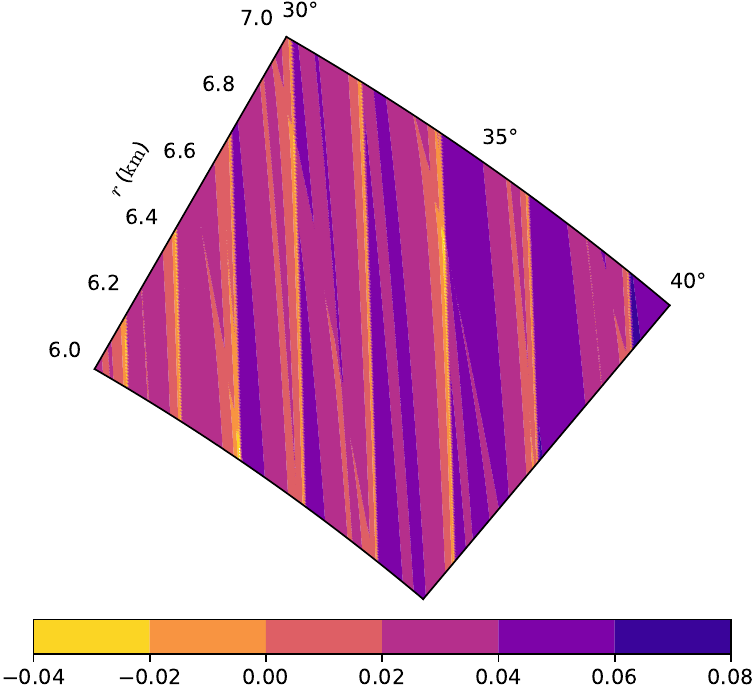}}
  \caption[MPR instability criterion in a magnified region of a magnetar core model with toroidal magnetic field]{\label{fig:K1HeatMapZoomedIn} (a) $K_1/c_{s,0}^2$ for a slice of a stellar core model seen in Figure~(\ref{fig:K1HeatMap}) ($B_0=5\times10^{17}$ G, $T=5\times10^{7}$ K); a further magnified section is shown in (b). This shows the fine details of the narrow unstable regions.}
\end{figure}

The unstable regions are localized in spheroidal shells where $B=(\mu_a^2-m_a^2)/(2en)$ for a fixed integer $n$ is satisfied-- as $n$ increases (e.g. with increasing density) these shells become more closely spaced, leading to the appearance of nearly continuously unstable regions deep in the core. This is observed in Figure~\ref{fig:K1HeatMap} as the field is decreased from the $B_0=5\times10^{17}$ G to the $B_0=2.5\times10^{17}$ G to the $B_0=1\times10^{17}$ G subfigures: the spacing between the unstable regions decreases, to the point that it is difficult to distinguish these regions in the $B_0=1\times10^{17}$ G subfigure. However, these locations are not required to be unstable, since for sufficiently strong $B$/sufficiently high temperature, they will be stable (c.f. Figure~\ref{fig:MPRFig} (a)--(b)). This leads to the highest field regions in the $B_0=5\times10^{17}$ subfigure of Figure~\ref{fig:K1HeatMap} being stable. Figure~\ref{fig:K1HeatMapZoomedIn} shows a magnified section of the $B_0=5\times10^{17}$ G, $T=5\times10^7$ K stellar model; the alternating pattern of stable and narrow unstable regions is more apparent here.

Figures~\ref{fig:K1HeatMap}--\ref{fig:K1HeatMapZoomedIn} show many unstable regions in the star of thickness $\lesssim$ a few meters. The spacing between these regions can be estimated as follows: since the unstable regions occur where new electron Landau levels start being populated, for neighbouring unstable regions we can write
\begin{equation}
\Delta n = 1 \approx\frac{(\mu_{\text{e}}+\Delta\mu_{\text{e}})^2-m_{\text{e}}^2}{2eB(1+\Delta B/B)}-\frac{\mu_{\text{e}}^2-m_{\text{e}}^2}{2eB}\approx \frac{(\mu_{\text{e}}+\Delta\mu_{\text{e}})^2}{2eB(1+\Delta B/B)}-n,
\label{eq:UnstableRegionSeparationEstimate1}
\end{equation}
where we can approximate the electrons as ultrarelativistic since $\mu_{\text{e}}\gg m_{\text{e}}$. $\Delta\mu_{\text{e}}$ and $\Delta B$ are the changes in $\mu_{\text{e}}$ and $B$ between the two unstable regions. Taking $\Delta B=(\text{d}B/\text{d}\varpi)\Delta\varpi$ and $\Delta\mu_{\text{e}}=(\text{d}\mu_{\text{e}}/\text{d}\varpi)\Delta\varpi$, Eq.~(\ref{eq:UnstableRegionSeparationEstimate1}) gives
\begin{equation}
\Delta\varpi=\left(2n\frac{\text{d}\mu_{\text{e}}}{\text{d}\varpi}-(n+1)\frac{\text{d}B}{\text{d}\varpi}\right)^{-1}.
\label{eq:UnstableRegionSeparationEstimate2}
\end{equation}
Working on the equator so $\varpi=r$, estimating from Figure~\ref{fig:CoreStructure} that $\rho\approx\rho_c(1-\varpi^2/R_{\star}^2)$, and also using the standard (and large number of occupied Landau levels result) that $\mu_{\text{e}}\approx(3\pi^2n_{\text{e}})^{1/3}\approx(3\pi^2Yf\rho/\overline{m})^{1/3}$ where we approximate $Y$ and $f$ as constant, we find
\begin{equation}
\Delta\varpi=-\frac{\rho}{\rho_c}\left(\frac{n+1}{\varpi}-\frac{\varpi}{R_{\star}^2}\left(\frac{5n}{3}+3\right)\right)^{-1},
\label{eq:UnstableRegionSeparationEstimate3}
\end{equation}
where the negative sign arises because the consecutive unstable regions for larger integers $n$ are located interior to each other. Thus $\Delta\varpi\sim-\varpi/n$, which matches what we observe: in the low-density regions far from the centre of the star, where $B$ can be larger and hence $n$ smaller, the spacing between potentially unstable regions $\Delta\varpi$ is larger (up to $\sim$km between them), whereas at shorter distances from the centre where $B$ is weaker/$n$ larger, the separation between neighbouring unstable regions can be of order meters to tens of meters. Thus the patterns of alternating stable-unstable regions extend into the higher density regions of the star, though resolving these becomes more difficult at high densities.

Following the argument of Eq.~(168) of~\citet{Akgun2008}, the growth time $\tau$ for the MPR instability can be estimated using the approximate dispersion relation
\begin{equation}
\omega^2=K_1k_x^2\rightarrow \tau=\frac{1}{|\text{Im}(\sqrt{K_1})k_x|},
\end{equation}
which applies for $k_z\gg k_x$. From Figure~\ref{fig:MPRFig}, in the unstable regions of parameter space $K_1\gtrsim -10^{-3}c^2$, so we obtain an approximate instability growth timescale
\begin{equation}
\tau\sim 1.6\times 10^{-3}\left(\frac{c}{\sqrt{1000|K_1|}}\right)\left(\frac{R^{-1}_{\star}}{k_x}\right)\text{ s}.
\label{eq:MPRGrowthTime}
\end{equation}
In the most unstable regions of the $B$--$\rho$--$T$ parameter space, this instability is so fast that it is unlikely to be mitigated by viscous dissipation. The required kinematic viscosity for the unstable modes to be dissipated is of the order
\begin{equation}
\nu \sim \frac{1}{k^2\tau}\approx 1.5\times 10^{15}\left(\frac{\sqrt{1000|K_1|}}{c}\right)\left(\frac{R^{-1}_{\star}}{k}\right)\left(\frac{k_x}{k}\right)\text{ cm}^2\text{ s}^{-1}.
\end{equation}
Even for $k_z\ll k\approx k_x$, this can be many orders of magnitude beyond the typical kinematic viscosities $\nu\sim10^6$--$10^7$ cm$^2$s$^{-1}$ expected in a normal fluid neutron star core, and also above the $\nu\sim10^7$--$10^8$ cm$^2$s$^{-1}$ viscosities expected in a superfluid-superconducting neutron star core~\citep{Schmitt2018}.

\begin{figure}
\center
\includegraphics[width=0.95\linewidth]{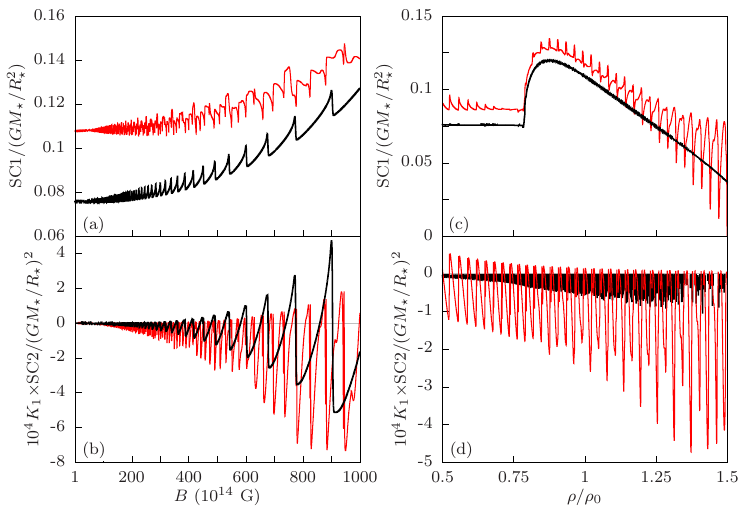}
\caption[Magnetic buoyancy criterion in an ultrastrong magnetic field and typical core densities 1]{\label{fig:MagneticBuoyancy1} The magnetic buoyancy stability criterion for $k_x=0$ defined in Eq.~(\ref{eq:SC1}) (SC1), and the stability criterion SC2 defined in~(\ref{eq:SC2}), as functions of $B$ (a,b) and $\rho$ (c,d). Each quantity is scaled by appropriate factors of $G$, $M_{\star}$ and $R_{\star}$ to make them dimensionless, while SC2 is multiplied by $K_1$ to eliminate divergent behaviour and an additional factor of $10^4$. In (a) and (b), the black and red curves correspond to $\mu_{\text{e}}=80$ MeV ($\rho\approx0.56\rho_0$) and $\mu_{\text{e}}=125$ MeV ($\rho\approx\rho_0$) respectively. In (c) and (d), the black and red curves correspond to $B=10^{15}$ and $5\times10^{16}$ G respectively. All curves are computed at $T=5\times10^8$ K. The $B=10^{15}$ G curve in (d) is multiplied by an additional factor of 2,000 (on top of the other factors) for display purposes.}
\end{figure}

Now that the behaviour $V^2$ and $K_1$ is understood, we examine the other stability criteria. In Figure~\ref{fig:MagneticBuoyancy1}, we show the $k_x=0$ stability criterion Eq.~(\ref{eq:SC1}) (SC1) and the first $k_x\neq 0$ criterion Eq.~(\ref{eq:SC2}) (SC2). In Figure~\ref{fig:MagneticBuoyancy1} we show the two $k_x\neq 0$ criteria Eq.~(\ref{eq:SC3}--\ref{eq:SC4}) (SC3,SC4). We note again that SC1, SC3 and SC4 are associated with magnetic buoyancy, while SC2 is a purely magnetohydrodynamic stability criterion. SC2, SC3 and SC4 have all been multiplied by a factor $K_1$ to eliminate the divergent behaviour for $K_1\rightarrow0$: since $K_1<0$ is already potentially unstable, such regions have been excised from the plots of SC2, SC3, and SC4.

\begin{figure}
\center
\includegraphics[width=0.95\linewidth]{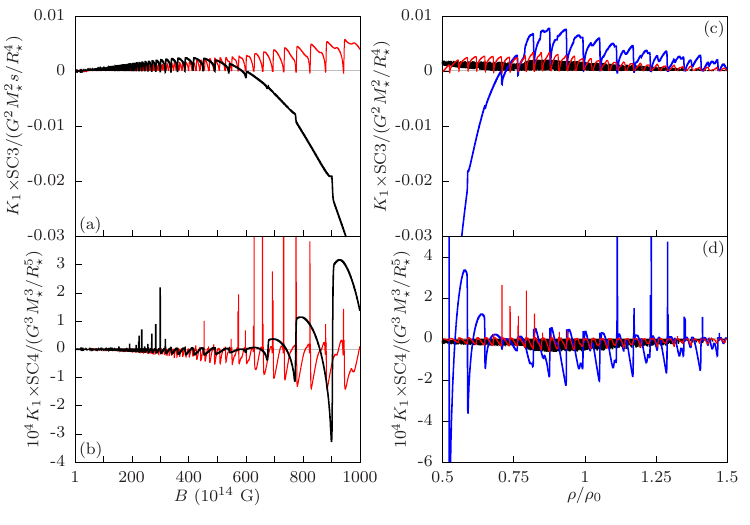}
\caption[Magnetic buoyancy criterion in an ultrastrong magnetic field and typical core densities 2]{\label{fig:MagneticBuoyancy2} The magnetic buoyancy stability criteria defined in Eq.~(\ref{eq:SC3}) (SC3) and~(\ref{eq:SC4}) (SC4), as functions of $B$ (a,b) and $\rho$ (c,d). Each quantity is scaled by appropriate factors of $G$, $M_{\star}$ and $R_{\star}$ to make them dimensionless, and both are multiplied by $K_1$ to eliminate divergent behaviour. SC4 is scaled by an additional factor of $10^4$. In (a) and (b), the black and red curves correspond to $\mu_{\text{e}}=80$ MeV ($\rho\approx0.56\rho_0$) and $\mu_{\text{e}}=125$ MeV ($\rho\approx\rho_0$) respectively. In (c) and (d), the black, red and blue curves correspond to $B=10^{15}$, $5\times10^{16}$ and $10^{17}$ G respectively. All curves are computed at $T=5\times10^8$ K. The $B=10^{15}$ G curves in (c) and (d) are multiplied by an additional factor of 100 and 10,000 respectively (on top of the other factors) for display purposes.}
\end{figure}

SC1 is always positive in the stellar model we have considered, and hence the fluid and field configuration we examine is magnetic buoyancy-stable for $k_x=0$ perturbations (i.e. purely radial perturbations). The stable stratification is large enough to suppress the instability i.e. $\tilde{N}$ is large enough to overwhelm the second term in Eq.~(\ref{eq:SC1}), which is proportional to the same term which results in negative values of $K_1$ and which could destabilize the system for sufficiently small $\tilde{N}$. This demonstrates the importance of stable stratification to MHD stability even in the strong-field case. SC2, SC3, and SC4 all involve $K_1$, which we have seen undergoes large fluctuations and can change sign for fields $\gtrsim 10^{15}$ G. Also note the similarities between Figure~\ref{fig:N2} and Figure~\ref{fig:MagneticBuoyancy1}--\ref{fig:MagneticBuoyancy2} panel (c), as the latter depend on scaled versions of $\tilde{N}^2$.

\begin{figure}
\center
\includegraphics[width=0.6\linewidth]{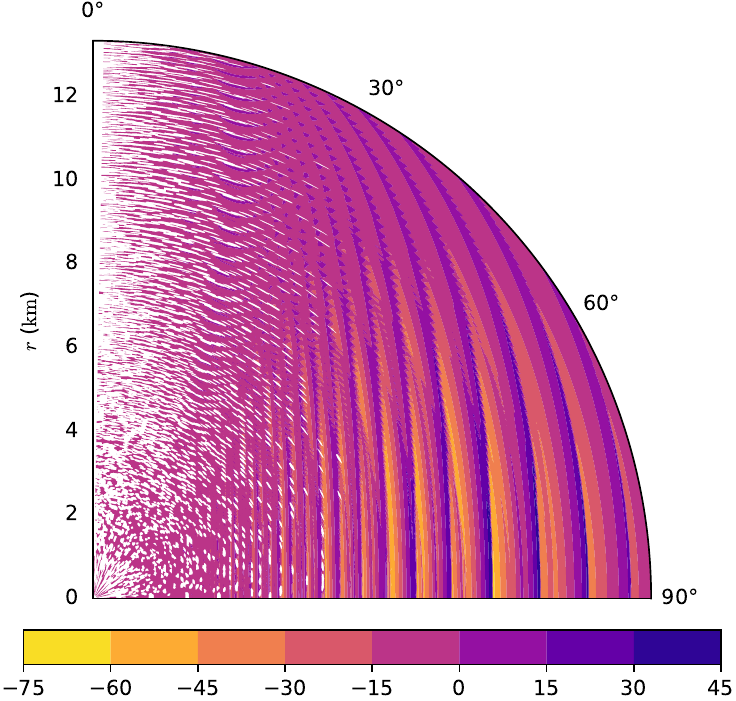}
\caption[Magnetic buoyancy criterion in a magnetar core model with toroidal magnetic field]{\label{fig:K1SC2HeatMap} $10^4K_1\times$SC2$/(GM_{\star}/R_{\star})^2$ for a the stellar core model  with $B_0=5\times10^{17}$ G and $T=5\times10^{8}$ K. Negative values correspond to potentially unstable regions. The empty regions correspond to where $K_1<0$ and the fluid is already unstable to the MPR instability-- such regions correspond to the unstable regions in the upper left plot of Figure~\ref{fig:K1HeatMap}, though not every $K_1<0$ region is resolved in each figure.}
\end{figure}

SC2, SC3, and SC4 can all be negative in large parts of the $B$--$\rho$ parameter space, contrasting with the limited regions of the parameter space in which $K_1$ can be negative which are localized at where additional Landau levels start filling. The region of parameter space where SC2 is negative increases as $B$ and $\rho$ are increased, as is demonstrated in Figure~\ref{fig:K1SC2HeatMap}, where $K_1\times$SC2 is plotted in a cross-section of the stellar model. SC3 is negative in limited regions where additional Landau levels start being populated, and also for regions where $B\gtrsim 6\times10^{16}$ G and $\rho\lesssim 0.8\rho_0$. The latter SC3 $<0$ regions are due to the second term in Eq.~(\ref{eq:kxNonzeroStabilityCriterionBuoyancy}), since the $\text{d}\ln B/\text{d}z=\text{d}\ln\rho/\text{d}z+\frac{1}{z}$ is the most negative at low densities (see Figure~\ref{fig:CoreStructure}) and the coefficient of this derivative becomes large for large $B$. SC4 is generally positive or marginally negative at $B\lesssim 2\times10^{16}$ G, and can be unstable for all densities examined.

The stability implications of these results can be further examined using the definition of the canonical energy per unit mass in the $k_x\neq0$ case. Upon taking $k_x\rightarrow 0$ as before and using $c_s^2\gg Bu_{\rho B}$ and $|g\xi^z/c_s^2|\ll 1$, Eq.~(\ref{eq:CanonicalEnergyMBKxNonzero3}) becomes
\begin{align}
\mathcal{E}_c[\xi]\approx{}&c^2_s\left|\partial_z\xi^z+ik_y\xi^y\right|^2
+K_1\left|\partial_z\xi^z+ik_y\xi^y\left(1-\frac{2Bu_{\rho B}}{K_1}\right)+\frac{L_2}{K_1}\xi^z\right|^2
+(\text{SC2})k_y^2\left|\xi^y\right|^2+(\text{SC3})\left|\xi^{z}\right|^2
\nonumber
\\
{}&\pm 2k_y\sqrt{(\text{SC2})(\text{SC3})-(\text{SC4})}\text{Re}\left[i\xi^y(\xi^z)^*\right].
\label{eq:CanonicalEnergykxNonzeroApproximation}
\end{align}
First, we note that in the absence of gravity, only the first three terms will be present. In the zero gravity, vacuum MHD case, $K_1$ would be positive definite and SC2 would be zero, and hence instability impossible. In the non-vacuum, strong field case examined in this paper, neither $K_1$ nor SC2 can be sufficiently negative such that they are larger in magnitude than $c_s^2$, but it is still possible for certain types of modes to have zero canonical energy and hence be unstable despite the strong stabilizing effect provided by the first term. For example, for incompressible modes the first term is zero, and hence such modes could be made unstable by $K_1$ and SC2 being negative. These instabilities would likely result in transfer of magnetic flux within the star to stabilize the fluid by decreasing or increasing the field strength in the unstable regions. They are analogous to the MPR instability in superconducting fluids in which flux tubes are clumped together to create regions of greater/lesser magnetic flux. Since the $k_x=0$ modes are stable, the unstable perturbations which transfer flux to try to stabilize the fluid must have a nonzero component along the magnetic field.

In the case where the buoyancy terms SC3 and SC4 are relevant, these can further destabilize the fluid, but like $K_1$ and SC2 will require a particular class of perturbation (e.g. incompressible) to overcome the stabilizing effect of the sound speed squared term in Eq.~(\ref{eq:CanonicalEnergykxNonzeroApproximation}). However, the pure MHD terms are much more significant for the stability than those associated with buoyancy except for the longest wavelength perturbations. The minimum relevant wave number for a perturbation is $k_{\text{min}}\approx 2\pi/R_{\star}\sim 4\times10^{-6}$ cm$^{-1}$, at which point $K_1k_{\text{min}}^2\sim$SC2$k_{\text{min}}^2\sim$SC3. For shorter wavelength perturbations the wave number dependence of the pure MHD terms mean that they will be much larger in magnitude than the magnetic buoyancy terms.

\section{Discussion and conclusion}
\label{sec:Conclusion}

In this paper we have extended the magnetohydrodynamic stability analysis of~\citet{Friedman1978} and~\citet{Glampedakis2007} to magnetizable fluids with $B\neq H$, and applied this to the study of MHD stability in the presence of extremely strong fields where QED effects and Landau quantization of fermions are relevant. This regime is important to magnetars, which have surface fields $\sim10^{15}$ G and likely stronger fields in their interiors. Using the canonical energy approach to studying stability, we determined sufficient local stability criteria for magnetic buoyancy, magnetosonic stability and the Muzikar--Pethick--Roberts (MPR) instability for magnetizable media with an electromagnetic Lagrangian depending on the magnetic field $B$, mass density $\rho$ and species fractions $Y_a$. The fluid perturbation theory and canonical energy results derived here are general and can be used for other applications where $H\neq B$ e.g. superfluid-superconducting neutron stars.

We find that the inclusion of strong-field effects to the MHD of a neutron star core leads to a possible Muzikar-Pethick-Roberts (MPR) instability associated with the Landau quantization of fermions for fields $B\gtrsim 10^{15}$ G and at typical core densities. These instabilities can have growth times on the order of $10^{-3}$ s, and are not stabilized by viscous dissipation under expected neutron star viscosities. The magnetic buoyancy instability associated with the strong-field, $B\neq H$ MHD studied here can also be active in the neutron star core, and the stability criteria associated with it are violated in broader regions of the $B$--$\rho$ parameter space than the narrow regions in which the MPR instability criterion is violated. However, these terms are only comparable in magnitude to the purely MHD terms associated with the MPR instability for long wavelength perturbations, and are unimportant for stability at shorter wavelengths. The vacuum Euler--Heisenberg Lagrangian was found to be unimportant to MHD stability overall at the magnetic fields $B\lesssim 10^{17}$ G we considered.

Perhaps the most important question regarding the MPR instability discussed in this paper is what implications it has on the evolution of magnetar fields. The alternating stable-unstable regions in $B$--$\rho$ space, demonstrated in a stellar model in Figures~\ref{fig:K1HeatMap}--\ref{fig:K1HeatMapZoomedIn}, suggest that the instability could lead to the formation of magnetic domains. This instability will likely be triggered throughout the star's life for two reasons: 1) previously stable regions become unstable as the star cools and the ``spikes'' in $u_{BB}$ become sharper (c.f. Figure~\ref{fig:MPRFig}); 2) field evolution that changes the field strength within the star such that certain regions now lie in the unstable parts of $B$--$\rho$ parameter space. One response to this destabilization is the transfer of magnetic flux from unstable regions until the local field is in the stable region of the $B$--$\rho$ parameter space. Since the unstable regions themselves generally span very narrow length scales of order $\lesssim$ meters, the magnetic field only needs to change by a small fraction of its strength to be stabilized, so unstable modes could transfer flux to stabilize the fluid. This suggests that the observational implications for these domains and the instability that could form them are limited, at least within the stellar core.

The formation of magnetic domains in strongly-magnetized neutron star matter has previously been discussed in~\citet{Blandford1982} and~\citet{Suh2010}. In these calculations it is the differential magnetic susceptibility $\chi$, corresponding here to $-u_{BB}$ ~\footnote{Though noting the other references perform the calculation of the differential susceptibility at fixed chemical potentials and not fixed density as used here.}, which encourages magnetic domain formation, and when $\chi>4\pi$ the system is thermodynamically unstable to domain formation. The MPR instability discussed in our paper also originates from $u_{BB}<-1/(4\pi)$, in contrast to the original MPR instability in superconducting fluids which depends on the $u_{\rho B}^2$ term in $K_1$. Magnetic domain formation encouraged by the instability discussed here suggests that smoothly-varying fields are not solutions to the background magnetohydrostatic equilibrium for sufficiently strong magnetic fields $\gtrsim 10^{15}$ G and at low temperatures.

Though we have made some simplifying assumptions for the background stellar model, our conclusions about the MPR instability arising from the Euler--Heisenberg--Fermi--Dirac Lagrangian should be robust. This is because the instability depends on $u_{BB}$, which is simply a function of $B$ and the $n_a$ (or the $\mu_a$), and is independent of the exact structure of the star (including the EOS) or the magnetic field direction. We thus do not expect our main argument about the generic instability of ultrastrong fields to change significantly when using a relativistic stellar model, different EOS and/or including the effect of the magnetic field on the background equilibrium model. Our restriction to toroidal fields does not affect our main results about the instability to MPR modes, since the same requirements on $K_1$ will appear independent of the direction of $B^i$: if $B^i$ was a poloidal field, only the magnetic buoyancy conditions, and the exact location of the unstable regions within the star, would be changed.

The next logical step is to apply this formalism to the crust~\citep{RauInPrep}, which due to lower densities and sound speeds may support similar instabilities as in the core even with lower field strengths. However, the stability analysis in the crust is fundamentally different due to the lack of a canonical energy procedure~\citep{Lyutikov2013}, so this is left to a follow-up paper. The canonical energy procedure can also be applied to a complete model of a neutron star core to study its global stability. An additional interesting area of future study would be to include a physically correct stellar exterior, which would modify the treatment of the surface terms examined in Appendix~\ref{app:HermiticityCSurface}, in which exterior vacuum was assumed. Such a study could examine the effect on the stability of different pulsar magnetosphere models and could potentially help judge their feasibility.

\section*{Acknowledgements}

We would like to thank the referee for useful comments. P.B.R. would also like to acknowledge the support of Cornell's Boochever Fellowship for spring 2020.

\section*{Data Availability}

The data underlying this article will be shared on reasonable request to the corresponding author.

\appendix

\section{Fluid perturbation theory}
\label{app:FluidPerturbationTheory}

The Eulerian perturbation $\delta Q$ and Lagrangian perturbation $\Delta  Q$ of quantity $Q$, which can be a scalar, vector or tensor, are related by
\begin{equation}
\Delta Q=\delta Q+\pounds_{\xi} Q,
\end{equation}
where $\Lie_{\xi}$ is the Lie derivative with respect to the Lagrangian displacement field $\xi^i=\Delta x^i$. We will occasionally denote the displacement field with respect to which the perturbations are considered using a subscript i.e. $\Delta_{\xi}Q$ is the Lagrangian perturbation of $Q$ in the case where $\xi^i$ is the Lagrangian displacement describing the position of the fluid elements with respect to their position in the unperturbed fluid. The Lie derivatives of a scalar $f$, vector $a^i$ and covariant vector $a_i$ are
\begin{subequations}
\begin{align}
\Lie_{\xi}f ={}& \xi^{j}\nabla_{j}f,
\\
\Lie_{\xi}a^i ={}& \xi^{j}\nabla_{j}a^i-a^j\nabla_{j}\xi^i,
\\
\Lie_{\xi}a_i ={}& \xi^{j}\nabla_{j}a_i+a_j\nabla_i\xi^j.
\end{align}
\end{subequations}
We now list the Eulerian and Lagrangian perturbations of relevant quantities-- for proof of these, see~\citet{Friedman1978}, \citet{Shapiro1983}\footnote{This reference uses the convention $\Delta Q=\delta Q+\xi^j\nabla_jQ$ so its results will only match ours for scalar $Q$.} and \citet{Glampedakis2007}:
\begin{align}
\Delta v^i={}&\partial_t\xi^i, \qquad\qquad \delta v^i = \partial_t\xi^i + v^j\nabla_j\xi^i - \xi^j\nabla_jv^i,
\\
\Delta \rho={}&-\rho\nabla_j\xi^j, \qquad \delta\rho = -\nabla_j(\rho\xi^j),
\\
\Delta s ={}&0, \qquad\qquad\qquad \delta s = -\xi^j\nabla_js,
\\
\Delta Y_a ={}&0, \qquad\qquad\qquad \delta Y_a = -\xi^j\nabla_jY_a,
\\
\Delta B^i={}&-B^i\nabla_j\xi^j, \qquad \delta B^i = -B^i\nabla_j\xi^j - \xi^j\nabla_jB^i + B^j\nabla_j\xi^i,
\label{eq:BPerturbations}
\end{align}
where we have assumed adiabatic perturbations ($\Delta s=0$) and slow species-converting reactions ($\Delta Y_a=0$). The unperturbed quantities in these equations are the background properties of the star. Indices are raised/lowered using the metric tensor $g_{ij}$, and raising/lowering indices of perturbed quantities also requires using
\begin{equation}
\Delta g_{ij}=\nabla_i\xi_j+\nabla_j\xi_i, \qquad \Delta g^{ij}=-(\nabla^i\xi^j+\nabla^j\xi^i), \qquad \delta g_{ij} = 0.
\label{eq:gijPerturbations}
\end{equation}
From the above, the Lagrangian perturbation of the Levi-Civita tensor is
\begin{equation}
\Delta \epsilon^{ijk} = -\epsilon^{ijk}\nabla_{\ell}\xi^{\ell}, \qquad \Delta \epsilon_{ijk} = \epsilon_{ijk}\nabla_{\ell}\xi^{\ell},
\end{equation}
which involves using the identity $\epsilon^{ijk}=1/\sqrt{g}[ijk]$, $\epsilon_{ijk}=\sqrt{g}[ijk]$ where $g$ is the metric determinant and $[ijk]$ is the Levi-Civita symbol. We also have the perturbed Poisson equation for the gravitational potential:
\begin{equation}
\nabla^2\delta\Phi=4\pi G\delta\rho.
\end{equation}

Since we consider a magnetic field-dependent pressure, we modify the usual definition for the pressure perturbation: the Lagrangian perturbation is defined as
\begin{equation}
\Delta P = \left.\frac{\partial P}{\partial\rho}\right|_{s,Y_a,B}\Delta\rho+
\left.\frac{\partial P}{\partial B}\right|_{\rho,s,Y_a}\Delta B + 
\left.\frac{\partial P}{\partial s}\right|_{\rho,Y_a,B}\Delta s+
\sum_a\left.\frac{\partial P}{\partial Y_a}\right|_{\rho,s,Y_a\neq Y_b,B}\Delta Y_a = \frac{\gamma P}{\rho}\Delta\rho+\zeta\Delta B,
\label{eq:PressureLagrangianPerturbation}
\end{equation}
where the adiabatic index of the perturbations $\gamma$ and the adiabatic, constant field sound speed are defined in Eq.~(\ref{eq:AdiabaticIndex}), and we have defined
\begin{equation}
\zeta \equiv \left.\frac{\partial P}{\partial B}\right|_{\rho,s,Y_a}.
\end{equation}
$\Delta B$ is derived using Eq.~(\ref{eq:BPerturbations},\ref{eq:gijPerturbations}):
\begin{equation}
\Delta B = -B\nabla_j\xi^j+B^j\hat{B}^k\nabla_k\xi_j.
\label{eq:DeltaBScalar}
\end{equation}
The Eulerian perturbation of $P$ is hence
\begin{equation}
\delta P = -\xi^j\nabla_j P -\gamma P\nabla_j\xi^j-\zeta B\nabla_j\xi^j,
\end{equation}
which can be rewritten in the suggestive form
\begin{equation}
\frac{\delta P}{P}=\gamma\frac{\delta\rho}{\rho}+\zeta\frac{\delta B}{P}+\gamma\xi^j\left(\nabla_j\ln\rho+\frac{\zeta B}{\gamma P}\nabla_j\ln B-\frac{1}{\gamma}\nabla_j\ln P\right),
\label{eq:PressureEulerianPerturbation2}
\end{equation}
from which one can define the Brunt--V\"{a}is\"{a}l\"{a} frequency in the case where $P$ depends on the magnetic field. Since we generally work at zero temperature, we ignore the entropy density in the remainder of this appendix, and no longer explicitly denote that it is being held constant.

We employ the following identities involving derivative and perturbation operators:
\begin{subequations}
\begin{align}
{}&\delta\partial_t = \partial_t\delta, \qquad \Delta \partial_t = \partial_t\Delta - \Lie_{\partial_t\xi},
\\
{}&\delta\nabla_i = \nabla_i\delta, \qquad \Delta(\nabla_if) = \nabla_i\Delta f,
\\
{}&\Delta(\nabla_ia_j) = \nabla_i\Delta a_j-a_k\nabla_i\nabla_j\xi^k+\xi^k(\nabla_k\nabla_i-\nabla_i\nabla_k)a_j, \qquad \Delta(\nabla_ia^j) = \nabla_i\Delta a^j+a^k\nabla_i\nabla_k\xi^j+\xi^k(\nabla_k\nabla_i-\nabla_i\nabla_k)a^j,
\\
{}&\Delta(\partial_t+\Lie_v)v^{i} = (\partial_t+\Lie_v)\Delta v^{i}, \qquad \Delta(\partial_t+\Lie_v)v_{i} = (\partial_t+\Lie_v)\Delta v_{i}.
\end{align}
\end{subequations}
Dividing Eq.~(\ref{eq:EulerEquation}) by $\rho$ and then taking its Lagrangian perturbation gives the perturbed Euler equation
\begin{align}
\partial_t^2\xi_i+2v^j\nabla_j\partial_t\xi_i+(v^j\nabla_j)^2\xi_i+\frac{1}{\rho}\nabla_j\xi^j\nabla_iP-\frac{1}{\rho}\nabla_i(\gamma P\nabla_j\xi^j){}&-\frac{1}{\rho}\nabla_i(\zeta B\nabla_j\xi^j)+\frac{1}{\rho}\nabla_i(\zeta B^j\hat{B}^k\nabla_k\xi_j)+\nabla_i\delta\Phi+\xi^j\nabla_j\nabla_i\Phi
\nonumber
\\
{}&+\nabla_i\xi^j(v^k\nabla_kv_j+\nabla_j\Phi)=\Delta\left(\frac{1}{\rho} \nabla^jT_{ij}^B\right).
\label{eq:PerturbedEuler1}
\end{align}

Following~\citet{Friedman1978}, after multiplication by $\rho$, we can rewrite Eq.~(\ref{eq:PerturbedEuler1}) as
\begin{equation}
A[\partial_t^2\xi_i]+B[\partial_t\xi_i]+C[\xi_i]=0,
\label{eq:MasterEquation1}
\end{equation}
where $A$, $B$ and $C$ are operators depending on the background quantities. $A$ and $B$ are identical to the operators of the same name appearing in~\citet{Friedman1978}, and are Hermitian/anti-Hermitian over the inner product
\begin{equation}
\langle \eta^i,\xi_i\rangle=\int\eta^{i*}\xi_idV,
\end{equation}
where $\eta^i$ is another Lagrangian displacement field distinct from $\xi^i$ and the volume integral is over the entire star. The $C$ operator is Hermitian-- the algebra showing this appears in Appendix~\ref{app:HermiticityC}. The canonical energy of the perturbations $E_c[\xi]$ is defined as
\begin{equation}
E_c[\xi]=\frac{1}{2}\langle\partial_t\xi^i,A[\partial_t\xi_i]\rangle+\frac{1}{2}\langle\xi^i,C[\xi_i]\rangle,
\end{equation}
which can be expressed in terms of physical parameters and the displacement field $\xi^i$ using Eq.~(\ref{eq:CCombined},\ref{eq:SurfaceI}).

\subsection{Hermiticity of $C$ operator}
\label{app:HermiticityC}

Previous results~\citep{Friedman1978,Glampedakis2007} mean that proving the Hermiticity of the operator $C$ is a matter of showing that the additional terms resulting from the $B$-dependence of $P$ and the inclusion of $\Lagr_{\text{EH}}$ are Hermitian. Using the following identity
\begin{equation}
\Delta_{\xi}(\nabla^jT_{ij})=\nabla^j\delta_{\xi} T_{ij}+\xi^k\nabla_k\nabla^jT_{ij}+\nabla_i\xi^k\nabla^jT_{kj},
\end{equation}
and the background Euler equation
\begin{equation}
0=\rho(v^j\nabla_j)v_i+\nabla_iP+\rho\nabla_i\Phi-\nabla^jT^B_{ij},
\label{eq:BackgroundEuler}
\end{equation}
Eq.~(\ref{eq:PerturbedEuler1}) can be expressed as
\begin{align}
\partial_t^2\xi_i+2v^j\nabla_j\partial_t\xi_i+(v^j\nabla_j)^2\xi_i+\frac{1}{\rho}\nabla_k\xi^k\nabla_iP-\frac{1}{\rho}\nabla_i(\gamma P\nabla_j\xi^j)-\frac{1}{\rho}\nabla_i(\zeta B\nabla_j\xi^j)+\frac{1}{\rho}\nabla_i(\zeta B^j\hat{B}^k\nabla_k\xi_j)-\frac{1}{\rho}\nabla_i\xi^j\nabla_jP+\nabla_i\delta\Phi+\xi^j\nabla_j\nabla_i\Phi
\nonumber
\\
=\frac{1}{\rho}\nabla_k\xi^k\nabla^jT^B_{ij}+\frac{1}{\rho}\nabla^j\delta_{\xi}T^B_{ij} + \frac{1}{\rho}\xi^k\nabla_k\nabla^jT^B_{ij}.
\label{eq:PerturbedEuler2}
\end{align}
Multiplying by $\rho\eta^i$, the resulting expression for $C$ can be split into two parts:
\begin{equation}
\eta^iC[\xi_i]\equiv\eta^iC_{0}[\xi_i]+\eta^iC_{\text{EM}}[\xi_i],
\label{eq:CSplit}
\end{equation}
where
\begin{subequations}
\begin{align}
\eta^iC_{0}[\xi_i]={}&\rho\eta^i(v^j\nabla_j)^2\xi_i+\eta^i\nabla_k\xi^k\nabla_iP-\eta^i\nabla_i(\gamma P\nabla_j\xi^j)-\eta^i\nabla_i\xi^j\nabla_jP+\rho\eta^i\nabla_i\delta\Phi+\rho\eta^i\xi^j\nabla_j\nabla_i\Phi,
\label{eq:C0NonHermitian}
\\
\eta^iC_{\text{EM}}[\xi_i]={}&-\eta^i\nabla_k\xi^k\nabla^jT^B_{ij}-\eta^i\nabla^j\delta_{\xi}T^B_{ij} - \eta^i\xi^k\nabla_k\nabla^jT^B_{ij}.
\label{eq:CEMNonHermitian}
\end{align}
\end{subequations}
The Hermiticity of $C_{0}$ was shown in~\citet{Friedman1978}, while the Hermiticity of $C_{\text{EM}}$ in the $H_i=B_i$ limit was shown in~\citet{Glampedakis2007}.

For reference, we present the Hermitian form of $\eta^iC_{0}[\xi_i]$:
\begin{align}
\eta^iC_{0}[\xi_i] ={}& -\rho v^j\nabla_j\xi_iv^k\nabla_k\eta^i+\nabla_iP\left(\xi^i\nabla_j\eta^j+\eta^i\nabla_j\xi^j\right)+\eta^i\xi^j(\nabla_i\nabla_jP+\rho\nabla_i\nabla_j\Phi)+\gamma P\nabla_j\xi^j\nabla_i\eta^i
-\frac{1}{4\pi G}\nabla_i\delta_{\xi}\Phi\nabla^i\delta_{\eta}\Phi
\nonumber
\\
{}&+\nabla_i\bigg[\rho\eta^kv^iv^j\nabla_j\xi_k-\eta^i\xi^j\nabla_jP-\gamma P\eta^i\nabla_j\xi^j+\rho\eta^i\delta_{\xi}\Phi+\frac{1}{4\pi G}\delta_{\xi}\Phi\nabla^i\delta_{\eta}\Phi\bigg].
\label{eq:C0}
\end{align}
To demonstrate the Hermiticity of $C_{\text{EM}}$, start by integrating by parts the terms containing $\zeta$ and $\delta_{\xi}T^B_{ij}$:
\begin{equation}
\eta^iC_{\text{EM}}[\xi_i]=\zeta B\nabla_j\xi^j\nabla_i\eta^i-\zeta B^j\hat{B}^k\nabla_k\xi_j\nabla_i\eta^i+\nabla_i\left(\zeta\Delta_{\xi} B\eta^i\right)-\eta^i\nabla_k\xi^k\nabla^jT^B_{ij}-\eta^i\xi^k\nabla_k\nabla^jT^B_{ij}-\nabla^j\left(\eta^i\delta_{\xi}T^B_{ij}\right)+\delta_{\xi}T^B_{ij}\nabla^j\eta^i.
\label{eq:CEMHermitization1}
\end{equation}
Writing $T^B_{ij}=-\mathcal{P}_Bg_{ij}+H_iB_j/4\pi$ using Eq.~(\ref{eq:MagneticPressure}), Eq.~(\ref{eq:CEMHermitization1}) becomes
\begin{subequations}
\begin{align}
\eta^iC_{\text{EM}}[\xi_i]={}&\eta^iC^a_{\text{EM}}[\xi_i]+\eta^iC^b_{\text{EM}}[\xi_i],
\\
\eta^iC^a_{\text{EM}}[\xi_i]\equiv{}&\eta^i\xi^k\nabla_k\nabla_i\mathcal{P}_B+\nabla_i\left(\eta^i\delta_{\xi}\mathcal{P}_B\right)+\nabla_k\xi^k\eta^i\nabla_i\mathcal{P}_B-\delta_{\xi}\mathcal{P}_B\nabla_i\eta^i,
\label{eq:CEMHermitization2A}
\\
\eta^iC^b_{\text{EM}}[\xi_i]\equiv{}&\zeta B\nabla_j\xi^j\nabla_i\eta^i-\zeta B^j\hat{B}^k\nabla_k\xi_j\nabla_i\eta^i+\nabla_i\left(\zeta\Delta_{\xi} B\eta^i\right)-\frac{1}{4\pi}\nabla_k\xi^k\eta^iB^j\nabla_jH_i-\frac{1}{4\pi}\eta^i\xi^k\nabla_k\left(B^j\nabla_jH_i\right)-\frac{1}{4\pi}\eta^iB^j\nabla_j\delta_{\xi}H_i
\nonumber
\\
{}&-\frac{1}{4\pi}\eta^i\delta_{\xi}B^j\nabla_jH_i,
\label{eq:CEMHermitization2B}
\end{align}
\end{subequations}
where $\nabla_jB^j=0=\nabla_j\delta_{\xi}B^j$ was used to obtain this form of $\eta^iC^b_{\text{EM}}[\xi_i]$. $\eta^iC^a_{\text{EM}}[\xi_i]$ can be rewritten using Eq.~(\ref{eq:DeltaBScalar}) and the definitions of Eq.~(\ref{eq:LagrangianPartialAbbreviations}):
\begin{align}
\delta_{\xi}\mathcal{P}_B=\Delta_{\xi}\mathcal{P}_B-\xi^j\nabla_j\mathcal{P}_B={}&-\xi^j\nabla_j\mathcal{P}_B-\frac{1}{4\pi}\left(B^2\nabla_j\xi^j-B_jB^k\nabla_k\xi^j\right)-\frac{\partial u_{\text{mat}}}{\partial B}B\nabla_j\xi^j+\frac{\partial u_{\text{mat}}}{\partial B}B^j\hat{B}^k\nabla_k\xi_j-u_{BB}B^2\nabla_j\xi^j
\nonumber
\\
{}&+u_{BB}B^jB^k\nabla_k\xi_j-u_{\rho B}\rho B\nabla_j\xi^j.
\label{eq:deltaP_B}
\end{align}
Abbreviating $\partial u_{\text{mat}}/\partial B|_{\rho,Y_a}\equiv u^{\text{mat}}_B$, $\eta^iC^a_{\text{EM}}[\xi_i]$ then becomes
\begin{align}
\eta^iC^a_{\text{EM}}[\xi_i]={}&\eta^i\xi^k\nabla_k\nabla_i\mathcal{P}_B+\nabla_i\mathcal{P}_B\left(\eta^i\nabla_k\xi^k+\xi^i\nabla_k\eta^k\right)+\frac{1}{4\pi}B^2\nabla_j\xi^j\nabla_i\eta^i+u^{\text{mat}}_BB\nabla_j\xi^j\nabla_i\eta^i+u_{BB}B^2\nabla_j\xi^j\nabla_i\eta^i+u_{\rho B}\rho B\nabla_j\xi^j\nabla_i\eta^i
\nonumber
\\
{}&+\nabla_i\left(\eta^i\delta_{\xi}\mathcal{P}_B\right)-u^{\text{mat}}_BB^j\hat{B}^k\nabla_k\xi_j\nabla_i\eta^i-u_{BB}B^jB^k\nabla_k\xi_j\nabla_i\eta^i-\frac{1}{4\pi}B_jB^k\nabla_k\xi^j\nabla_i\eta^i.
\label{eq:CEMHermitizationA1}
\end{align}
The first six terms on the right-hand side of this equation are symmetric and the seventh is a total divergence term. Define the remaining, final three terms as $\eta^iC^{a,2}_{\text{EM}}[\xi_i]$
\begin{equation}
\eta^iC^{a,2}_{\text{EM}}[\xi_i]\equiv-u^{\text{mat}}_BB^j\hat{B}^k\nabla_k\xi_j\nabla_i\eta^i-u_{BB}B^jB^k\nabla_k\xi_j\nabla_i\eta^i-\frac{1}{4\pi}B_jB^k\nabla_k\xi^j\nabla_i\eta^i.
\label{eq:CEMHermitizationA2}
\end{equation}

Returning to Eq.~(\ref{eq:CEMHermitization2B}): expand the fourth term on the right-hand side, integrate by parts the sixth term and use $\nabla_jB^j=0$, then use Eq.~(\ref{eq:BPerturbations}) to replace $\delta_{\xi} B_i$ in the final term. This gives
\begin{align}
\eta^iC^b_{\text{EM}}[\xi_i]={}&\zeta B\nabla_j\xi^j\nabla_i\eta^i-\zeta B^j\hat{B}^k\nabla_k\xi_j\nabla_i\eta^i+\nabla_i\left(\zeta\Delta_{\xi} B\eta^i\right)-\frac{1}{4\pi}\eta^iB^j\xi^k\nabla_k\nabla_jH_i-\frac{1}{4\pi}\nabla_j\left(\eta^iB^j\delta_{\xi}H_i\right)+\frac{1}{4\pi}\delta_{\xi}H_iB^j\nabla_j\eta^i
\nonumber
\\
{}&-\frac{1}{4\pi}B^k\nabla_k\xi^j\eta^i\nabla_jH_i
\label{eq:CEMHermitizationB1}
\end{align}
Since $u_M$ only depends on the magnitude of $B^i$, we can use Eq.~(\ref{eq:duMdBi},\ref{eq:BPerturbations}) and
\begin{align}
\frac{\partial^2 u_M}{\partial B^i\partial B^j}={}&\left(g_{ij}-\hat{B}_i\hat{B}_j\right)\frac{1}{B}u_B+\hat{B}_i\hat{B}_ju_{BB},
\label{eq:d2uMdB2ij}
\end{align}
to write the Eulerian perturbation of $H_i$ as
\begin{align}
\delta_{\xi}H_i=\delta_{\xi}B_i+4\pi\delta_{\xi}\frac{\partial u_M}{\partial B^i}={}&\delta_{\xi}B_i+\xi^j\nabla_j(B_i-H_i)-4\pi\rho \frac{\partial^2u_M}{\partial\rho\partial B^i}\nabla_j\xi^j-4\pi\frac{\partial^2u_M}{\partial B^i\partial B^k}B^k\nabla_j\xi^j+4\pi\frac{\partial^2u_M}{\partial B^i\partial B^k}B^j\nabla_j\xi^k
\nonumber
\\
={}&-B_i\nabla_j\xi^j+B^j\nabla_j\xi_i-\xi^j\nabla_jH_i-4\pi(g_{ij}-\hat{B}_i\hat{B}_j)u_B\hat{B}^k\nabla_k\xi^j-4\pi u_{BB}B_i\nabla_j\xi^j+4\pi u_{BB}\hat{B}_i\hat{B}_jB^k\nabla_k\xi^j
\nonumber
\\
{}&-4\pi u_{\rho B}\rho\hat{B}_i\nabla_j\xi^j.
\label{eq:deltaHi}
\end{align}
Inserting this in the second-last term in Eq.~(\ref{eq:CEMHermitizationB1}), we obtain
\begin{align}
\eta^iC^b_{\text{EM}}[\xi_i]={}&\zeta B\nabla_j\xi^j\nabla_i\eta^i+\frac{1}{4\pi}B^j\nabla_j\eta^iB^k\nabla_k\xi_i+(g_{ij}-\hat{B}_i\hat{B}_j)u_BB\hat{B}^k\nabla_k\xi^j\hat{B}^{\ell}\nabla_{\ell}\eta^i+u_{BB}\hat{B}_iB^j\nabla_j\eta^i\hat{B}_{\ell}B^k\nabla_k\xi^{\ell}-\zeta B^j\hat{B}^k\nabla_k\xi_j\nabla_i\eta^i
\nonumber
\\
{}&+\nabla_i\left(\zeta \Delta_{\xi}B\eta^i\right)-\frac{1}{4\pi}\nabla_j\left(B^j\eta^i\delta_{\xi}H_i\right)-\frac{1}{4\pi}B^iB^j\nabla_j\eta_i\nabla_k\xi^k-u_{BB}B^iB^j\nabla_j\eta_i
\nabla_k\xi^k\nonumber
\\
{}&-u_{\rho B}\rho\hat{B}_iB^j\nabla_j\eta^i\nabla_k\xi^k-\frac{1}{4\pi}\eta^iB^j\xi^k\nabla_k\nabla_jH_i-\frac{1}{4\pi}B^k\nabla_k\xi^j\eta^i\nabla_jH_i
-\frac{1}{4\pi}B^j\nabla_j\eta^i\xi^k\nabla_kH_i.
\label{eq:CEMHermitizationB2}
\end{align}
The first four terms of this equation are Hermitian. Integrating the final term by parts and adding Eq.~(\ref{eq:CEMHermitizationA2}) to this gives, after combining the total divergence terms,
\begin{align}
\eta^iC^b_{\text{EM}}[\xi_i]+\eta^iC^{a,2}_{\text{EM}}[\xi_i]={}&\zeta B\nabla_j\xi^j\nabla_i\eta^i+\frac{1}{4\pi}B^j\nabla_j\eta^iB^k\nabla_k\xi_i+(g_{ij}-\hat{B}_i\hat{B}_j)u_BB\hat{B}^k\nabla_k\xi^j\hat{B}^{\ell}\nabla_{\ell}\eta^i+u_{BB}\hat{B}_iB^j\nabla_j\eta^i\hat{B}_{\ell}B^k\nabla_k\xi^{\ell}
\nonumber
\\
{}&-u_{BB}B^iB^j\left(\nabla_i\eta_j\nabla_k\xi^k+\nabla_i\xi_j\nabla_k\eta^k\right)-\frac{1}{4\pi}B^iB^j\left(\nabla_i\eta_j\nabla_k\xi^k+\nabla_i\xi_j\nabla_k\eta^k\right)-\zeta B^j\hat{B}^k\nabla_k\xi_j\nabla_i\eta^i
\nonumber
\\
{}&-u_{\rho B}\rho\hat{B}_iB^j\nabla_j\eta^i\nabla_k\xi^k-u^{\text{mat}}_BB^j\hat{B}^k\nabla_k\xi_j\nabla_i\eta^i+\nabla_i\left(\zeta\Delta_{\xi}B\eta^i-\frac{1}{4\pi}B^i\eta^j\left[\delta_{\xi}H_j+\xi^k\nabla_kH_j\right]\right)
\nonumber
\\
{}&+\frac{1}{4\pi}\eta^iB^j\xi^k\left(\nabla_k\nabla_j-\nabla_j\nabla_k\right)H_i.
\label{eq:CEMHermitizationB3}
\end{align}
The final term is zero since
\begin{equation}
\left(\nabla_j\nabla_k-\nabla_k\nabla_j\right)H_i=H_{\ell}R^{\ell}_{ijk}=0,
\end{equation}
and the Riemann tensor $R^{\ell}_{ijk}$ is zero in flat spacetime. 

The first six terms in Eq.~(\ref{eq:CEMHermitizationB3}) are Hermitian. To proceed, we need to determine what $\zeta$ is in terms of $u_B$, $u_{BB}$, $u_{\rho B}$ and $u^{\text{mat}}_B$. Using Eq.~(\ref{eq:dP},\ref{eq:uB},\ref{eq:urhoB}) and $\rho=m_{\text{N}}n_{\text{b}}$, we can show that
\begin{equation}
\zeta=\left.\frac{\partial P}{\partial B}\right|_{\rho,Y_a}=-u_B^{\text{mat}}+\rho u_{\rho B}.
\end{equation}
This eliminates one of the remaining non-Hermitian terms and makes the other Hermitian:
\begin{align}
\eta^iC^b_{\text{EM}}[\xi_i]+\eta^iC^{a,2}_{\text{EM}}[\xi_i]={}&-u^{\text{mat}}_BB\nabla_j\xi^j\nabla_i\eta^i+u_{\rho B}\rho B\nabla_j\xi^j\nabla_i\eta^i+\frac{BH}{4\pi}\hat{B}^j\nabla_j\eta^i\hat{B}^k\nabla_k\xi^i+\left(u_{BB}-\frac{u_B}{B}\right)\hat{B}_iB^j\nabla_j\eta^i\hat{B}_{\ell}B^k\nabla_k\xi^{\ell}
\nonumber
\\
{}&-\left(u_{BB}+\frac{\rho u_{\rho B}}{B}\right)B^iB^j\left(\nabla_i\eta_j\nabla_k\xi^k+\nabla_i\xi_j\nabla_k\eta^k\right)-\frac{1}{4\pi}B^iB^j\left(\nabla_i\eta_j\nabla_k\xi^k+\nabla_i\xi_j\nabla_k\eta^k\right)
\nonumber
\\
{}&+\nabla_i\left(\zeta\Delta_{\xi}B\eta^i-\frac{1}{4\pi}B^i\eta^j\left[\delta_{\xi}H_j+\xi^k\nabla_kH_j\right]\right).
\label{eq:CEMHermitizationB4}
\end{align}
This is fully Hermitian, but there is a final step that we take. We write the second-last term in Eq.~(\ref{eq:CEMHermitizationB4}) as
\begin{equation}
-\frac{1}{4\pi}B^iB^j\left(\nabla_i\eta_j\nabla_k\xi^k+\nabla_i\xi_j\nabla_k\eta^k\right)=\frac{u_B}{B}B^iB^j\left(\nabla_i\eta_j\nabla_k\xi^k+\nabla_i\xi_j\nabla_k\eta^k\right)-\frac{1}{4\pi}B^iH^j\left(\nabla_i\eta_j\nabla_k\xi^k+\nabla_i\xi_j\nabla_k\eta^k\right),
\end{equation}
the second term of which is no longer Hermitian. Noting the second term on the right-hand side of Eq.~(\ref{eq:C0}), using the background Euler equation Eq.~(\ref{eq:BackgroundEuler}), equals
\begin{equation}
\nabla_iP\left(\xi^i\nabla_j\eta^j+\eta^i\nabla_j\xi^j\right)=\left[-\rho\left(v^j\nabla_jv_i+\nabla_i\Phi\right)-\nabla_i\mathcal{P}_B+\frac{1}{4\pi}B^k\nabla_kH_i\right]\left(\xi^i\nabla_j\eta^j+\eta^i\nabla_j\xi^j\right),
\label{eq:C0TermRearrangement}
\end{equation}
we can add Eq.~(\ref{eq:C0TermRearrangement}) to Eq.~(\ref{eq:CEMHermitizationB4}) , giving
\begin{align}
\eta^iC^b_{\text{EM}}[\xi_i]+\eta^iC^{a2}_{\text{EM}}[\xi_i]+\nabla_iP\left(\xi^i\nabla_j\eta^j+\eta^i\nabla_j\xi^j\right)={}&-\rho\left[v^j\nabla_jv_i+\nabla_i\Phi\right]\left(\xi^i\nabla_j\eta^j+\eta^i\nabla_j\xi^j\right)-\nabla_i\mathcal{P}_B\left(\xi^i\nabla_j\eta^j+\eta^i\nabla_j\xi^j\right)
\nonumber
\\
{}&-u^{\text{mat}}_BB\nabla_j\xi^j\nabla_i\eta^i+u_{\rho B}\rho B\nabla_j\xi^j\nabla_i\eta^i+\frac{BH}{4\pi}\hat{B}^j\nabla_j\eta^i\hat{B}^k\nabla_k\xi^i
\nonumber
\\
{}&+\left(u_{BB}-\frac{u_B}{B}\right)\hat{B}_iB^j\nabla_j\eta^i\hat{B}_{\ell}B^k\nabla_k\xi^{\ell}
\nonumber
\\
{}&+\left(\frac{u_B}{B}-u_{BB}-\frac{\rho u_{\rho B}}{B}\right)B^iB^j\left(\nabla_i\eta_j\nabla_k\xi^k+\nabla_i\xi_j\nabla_k\eta^k\right)
\nonumber
\\
{}&+\frac{1}{4\pi}\nabla_i\eta^iB^j(\xi^k\nabla_jH_k-H_k\nabla_j\xi^k)+\frac{1}{4\pi}\nabla_i\xi^iB^j(\eta^k\nabla_jH_k-H_k\nabla_j\eta^k)
\nonumber
\\
{}&+\nabla_i\left(\zeta\Delta_{\xi}B\eta^i-\frac{1}{4\pi}B^i\eta^j\left[\delta_{\xi}H_j+\xi^k\nabla_kH_j\right]\right).
\label{eq:CEMHermitizationB5}
\end{align}
Eq.~(\ref{eq:CEMHermitizationB5}) is entirely symmetric plus total divergence terms. 

Combining Eq.~(\ref{eq:C0},\ref{eq:CEMHermitizationA2},\ref{eq:CEMHermitizationB5}) gives the explicitly Hermitian expression for $\eta^iC[\xi_i]$
\begin{align}
\eta^iC[\xi_i]={}& -\rho v^j\nabla_j\xi_iv^k\nabla_k\eta^i+\eta^i\xi^j\nabla_i\nabla_j\left(P+\mathcal{P}_B\right)+\rho\eta^i\xi^j\nabla_i\nabla_j\Phi-\frac{1}{4\pi G}\nabla_i\delta_{\xi}\Phi\nabla^i\delta_{\eta}\Phi+\left(\gamma P+\frac{B^2}{4\pi}+u_{BB}B^2+2\rho u_{B\rho}B\right)\nabla_j\xi^j\nabla_i\eta^i
\nonumber
\\
{}&+\frac{BH}{4\pi}\hat{B}^j\nabla_j\eta^i\hat{B}^k\nabla_k\xi^i+\left(u_{BB}-\frac{u_B}{B}\right)\hat{B}_iB^j\nabla_j\eta^i\hat{B}_{\ell}B^k\nabla_k\xi^{\ell}+\left(\frac{u_B}{B}-u_{BB}-\frac{\rho u_{\rho B}}{B}\right)B^iB^j\left(\nabla_i\eta_j\nabla_k\xi^k+\nabla_i\xi_j\nabla_k\eta^k\right)
\nonumber
\\
{}&+\frac{1}{4\pi}\nabla_i\eta^iB^j(\xi^k\nabla_jH_k-H_k\nabla_j\xi^k)+\frac{1}{4\pi}\nabla_i\xi^iB^j(\eta^k\nabla_jH_k-H_k\nabla_j\eta^k)-\rho\left(v^k\nabla_kv_i+\nabla_i\Phi\right)\left[\xi^i\nabla_j\eta^j+\eta^i\nabla_j\xi^j\right]
\nonumber
\\
{}&+\nabla_i\bigg[\rho\eta^kv^iv^j\nabla_j\xi_k+\rho\eta^i\delta_{\xi}\Phi+\frac{1}{4\pi G}\delta_{\xi}\Phi\nabla^i\delta_{\eta}\Phi+\eta^i\delta_{\xi}\left(P+\mathcal{P}_B\right)-\frac{1}{4\pi}B^i\eta^j\left(\delta_{\xi}H_j+\xi^k\nabla_kH_j\right)\bigg].
\label{eq:CCombined}
\end{align}
Since we will always be integrating $\eta^iC[\xi_i]$ over the volume of the star, the total divergence terms can be converted into surface terms. We now show that they are also Hermitian.

\subsection{Surface terms}
\label{app:HermiticityCSurface}

From Eq.~(\ref{eq:CCombined}), the combined total divergence term in $\eta^iC[\xi_i]$ equals $\nabla_iS^i$ where
\begin{equation}
S^i\equiv\rho\eta^kv^iv^j\nabla_j\xi_k+\rho\eta^i\delta_{\xi}\Phi+\frac{1}{4\pi G}\delta_{\xi}\Phi\nabla^i\delta_{\eta}\Phi+\eta^i\delta_{\xi}\left(P+\mathcal{P}_B\right)-\frac{1}{4\pi}B^i\eta^j(\delta_{\xi}H_j+\xi^k\nabla_kH_j).
\label{eq:SurfaceVector1}
\end{equation}
Making the simplifying assumption that the exterior of the star is vacuum threaded by a magnetic field, we have $\rho=0$ and $P=0$ at the surface, so we drop the first and fourth  terms in Eq.~(\ref{eq:SurfaceVector1}). The remaining terms can be rewritten as
\begin{equation}
S^i=\eta^i\delta_{\xi}\left(P+\mathcal{P}_B\right)-\frac{1}{4\pi}B^i\eta^j\left(\delta_{\xi}H_j+\xi^k\nabla_kH_j\right)+\frac{1}{4\pi G}\delta_{\xi}\Phi\nabla^i\delta_{\eta}\Phi.
\end{equation}
The $\delta\Phi$-dependent term here can also be shown to be Hermitian~\citep{Lynden-Bell1967}. We thus define
\begin{equation}
\tilde{S}^i\equiv\eta^i\delta_{\xi}\left(P+\mathcal{P}_B\right)-\frac{1}{4\pi}B^i\eta^j\left(\delta_{\xi}H_j+\xi^k\nabla_kH_j\right),
\label{eq:SurfaceVector2}
\end{equation}
whose Hermiticity we will now demonstrate.

Following a similar procedure to~\citet{Glampedakis2007}, to show that $\tilde{S}^i$ is Hermitian requires defining a surface unit normal vector $\hat{n}_i$ with the Lagrangian perturbation
\begin{equation}
\Delta\hat{n}_i=\hat{n}_i\hat{n}_j\hat{n}^k\nabla_k\xi^j.
\end{equation}
We also have the boundary conditions resulting from Gauss' Law for magnetism and Faraday's Law,
\begin{align}
\hat{n}_{i}\langle B^i\rangle{}&=0,
\label{eq:GaussMagnetismBoundary}
\\
\epsilon^{ijk}\hat{n}_{j}\langle E_k\rangle{}&=\frac{1}{c}(\hat{n}_jv^j)\langle B^i\rangle,
\label{eq:FaradayBoundary}
\end{align}
where $\langle F\rangle=F^{x}-F$ is the difference between the exterior quantity $F^{x}$ and the interior quantity $F$ at the boundary. Through taking the Lagrangian perturbation of Eq.~(\ref{eq:FaradayBoundary}), we find
\begin{equation}
\hat{n}_j\xi^jB^i_{x}-\hat{n}_jB^j\xi^i+\epsilon^{ijk}\hat{n}_j\delta A^{x}_{k}=0,
\label{eq:FaradayBoundary2}
\end{equation}
where $A_k$ is the vector potential.

We also enforce continuity of the component of the fluid-magnetic stress tensor $T_{ij}$
\begin{equation}
T_{ij}=-Pg_{ij}+T^B_{ij},
\end{equation}
normal to the stellar surface. Continuity of $T_{ij}$ normal to the stellar surface implies, using Eq.~(\ref{eq:GaussMagnetismBoundary}),
\begin{equation}
\langle T_{ij}\hat{n}^j\rangle=0\rightarrow \hat{n}_{i}\left\langle P+\mathcal{P}_B\right\rangle=\frac{1}{4\pi}(\hat{n}_jB^j)\langle H_i\rangle.
\label{eq:TijContinuity}
\end{equation}
Contracting with the unit normal vector gives
\begin{equation}
\left\langle P+\mathcal{P}_B\right\rangle=\frac{1}{4\pi}(\hat{n}_jB^j)\hat{n}^i\langle H_i\rangle,
\end{equation}
and hence inserting this back into Eq.~(\ref{eq:TijContinuity}), we find
\begin{equation}
(\hat{n}_jB^j)(g_{ik}-\hat{n}_{i}\hat{n}_k)\langle H^k\rangle=0.
\label{eq:TijContinuity2}
\end{equation}
Taking the Lagrangian perturbation of Eq.~(\ref{eq:TijContinuity2}) gives
\begin{align}
(\hat{n}_jB^j)\left[(g_{ik}-\hat{n}_{i}\hat{n}_k)\Delta_{\xi}\langle H^k\rangle+(\nabla_i\xi_k+\nabla_k\xi_i-2\hat{n}_i\hat{n}_k\hat{n}_{\ell}\hat{n}^m\nabla_m\xi^{\ell})\langle H^k\rangle\right]=0,
\\
(\hat{n}_jB^j)(\delta_{i}^{k}-\hat{n}_{i}\hat{n}^k)\Delta_{\xi}\langle H_k\rangle=0,
\label{eq:PerturbedTijContinuity2}
\end{align}
where Eq.~(\ref{eq:TijContinuity2}) was used to simplify Eq.~(\ref{eq:PerturbedTijContinuity2}).

Now converting the Eulerian perturbation to a Lagrangian one in the first term of Eq.~(\ref{eq:SurfaceVector2}) and using Eq.~(\ref{eq:TijContinuity}) plus $\rho_x=0=P_x$ gives
\begin{align}
\tilde{S}^i={}&\eta^i\xi^j\nabla_j\left[\left\langle P+\mathcal{P}_B\right\rangle-\frac{1}{4\pi}\hat{n}_kB^k\hat{n}^{\ell}\langle H_{\ell}\rangle\right]+\eta^i\delta_{\xi}\left(-\frac{1}{8\pi}B^2_x-u_M^x+\frac{1}{4\pi}H^x_kB^k_x\right)
\nonumber
\\
{}&-\frac{1}{4\pi}\eta^i\delta_{\xi}\left(\hat{n}_jB^j\hat{n}^k\langle H_k\rangle\right)-\frac{1}{4\pi}B^i\eta^j(\delta_{\xi}H_j+\xi^k\nabla_kH_j).
\label{eq:SurfaceVector3}
\end{align}
Eq.~(\ref{eq:TijContinuity}) also implies
\begin{equation}
\epsilon^{i\ell m}\hat{n}_{\ell}\nabla_m\left[\left\langle P+\mathcal{P}_B\right\rangle-\frac{1}{4\pi}(\hat{n}_jB^j)\hat{n}^k\langle H_k\rangle\right]=0,
\label{eq:InterfaceCurlCondition1}
\end{equation}
and hence contracting with $\epsilon_{npi}\xi^p$ gives
\begin{equation}
\hat{n}_i\xi^j\nabla_j\left[...\right]=\hat{n}_j\xi^j\nabla_i\left[...\right],
\label{eq:InterfaceCurlCondition2}
\end{equation}
where $[...]$ is the argument inside square brackets in Eq.~(\ref{eq:InterfaceCurlCondition1}). We note that
\begin{equation}
\delta_{\xi}\left(-\frac{1}{8\pi}B^2_x-u_M^x+\frac{1}{4\pi}H^x_kB^k_x\right)=\left(-\frac{1}{4\pi}B^x_j-u_j^{B,x}\right)\delta_{\xi}B_x^j+\frac{1}{4\pi}H^x_j\delta_{\xi}B^j_x+\frac{1}{4\pi}B_x^j\delta_{\xi}H_j^x=\frac{1}{4\pi}B_x^j\delta_{\xi}H_j^x,
\label{eq:SurfaceVectorTerm1}
\end{equation}
and that Amp\`{e}re's Law outside of the star implies
\begin{equation}
\epsilon^{ijk}\nabla_{j}H^x_k=0\rightarrow \nabla_jH^x_k=\nabla_kH^x_j,
\label{eq:ExteriorAmperesLaw}
\end{equation}
so
\begin{equation}
\delta_{\xi}H_j+\xi^k\nabla_kH_j=-\delta_{\xi}\langle H_j\rangle-\delta H_j^x-\xi^k\nabla_k\langle H_j\rangle+\frac{1}{2}\xi^k\left(\nabla_jH^x_k+\nabla_kH^x_j\right).
\label{eq:SurfaceVectorTerm2}
\end{equation}
Integrating the divergence of Eq.~(\ref{eq:SurfaceVector3}) over the volume of the star and using the divergence theorem, then using Eq.~(\ref{eq:FaradayBoundary2},\ref{eq:InterfaceCurlCondition2},\ref{eq:SurfaceVectorTerm1},\ref{eq:SurfaceVectorTerm2}) gives
\begin{align}
\int\nabla_i\tilde{S}^i\text{d}V=\oint \text{d}S\Bigg({}&\hat{n}_i\eta^i\hat{n}_j\xi^j\hat{n}^k\nabla_k\left[\left\langle P+\mathcal{P}_B\right\rangle-\frac{1}{4\pi}\hat{n}_{\ell}B^{\ell}\hat{n}^m\langle H_m\rangle\right]-\frac{1}{8\pi}\hat{n}_i B^i\eta^j\xi^k\left(\nabla_j H^x_k+\nabla_k H^x_j\right)
\nonumber
\\
{}&-\frac{1}{4\pi}\epsilon^{jik}\hat{n}_i\delta_{\eta}A_k^x\delta_{\xi}H^x_j+\frac{1}{4\pi}\hat{n}_iB^i\eta^j\left(\delta_{\xi}\langle H_j\rangle+\xi^k\nabla_k\langle H_j\rangle\right)-\frac{1}{4\pi}\hat{n}_i\eta^i\delta_{\xi}\left(\hat{n}_jB^j\hat{n}^k\langle H_k\rangle\right)\Bigg).
\label{eq:SurfaceVector4}
\end{align}
The terms in the top line are clearly symmetric under $\xi^i\leftrightarrow\eta^i$. The term containing $A^x_k$ can be rearranged using
\begin{equation}
\oint \text{d}S\epsilon^{jik}\hat{n}_i\delta_{\eta}A_k^x\delta_{\xi}H^x_j=-\int_x\text{d}V\nabla_{i}\left(\epsilon^{ijk}\delta_{\eta}A_k^x\delta_{\xi}H^x_j\right)=-\int_x\text{d}V\left(\delta_{\eta}B_x^k\delta_{\xi}H^x_k-\delta_{\eta}A_i^x\epsilon^{ijk}\nabla_j\delta_{\xi}H^x_k\right).
\label{eq:SurfaceVectorATerm}
\end{equation}
where the volume integral is now over the exterior of the star, denoted by a subscript $x$. By Eq.~(\ref{eq:ExteriorAmperesLaw}) the second term in this integral is zero, and since (noting that $\rho_x=s_x=Y_a^x=0$)
\begin{equation}
\delta_{\xi}H^x_k=\delta_{\xi}B^x_k+4\pi\delta_{\xi}\frac{\partial u_M^x}{\partial B^k_x}=\delta_{\xi}B^x_k+4\pi\frac{\partial^2 u_M^x}{\partial B^k_x\partial B^j_x}\delta_{\xi}B^j_x,
\end{equation}
the first term in Eq.~(\ref{eq:SurfaceVectorATerm}) is also symmetric. Looking at the remaining terms in Eq.~(\ref{eq:SurfaceVector3}), using Eq.~(\ref{eq:TijContinuity2},\ref{eq:PerturbedTijContinuity2}) we can show that
\begin{align}
{}&\frac{1}{4\pi}\oint \text{d}S\left[\hat{n}_iB^i\eta^j\left(\delta_{\xi}\langle H_j\rangle+\xi^k\nabla_k\langle H_j\rangle\right)-\hat{n}_i\eta^i\delta_{\xi}\left(\hat{n}_jB^j\hat{n}^k\langle H_k\rangle\right)\right]
\nonumber
\\
{}&=\frac{1}{4\pi}\oint \text{d}S\left[\hat{n}_iB^i\eta^j\left(\Delta_{\xi}\langle H_j\rangle-\langle H_k\rangle\nabla_j\xi^k\right)-\hat{n}_i\eta^i\hat{n}_jB^j\hat{n}^k\Delta_{\xi}\langle H_k\rangle-\hat{n}_i\eta^i\langle H_k\rangle\Delta_{\xi}\left(\hat{n}^k\hat{n}_jB^j\right)+\hat{n}_i\eta^i\xi^{\ell}\nabla_{\ell}\left(\hat{n}_jB^j\hat{n}^k\langle H_k\rangle\right)\right]
\nonumber
\\
{}&=\frac{1}{4\pi}\oint \text{d}S\left[\hat{n}_iB^i\eta^j(\delta^k_j-\hat{n}_j\hat{n}^k)\Delta_{\xi}\langle H_k\rangle-\hat{n}_iB^i\langle H_k\rangle\eta^j\nabla_j\xi^k+\eta^i\langle H_i\rangle\hat{n}_jB^j\nabla_{\ell}\xi^{\ell}+\hat{n}_i\eta^i\xi^{\ell}\nabla_{\ell}\left(\hat{n}_jB^j\hat{n}^k\langle H_k\rangle\right)\right]
\nonumber
\\
{}&=\frac{1}{4\pi}\oint \text{d}S\left[-\hat{n}_iB^i\langle H_k\rangle\eta^j\nabla_j\xi^k+\hat{n}_j\eta^j\nabla_{\ell}\left(\xi^{\ell}\hat{n}_iB^i\hat{n}^k\langle H_k\rangle\right)\right]
\nonumber
\\
{}&=\frac{1}{4\pi}\oint \text{d}S\hat{n}_i\left[\frac{1}{2}\nabla_j\left(\left(\eta^i\xi^j+\eta^j\xi^i\right)\hat{n}_{\ell}B^{\ell}\hat{n}^k\langle H_k\rangle\right)-B^i\langle H_k\rangle\left(\eta^j\nabla_j\xi^k+\xi^j\nabla_j\eta^k\right)\right],
\label{eq:FinalSurfaceTerm}
\end{align}
which is also clearly symmetric. Note that we used that all components of the Riemann tensor $R^{i}_{\ jk\ell}$ are zero to be able to symmetrize the first term in the last line of Eq.~(\ref{eq:FinalSurfaceTerm}). Hence the entirety of the total divergence term in $C$ is Hermitian when we impose that the exterior of the star is vacuum. We can thus define
\begin{align}
\mathcal{I}(\eta^i,\xi_i)=\int \nabla_iS^i\text{d}V={}&\oint \text{d}S\Bigg(\hat{n}_i\eta^i\hat{n}_j\xi^j\hat{n}^k\nabla_k\left[\left\langle P+\mathcal{P}_B\right\rangle-\frac{\hat{n}_{\ell}B^{\ell}}{4\pi}\hat{n}^m\langle H_m\rangle\right]-\frac{\hat{n}_i B^i}{8\pi}\eta^j\xi^k\left(\nabla_j H^x_k+\nabla_k H^x_j\right)
\nonumber
\\
{}&\quad\qquad+\frac{1}{8\pi}\hat{n}_i\nabla_j\left[\left(\eta^i\xi^j+\eta^j\xi^i\right)\hat{n}_{\ell}B^{\ell}\hat{n}^k\langle H_k\rangle\right]-\frac{1}{4\pi}\hat{n}_iB^i\langle H_k\rangle\left(\eta^j\nabla_j\xi^k+\xi^j\nabla_j\eta^k\right)+\frac{1}{4\pi G}\nabla_i\left(\delta_{\xi}\Phi\nabla^i\delta_{\eta}\Phi\right)\Bigg)
\nonumber
\\
{}&+\frac{1}{4\pi}\int_x\text{d}V\left(g_{jk}+4\pi \frac{\partial^2 u_M^x}{\partial B^j_x\partial B^k_x}\right)\delta_{\eta}B_x^j\delta_{\xi}B_x^k.
\label{eq:SurfaceI}
\end{align}

\section{Thermodynamic derivatives of Euler--Heisenberg--Fermi--Dirac Lagrangian}
\label{app:Thermodynamics}

In this appendix we give explicit expressions for partial derivatives of $P$ and $\Lagr_{\text{EH}}$ as used in the definitions of $u_B$, $u_{BB}$, etc. From Eq.~(\ref{eq:PressureNeutrons}), we have
\begin{subequations}
\begin{align}
{}&\left.\frac{\partial P_{\text{f},\text{n}}}{\partial\mu_{\text{n}}}\right|_{B,\phi}=\frac{(\mu_{\text{n}}^{*2}-m_*^2)^{3/2}}{3\pi^2},
\\
{}&\left.\frac{\partial^2 P_{\text{f},\text{n}}}{\partial\mu_{\text{n}}^2}\right|_{B,\phi}=\frac{\mu_{\text{n}}^*\sqrt{\mu_{\text{n}}^{*2}-m_*^2}}{\pi^2}.
\end{align}
\end{subequations}
From Eq.~(\ref{eq:PressureProtons}--\ref{eq:PressureLeptons}) and Eq.~(\ref{eq:EHLagrangianVacuum}), and remembering to take $m_a\rightarrow m_*$, $\mu_a\rightarrow\mu_a^*$ on the right-hand side of these equations when $a=\text{p}$,
\begin{subequations}
\begin{align}
{}&\left.\frac{\partial P_{\text{f},a}}{\partial\mu_a}\right|_{B,\phi}=n_a=\frac{eB}{2\pi^2}\sum_{n=0}^{n_{\text{max}}}\gamma_n\sqrt{\mu_a^2-m_a^2-2eBn},
\\
{}&\left.\frac{\partial P_{\text{f},a}}{\partial B}\right|_{\mu_a,\phi}=\frac{e}{4\pi^2}\sum_{n=0}^{n_{\text{max}}}\gamma_n\left[\mu_a\sqrt{\mu_a^2-m_a^2-2eBn}-(m_a^2+4eBn)\ln\left(\frac{\mu_a+\sqrt{\mu_a^2-m_a^2-2eBn}}{\sqrt{m_a^2+2eBn}}\right)\right],
\\
{}&\left.\frac{\partial^2 P_{\text{f},a}}{\partial\mu_a^2}\right|_{B,\phi}=\frac{eB}{2\pi^2}\sum_{n=0}^{n_{\text{max}}}\gamma_n\frac{\mu_a}{\sqrt{\mu_a^2-m_a^2-2eBn}},
\label{eq:d2Pdmu2}
\\
{}&\left.\frac{\partial P_{\text{f},a}}{\partial B\partial\mu_a}\right|_{\phi}=\frac{eB}{2\pi^2}\sum_{n=0}^{n_{\text{max}}}\gamma_n\frac{\mu_a^2-m_a^2-3eBn}{\sqrt{\mu_a^2-m_a^2-2eBn}},
\\
{}&\left.\frac{\partial^2 P_{\text{f},a}}{\partial B^2}\right|_{\mu_a,\phi}=\frac{e^2}{2\pi^2}\sum_{n=0}^{n_{\text{max}}}\gamma_nn\left[\frac{eBn\mu_a}{(m_a^2+2eBn)\sqrt{\mu_a^2-m_a^2-2eBn}}-2\ln\left(\frac{\mu_a+\sqrt{\mu_a^2-m_a^2-2eBn}}{\sqrt{m_a^2+2eBn}}\right)\right],
\label{eq:d2PdB2}
\\
{}&\left.\frac{\partial^2 P_{\text{f},\text{p}}}{\partial B\partial m_*}\right|_{\mu_a,\phi}=\frac{em_*}{2\pi^2}\sum_{n=0}^{n_{\text{max}}}\gamma_n\left[\frac{eBn\mu_{\text{p}}^*}{(m_*^2+2eBn)\sqrt{\mu_{\text{p}}^{*2}-m_*^2-2eBn}}-\ln\left(\frac{\mu_{\text{p}}^*+\sqrt{\mu_{\text{p}}^{*2}-m_*^2-2eBn}}{\sqrt{m_*^2+2eBn}}\right)\right],
\\
{}&\frac{\partial\Lagr_{\text{EH}}}{\partial B}=-\frac{em_{\text{e}}^2}{8\pi^2}\int^{\infty}_0\frac{dx}{x^2}e^{-x}\left[\text{coth}\left(x\frac{eB}{m_{\text{e}}^2}\right)-x\frac{eB}{m_{\text{e}}^2}\text{csch}^2\left(x\frac{eB}{m_{\text{e}}^2}\right)-\frac{2}{3}\left(x\frac{eB}{m_{\text{e}}^2}\right)\right],
\\
{}&\frac{\partial^2\Lagr_{\text{EH}}}{\partial B^2}=-\frac{e^2}{4\pi^2}\int^{\infty}_0\frac{dx}{x}e^{-x}\left[\text{csch}^2\left(x\frac{eB}{m_{\text{e}}^2}\right)\left\{x\frac{eB}{m_{\text{e}}^2}\text{coth}\left(x\frac{eB}{m_{\text{e}}^2}\right)-1\right\}-\frac{1}{3}\right].
\end{align}
\end{subequations}
For $n=n_{\text{max}}$, the second order partial derivatives of $P_{\text{f},a}$ for charged fermions are divergent, and so finite temperatures must be used for $n\geq n_{\text{max}}$ such that these divergences are regularized. Defining $m_{a,B}\equiv\sqrt{m_a^2+2eBn}$, we have
\begin{subequations}
\begin{align}
{}&\left.\frac{\partial^2 P_{\text{f},a}}{\partial\mu_a^2}\right|_{B,\mu_b\neq\mu_a,\phi}=\frac{eB\beta^2}{2\pi^2}\sum_{n}^{\infty}\gamma_n\int^{\infty}_{m_{a,B}}dE\sqrt{E^2-m_{a,B}^2}\exp(\beta(E-\mu_a))(\exp(\beta(E-\mu_a))-1)f_a^3(E,B,n,\mu_a),
\\
{}&\left.\frac{\partial^2 P_{\text{f},a}}{\partial B\partial\mu_a}\right|_{\mu_b\neq\mu_a,\phi}=\frac{e\beta}{2\pi^2}\sum_{n}^{\infty}\gamma_n\int^{\infty}_{m_{a,B}}dE\sqrt{E^2-m_{a,B}^2}\Bigg[\left(1-\frac{eBn}{E^2}\right)\exp(\beta(E-\mu_a))f_a^2(E,B,n,\mu_a)
\nonumber
\\
{}&\qquad\qquad\qquad\qquad\qquad\qquad\qquad\qquad\qquad\qquad\qquad\quad-\beta eBn\frac{1}{E}\exp(\beta(E-\mu_a))(\exp(\beta(E-\mu_a))-1)f_a^3(E,B,n,\mu_a)\Bigg],
\\
{}&\left.\frac{\partial^2 P_{\text{f},a}}{\partial B^2}\right|_{\mu_b\neq\mu_a,\phi}=-\frac{e^2}{2\pi^2}\sum_{n}^{\infty}\gamma_n\int^{\infty}_{m_{a,B}}dE\frac{\sqrt{E^2-m_{a,B}^2}}{E^2}\Bigg[\frac{2E^2-3eBn}{E^2}f_a(E,B,n,\mu_a)
-\beta\frac{2E^2-3eBn}{E}\exp(\beta(E-\mu_a))f_a^2(E,B,n,\mu_a)
\nonumber
\\
{}&\qquad\qquad\qquad\qquad\qquad\qquad\qquad\qquad\qquad\qquad\qquad\qquad-\beta^2 eBn \exp(\beta(E-\mu_a))(\exp(\beta(E-\mu_a))-1)f_a^3(E,B,n,\mu_a)\Bigg],
\label{eq:d2PdB2FiniteT}
\\
{}&\left.\frac{\partial^2 P_{\text{f},\text{p}}}{\partial B\partial m_*}\right|_{\mu_a,\phi}=-\frac{em_*}{2\pi^2}\sum_{n}^{\infty}\gamma_n\int^{\infty}_{m_{\text{p},B}}dE\frac{\sqrt{E^2-m_{\text{p},B}^2}}{E^2}\Bigg[\frac{E^2-3eBn}{E^2}f_{\text{p}}(E)
+\beta\frac{E^2-3eBn}{E}\exp(\beta(E-\mu_a))f_{\text{p}}^2(E)
\nonumber
\\
{}&\qquad\qquad\qquad\qquad\qquad\qquad\qquad\qquad\qquad\qquad\qquad-\beta^2 eBn \exp(\beta(E-\mu_a))(\exp(\beta(E-\mu_{\text{p}}^*))-1)f_{\text{p}}^3(E)\Bigg].
\end{align}
\end{subequations}
For the temperatures used in this paper, the contribution to these derivatives from $n>n_{\text{max}}$ are very small compared to those for $n\leq n_{\text{max}}$, and so only a few Landau levels above $n_{\text{max}}$ are needed in the finite temperature calculation.

\bibliography{library,textbooks,dynamosrefs,librarySpecial}

\begin{thebibliography}{}
\makeatletter
\relax
\def\mn@urlcharsother{\let\do\@makeother \do\$\do\&\do\#\do\^\do\_\do\%\do\~}
\def\mn@doi{\begingroup\mn@urlcharsother \@ifnextchar [ {\mn@doi@}
  {\mn@doi@[]}}
\def\mn@doi@[#1]#2{\def\@tempa{#1}\ifx\@tempa\@empty \href
  {http://dx.doi.org/#2} {doi:#2}\else \href {http://dx.doi.org/#2} {#1}\fi
  \endgroup}
\def\mn@eprint#1#2{\mn@eprint@#1:#2::\@nil}
\def\mn@eprint@arXiv#1{\href {http://arxiv.org/abs/#1} {{\tt arXiv:#1}}}
\def\mn@eprint@dblp#1{\href {http://dblp.uni-trier.de/rec/bibtex/#1.xml}
  {dblp:#1}}
\def\mn@eprint@#1:#2:#3:#4\@nil{\def\@tempa {#1}\def\@tempb {#2}\def\@tempc
  {#3}\ifx \@tempc \@empty \let \@tempc \@tempb \let \@tempb \@tempa \fi \ifx
  \@tempb \@empty \def\@tempb {arXiv}\fi \@ifundefined
  {mn@eprint@\@tempb}{\@tempb:\@tempc}{\expandafter \expandafter \csname
  mn@eprint@\@tempb\endcsname \expandafter{\@tempc}}}

\bibitem[\protect\citeauthoryear{Acheson}{Acheson}{1979}]{Acheson1979}
Acheson D.~J.,  1979, \mn@doi [Sol. Phys.] {10.1007/BF00150129}, 62, 23

\bibitem[\protect\citeauthoryear{Akg{\"{u}}n \& Wasserman}{Akg{\"{u}}n \&
  Wasserman}{2008}]{Akgun2008}
Akg{\"{u}}n T.,  Wasserman I.,  2008, \mn@doi [Mon. Not. R. Astron. Soc.]
  {10.1111/j.1365-2966.2007.12660.x}, 383, 1551

\bibitem[\protect\citeauthoryear{Akg{\"{u}}n, Reisenegger, Mastrano  \&
  Marchant}{Akg{\"{u}}n et~al.}{2013}]{Akgun2013}
Akg{\"{u}}n T.,  Reisenegger A.,  Mastrano A.,   Marchant P.,  2013, \mn@doi
  [Mon. Not. R. Astron. Soc.] {10.1093/mnras/stt913}, 433, 2445

\bibitem[\protect\citeauthoryear{Bardeen, Friedman, Schutz  \& Sorkin}{Bardeen
  et~al.}{1977}]{Bardeen1977}
Bardeen J.~M.,  Friedman J.~L.,  Schutz B.~F.,   Sorkin R.,  1977, \mn@doi
  [Astrophys. J.] {10.1017/CBO9781107415324.004}, 217, L49

\bibitem[\protect\citeauthoryear{Beloborodov}{Beloborodov}{2017}]{Beloborodov2017}
Beloborodov A.~M.,  2017, \mn@doi [Astrophys. J.] {10.3847/2041-8213/aa78f3},
  843, L26

\bibitem[\protect\citeauthoryear{Bernstein, Frieman, Kruskal  \&
  Kulsrud}{Bernstein et~al.}{1958}]{Bernstein1958}
Bernstein I.~B.,  Frieman E.~A.,  Kruskal M.~D.,   Kulsrud R.~M.,  1958,
  \mn@doi [Proc. R. Soc. London A] {10.1098/rspa.1958.0023}, 244, 17

\bibitem[\protect\citeauthoryear{Bilous et~al.,}{Bilous
  et~al.}{2019}]{Bilous2019}
Bilous A.~V.,  et~al., 2019, \mn@doi [Astrophys. J. Lett.]
  {10.3847/2041-8213/ab53e7}, 887, L23

\bibitem[\protect\citeauthoryear{Blandford \& Hernquist}{Blandford \&
  Hernquist}{1982}]{Blandford1982}
Blandford R.~D.,  Hernquist L.,  1982, \mn@doi [J. Phys. C Solid State Phys.]
  {10.1088/0022-3719/15/30/017}, 15, 6233

\bibitem[\protect\citeauthoryear{Braithwaite}{Braithwaite}{2009}]{Braithwaite2009}
Braithwaite J.,  2009, \mn@doi [Mon. Not. R. Astron. Soc.]
  {10.1111/j.1365-2966.2008.14034.x}, 397, 763

\bibitem[\protect\citeauthoryear{Braithwaite \& Nordlund}{Braithwaite \&
  Nordlund}{2006}]{Braithwaite2006}
Braithwaite J.,  Nordlund {\AA}.,  2006, \mn@doi [Astron. Astrophys.]
  {10.1051/0004-6361:20041980}, 450, 1077

\bibitem[\protect\citeauthoryear{Braithwaite \& Spruit}{Braithwaite \&
  Spruit}{2004}]{Braithwaite2004}
Braithwaite J.,  Spruit H.~C.,  2004, \mn@doi [Nature] {10.1038/nature02934},
  431, 819

\bibitem[\protect\citeauthoryear{Broderick, Prakash  \& Lattimer}{Broderick
  et~al.}{2000}]{Broderick2000}
Broderick A.,  Prakash M.,   Lattimer J.~M.,  2000, \mn@doi [Astrophys. J.]
  {10.1086/309010}, 537, 351

\bibitem[\protect\citeauthoryear{Carter}{Carter}{1973}]{Carter1973}
Carter B.,  1973, \mn@doi [Commun. Math. Phys.] {10.1007/BF01645505}, 30, 261

\bibitem[\protect\citeauthoryear{Chamel \& Stoyanov}{Chamel \&
  Stoyanov}{2020}]{Chamel2020a}
Chamel N.,  Stoyanov Z.~K.,  2020, \mn@doi [Phys. Rev. C]
  {10.1103/PhysRevC.101.065802}, 101, 65802

\bibitem[\protect\citeauthoryear{Chamel et~al.,}{Chamel
  et~al.}{2012}]{Chamel2012}
Chamel N.,  et~al., 2012, \mn@doi [Phys. Rev. C] {10.1103/PhysRevC.86.055804},
  86, 055804

\bibitem[\protect\citeauthoryear{Chodos, Everding  \& Owen}{Chodos
  et~al.}{1990}]{Chodos1990}
Chodos A.,  Everding K.,   Owen D.~A.,  1990, \mn@doi [Phys. Rev. D]
  {10.1103/PhysRevD.42.2881}, 42, 2881

\bibitem[\protect\citeauthoryear{Cumming, Arras  \& Zweibel}{Cumming
  et~al.}{2004}]{Cumming2004}
Cumming A.,  Arras P.,   Zweibel E.,  2004, \mn@doi [Astrophys. J.]
  {10.1086/421324}, 609, 999

\bibitem[\protect\citeauthoryear{Duez, Braithwaite  \& Mathis}{Duez
  et~al.}{2010}]{Duez2010}
Duez V.,  Braithwaite J.,   Mathis S.,  2010, \mn@doi [Astrophys. J. Lett.]
  {10.1088/2041-8205/724/1/L34}, 724, 34

\bibitem[\protect\citeauthoryear{Easson \& Pethick}{Easson \&
  Pethick}{1977}]{Easson1977}
Easson I.,  Pethick C.~J.,  1977, \mn@doi [Phys. Rev. D]
  {10.1103/PhysRevD.16.275}, 16, 275

\bibitem[\protect\citeauthoryear{Elmfors, Persson  \& Skagerstam}{Elmfors
  et~al.}{1993}]{Elmfors1993}
Elmfors P.,  Persson D.,   Skagerstam B.~S.,  1993, \mn@doi [Phys. Rev. Lett.]
  {10.1017/CBO9781107415324.004}, 71, 480

\bibitem[\protect\citeauthoryear{Flowers \& Ruderman}{Flowers \&
  Ruderman}{1977}]{Flowers1977}
Flowers E.,  Ruderman M.~A.,  1977, \mn@doi [Astrophys. J.] {10.1086/155359},
  215, 302

\bibitem[\protect\citeauthoryear{Friedman \& Schutz}{Friedman \&
  Schutz}{1975}]{Friedman1975}
Friedman J.~L.,  Schutz B.~F.,  1975, \mn@doi [Astrophys. J.]
  {10.1017/CBO9781107415324.004}, 200, 204

\bibitem[\protect\citeauthoryear{Friedman \& Schutz}{Friedman \&
  Schutz}{1978}]{Friedman1978}
Friedman J.~L.,  Schutz B.~F.,  1978, \mn@doi [Astrophys. J.] {10.1086/156098},
  221, 937

\bibitem[\protect\citeauthoryear{Glampedakis \& Andersson}{Glampedakis \&
  Andersson}{2007}]{Glampedakis2007}
Glampedakis K.,  Andersson N.,  2007, \mn@doi [Mon. Not. R. Astron. Soc.]
  {10.1111/j.1365-2966.2007.11625.x}, 377, 630

\bibitem[\protect\citeauthoryear{Glendenning}{Glendenning}{1997}]{Glendenning1997}
Glendenning N.~K.,  1997, {Compact} {Stars}.
Springer, New York

\bibitem[\protect\citeauthoryear{Gough \& Tayler}{Gough \&
  Tayler}{1966}]{Gough1966}
Gough D.~O.,  Tayler R.~J.,  1966, \mn@doi [Mon. Not. R. Astron. Soc.]
  {10.1093/mnras/133.1.85}, 133, 85

\bibitem[\protect\citeauthoryear{Gourgouliatos, Cumming, Reisenegger, Armaza,
  Lyutikov  \& Valdivia}{Gourgouliatos et~al.}{2013}]{Gourgouliatos2013}
Gourgouliatos K.~N.,  Cumming A.,  Reisenegger A.,  Armaza C.,  Lyutikov M.,
  Valdivia J.~A.,  2013, \mn@doi [Mon. Not. R. Astron. Soc.]
  {10.1093/mnras/stt1195}, 434, 2480

\bibitem[\protect\citeauthoryear{Heisenberg \& Euler}{Heisenberg \&
  Euler}{1936}]{Heisenberg1936}
Heisenberg W.,  Euler H.,  1936, \mn@doi [Zeitschrift f{\"{u}}r Phys.]
  {10.1007/BF01343663}, 98, 714

\bibitem[\protect\citeauthoryear{Heyl \& Hernquist}{Heyl \&
  Hernquist}{2005}]{Heyl2005}
Heyl J.~S.,  Hernquist L.,  2005, \mn@doi [Astrophys. J.] {10.1086/425974},
  618, 463

\bibitem[\protect\citeauthoryear{Kantor \& Gusakov}{Kantor \&
  Gusakov}{2014}]{Kantor2014}
Kantor E.~M.,  Gusakov M.~E.,  2014, \mn@doi [Mon. Not. R. Astron. Soc. Lett.]
  {10.1093/mnrasl/slu061}, 442, 90

\bibitem[\protect\citeauthoryear{Kaspi \& Beloborodov}{Kaspi \&
  Beloborodov}{2017}]{Kaspi2017}
Kaspi V.~M.,  Beloborodov A.~M.,  2017, \mn@doi [Annu. Rev. Astron. Astrophys.]
  {10.1146/annurev-astro-081915-023329}, 55, 261

\bibitem[\protect\citeauthoryear{Lai \& Shapiro}{Lai \&
  Shapiro}{1991}]{Lai1991}
Lai D.,  Shapiro S.~L.,  1991, \mn@doi [Astrophys. J.] {10.1086/170831}, 383,
  745

\bibitem[\protect\citeauthoryear{Landau \& Lifshitz}{Landau \&
  Lifshitz}{1960}]{Landau1960}
Landau L.~D.,  Lifshitz E.~M.,  1960, {Electrodynamics} of {Continuous}
  {Media}, 2 edn.
Pergamon Press, Oxford

\bibitem[\protect\citeauthoryear{Lander \& Jones}{Lander \&
  Jones}{2009}]{Lander2009}
Lander S.~K.,  Jones D.~I.,  2009, \mn@doi [Mon. Not. R. Astron. Soc.]
  {10.1111/j.1365-2966.2009.14667.x}, 395, 2162

\bibitem[\protect\citeauthoryear{Lu \& Kumar}{Lu \& Kumar}{2018}]{Lu2018}
Lu W.,  Kumar P.,  2018, \mn@doi [Mon. Not. R. Astron. Soc.]
  {10.1093/mnras/sty716}, 477, 2470

\bibitem[\protect\citeauthoryear{Lynden-Bell \& Ostriker}{Lynden-Bell \&
  Ostriker}{1967}]{Lynden-Bell1967}
Lynden-Bell D.,  Ostriker J.~P.,  1967, \mn@doi [Mon. Not. R. Astron. Soc.]
  {10.1093/mnras/136.3.293}, 136, 293

\bibitem[\protect\citeauthoryear{Lyubarsky}{Lyubarsky}{2014}]{Lyubarsky2014}
Lyubarsky Y.,  2014, \mn@doi [Mon. Not. R. Astron. Soc. Lett.]
  {10.1093/mnrasl/slu046}, 442, L9

\bibitem[\protect\citeauthoryear{Lyubarsky}{Lyubarsky}{2020}]{Lyubarsky2020}
Lyubarsky Y.,  2020, \mn@doi [Astrophys. J.] {10.3847/1538-4357/ab97b5}, 897, 1

\bibitem[\protect\citeauthoryear{Lyutikov}{Lyutikov}{2006}]{Lyutikov2006}
Lyutikov M.,  2006, \mn@doi [Mon. Not. R. Astron. Soc.]
  {10.1111/j.1365-2966.2006.10069.x}, 367, 1594

\bibitem[\protect\citeauthoryear{Lyutikov}{Lyutikov}{2013}]{Lyutikov2013}
Lyutikov M.,  2013, \mn@doi [Phys. Rev. E] {10.1103/PhysRevE.88.053103}, 88, 1

\bibitem[\protect\citeauthoryear{Mao, Iwamoto  \& Li}{Mao
  et~al.}{2003}]{Mao2003}
Mao G.-J.,  Iwamoto A.,   Li Z.-X.,  2003, \mn@doi [Chinese J. Astron.
  Astrophys.] {10.1088/1009-9271/3/4/359}, 3, 359

\bibitem[\protect\citeauthoryear{Markey \& Tayler}{Markey \&
  Tayler}{1973}]{Markey1973}
Markey P.,  Tayler R.~J.,  1973, \mn@doi [Mon. Not. R. Astron. Soc.]
  {10.1093/mnras/163.1.77}, 163, 77

\bibitem[\protect\citeauthoryear{Mereghetti, Pons  \& Melatos}{Mereghetti
  et~al.}{2015}]{Mereghetti2015}
Mereghetti S.,  Pons J.~A.,   Melatos A.,  2015, \mn@doi [Space Sci. Rev.]
  {10.1007/s11214-015-0146-y}, 191, 315

\bibitem[\protect\citeauthoryear{Metzger, Margalit  \& Sironi}{Metzger
  et~al.}{2019}]{Metzger2019}
Metzger B.~D.,  Margalit B.,   Sironi L.,  2019, \mn@doi [Mon. Not. R. Astron.
  Soc.] {10.1093/mnras/stz700}, 485, 4091

\bibitem[\protect\citeauthoryear{Mitchell, Braithwaite, Reisenegger, Spruit,
  Valdivia  \& Langer}{Mitchell et~al.}{2015}]{Mitchell2015}
Mitchell J.~P.,  Braithwaite J.,  Reisenegger A.,  Spruit H.,  Valdivia J.~A.,
   Langer N.,  2015, \mn@doi [Mon. Not. R. Astron. Soc.]
  {10.1093/mnras/stu2514}, 447, 1213

\bibitem[\protect\citeauthoryear{Muzikar \& Pethick}{Muzikar \&
  Pethick}{1981}]{Muzikar1981}
Muzikar P.,  Pethick C.~J.,  1981, \mn@doi [Phys. Rev. B]
  {10.1103/PhysRevB.24.2533}, 24, 2533

\bibitem[\protect\citeauthoryear{Newcomb}{Newcomb}{1961}]{Newcomb1961}
Newcomb W.~A.,  1961, \mn@doi [Phys. Fluids] {10.1063/1.1706342}, 4, 391

\bibitem[\protect\citeauthoryear{Parker}{Parker}{1955}]{Parker1955}
Parker E.~N.,  1955, \mn@doi [Astrophys. J.] {10.1017/CBO9781107415324.004},
  121, 491

\bibitem[\protect\citeauthoryear{Passamonti, Andersson  \& Ho}{Passamonti
  et~al.}{2016}]{Passamonti2016}
Passamonti A.,  Andersson N.,   Ho W. C.~G.,  2016, \mn@doi [Mon. Not. R.
  Astron. Soc.] {10.1093/mnras/stv2149}, 455, 1489

\bibitem[\protect\citeauthoryear{Persson \& Zeitlin}{Persson \&
  Zeitlin}{1995}]{Persson1995}
Persson D.,  Zeitlin V.,  1995, \mn@doi [Phys. Rev. D]
  {10.1103/PhysRevD.51.2026}, 51, 2026

\bibitem[\protect\citeauthoryear{{Popov} \& {Postnov}}{{Popov} \&
  {Postnov}}{2010}]{Popov2010}
{Popov} S.~B.,  {Postnov} K.~A.,  2010, in {Harutyunian} H.~A.,  {Mickaelian}
  A.~M.,   {Terzian} Y.,  eds, Evolution of Cosmic Objects through their
  Physical Activity. National Academy of Scieces of the Republic of Armenia,
  Yerevan, pp 129--132

\bibitem[\protect\citeauthoryear{Potekhin}{Potekhin}{1999}]{Potekhin1999a}
Potekhin A.~Y.,  1999, Astron. Astrophys., 351, 787

\bibitem[\protect\citeauthoryear{Potekhin \& Chabrier}{Potekhin \&
  Chabrier}{2018}]{Potekhin2018}
Potekhin A.~Y.,  Chabrier G.,  2018, \mn@doi [Astron. Astrophys.]
  {10.1051/0004-6361/201731866}, 609, 1

\bibitem[\protect\citeauthoryear{Potekhin \& Yakovlev}{Potekhin \&
  Yakovlev}{2001}]{Potekhin2001}
Potekhin A.~Y.,  Yakovlev D.~G.,  2001, Astron. Astrophys., 374, 213

\bibitem[\protect\citeauthoryear{Potekhin, Pons  \& Page}{Potekhin
  et~al.}{2015}]{Potekhin2015}
Potekhin A.~Y.,  Pons J.~A.,   Page D.,  2015, \mn@doi [Space Sci. Rev.]
  {10.1007/s11214-015-0180-9}, 191, 239

\bibitem[\protect\citeauthoryear{Rau \& Wasserman}{Rau \&
  Wasserman}{2018}]{Rau2018}
Rau P.~B.,  Wasserman I.,  2018, \mn@doi [Mon. Not. R. Astron. Soc.]
  {10.1093/mnras/sty2458}, 481, 4427

\bibitem[\protect\citeauthoryear{Rau \& Wasserman}{Rau \&
  Wasserman}{2021}]{RauInPrep}
Rau P.~B.,  Wasserman I.,  2021, Magnetohydrodynamic stability of magnetars in
  the ultrastrong field regime {II}: {Th}e crust, (in preparation)

\bibitem[\protect\citeauthoryear{Reisenegger}{Reisenegger}{2009}]{Reisenegger2009}
Reisenegger A.,  2009, \mn@doi [Astron. Astrophys.]
  {10.1051/0004-6361/200810895}, 499, 557

\bibitem[\protect\citeauthoryear{Reisenegger \& Goldreich}{Reisenegger \&
  Goldreich}{1992}]{Reisenegger1992}
Reisenegger A.,  Goldreich P.,  1992, \mn@doi [Astrophys. J.] {10.1086/171645},
  395, 240

\bibitem[\protect\citeauthoryear{Riley et~al.,}{Riley et~al.}{2019}]{Riley2019}
Riley T.~E.,  et~al., 2019, \mn@doi [Astrophys. J.] {10.3847/2041-8213/ab481c},
  887, L21

\bibitem[\protect\citeauthoryear{Roberts}{Roberts}{1981}]{Roberts1981}
Roberts P.~H.,  1981, \mn@doi [Q. J. Mech. Appl. Math.]
  {10.1093/qjmam/34.3.327}, 34, 327

\bibitem[\protect\citeauthoryear{Schmitt \& Shternin}{Schmitt \&
  Shternin}{2018}]{Schmitt2018}
Schmitt A.,  Shternin P.,  2018, in Rezzolla L.,  Pizzochero P.,  Jones D.~I.,
  Rea N.,   Vida{\~{n}}a I.,  eds, , The {Physics} and {Astrophysics} of
  {Neutron} {Stars}.
Springer, Heidelberg, Chapt.~9, pp 455--574

\bibitem[\protect\citeauthoryear{Schubert}{Schubert}{1968}]{Schubert1968}
Schubert G.,  1968, \mn@doi [Astrophys. J.] {10.1086/149508}, 151, 1099

\bibitem[\protect\citeauthoryear{Shapiro \& Teukolsky}{Shapiro \&
  Teukolsky}{1983}]{Shapiro1983}
Shapiro S.~L.,  Teukolsky S.~A.,  1983, {Black} {Holes}, {White} {Dwarfs} and
  {Neutron} {Stars}: {The} {Physics} of {Compact} {Objects}.
John Wiley \& Sons, New York

\bibitem[\protect\citeauthoryear{Sinha, Mukhopadhyay  \& Sedrakian}{Sinha
  et~al.}{2013}]{Sinha2013}
Sinha M.,  Mukhopadhyay B.,   Sedrakian A.,  2013, \mn@doi [Nucl. Phys. A]
  {10.1016/j.nuclphysa.2012.12.076}, 898, 43

\bibitem[\protect\citeauthoryear{Suh \& Mathews}{Suh \&
  Mathews}{2010}]{Suh2010}
Suh I.~S.,  Mathews G.~J.,  2010, \mn@doi [Astrophys. J.]
  {10.1088/0004-637X/717/2/843}, 717, 843

\bibitem[\protect\citeauthoryear{Taub}{Taub}{1969}]{Taub1969}
Taub A.~H.,  1969, \mn@doi [Commun. Math. Phys.] {10.1007/BF01073578}, 15, 235

\bibitem[\protect\citeauthoryear{Tayler}{Tayler}{1973}]{Tayler1973}
Tayler R.~J.,  1973, \mn@doi [Mon. Not. R. Astron. Soc.]
  {10.1093/mnras/161.4.365}, 161, 365

\bibitem[\protect\citeauthoryear{{The CHIME/FRB Collaboration:} et~al.,}{{The
  CHIME/FRB Collaboration:} et~al.}{2020}]{CHIMEFRB2020}
{The CHIME/FRB Collaboration:} et~al., 2020, \mn@doi [Nature]
  {10.1038/s41586-020-2863-y}, 587, 54

\bibitem[\protect\citeauthoryear{Thompson \& Duncan}{Thompson \&
  Duncan}{1995}]{Thompson1995}
Thompson C.,  Duncan R.~C.,  1995, \mn@doi [Mon. Not. R. Astron. Soc.]
  {10.1093/mnras/275.2.255}, 275, 255

\bibitem[\protect\citeauthoryear{Turolla, Zane  \& Watts}{Turolla
  et~al.}{2015}]{Turolla2015}
Turolla R.,  Zane S.,   Watts A.~L.,  2015, \mn@doi [Reports Prog. Phys.]
  {10.1088/0034-4885/78/11/116901}, 78, 116901

\bibitem[\protect\citeauthoryear{Uryu, Yoshida, Gourgoulhon, Markakis,
  Fujisawa, Tsokaros, Taniguchi  \& Eriguchi}{Uryu et~al.}{2019}]{Uryu2019}
Uryu K.,  Yoshida S.,  Gourgoulhon E.,  Markakis C.,  Fujisawa K.,  Tsokaros
  A.,  Taniguchi K.,   Eriguchi Y.,  2019, \mn@doi [Phys. Rev. D]
  {10.1103/PhysRevD.100.123019}, 100, 123019

\bibitem[\protect\citeauthoryear{Walecka}{Walecka}{1995}]{Walecka1995}
Walecka J.~D.,  1995, {Theoretical} {Nuclear} and {Subnuclear} {Physics}.
Oxford University Press, New York

\bibitem[\protect\citeauthoryear{Wright}{Wright}{1973}]{Wright1973}
Wright G. A.~E.,  1973, \mn@doi [Mon. Not. R. Astron. Soc.]
  {10.1093/mnras/162.4.339}, 162, 339

\bibitem[\protect\citeauthoryear{Yoshida, Yoshida  \& Eriguchi}{Yoshida
  et~al.}{2006}]{Yoshida2006}
Yoshida S.,  Yoshida S.,   Eriguchi Y.,  2006, \mn@doi [Astrophys. J.]
  {10.1086/507513}, 651, 462

\bibitem[\protect\citeauthoryear{Yu \& Weinberg}{Yu \& Weinberg}{2017}]{Yu2017}
Yu H.,  Weinberg N.~N.,  2017, \mn@doi [Mon. Not. R. Astron. Soc.]
  {10.1093/mnras/stw2552}, 464, 2622

\makeatother
\end{thebibliography}

\bsp	
\label{lastpage}
\end{document}